\documentclass{aa}  
\pdfoutput=1

\usepackage{amsmath}
\usepackage{graphicx}
\usepackage{caption}
\usepackage[varg]{txfonts}
\usepackage{gensymb}
\usepackage{longtable, lscape}
\usepackage{natbib}
\usepackage{lineno}
\usepackage{xcolor}
\usepackage{hyperref} 
\hypersetup{backref=true,       
    pagebackref=true,               
    hyperindex=true,                
    colorlinks=true,                
    breaklinks=true,                
    urlcolor= black,                
    linkcolor= blue,                
    bookmarks=true,                 
    bookmarksopen=false,
    filecolor=black,
    citecolor=blue,
    linkbordercolor=blue
}

\newcommand\blfootnote[1]{
  \begingroup
  \renewcommand\thefootnote{}\footnote{#1}
  \addtocounter{footnote}{-1}
  \endgroup
}

\begin{document}

	\title{Hard X-ray properties of radio-selected blazars}
   	\author{M. Langejahn\inst{1},
          M. Kadler\inst{1},
          J. Wilms\inst{2},
          E. Litzinger\inst{1, 2},
          M. Kreter\inst{3},
          N. Gehrels\inst{4}$^\dagger$,
          W.~H.~Baumgartner\inst{5},
          C.~B.~Markwardt\inst{4},
          J. Tueller\inst{4}$^\dagger$ 
          }
	\authorrunning{M. Langejahn et al.}
          
   \institute{Lehrstuhl f\"ur Astronomie, Universit\"at W\"urzburg, Emil-Fischer Str.~31, 97074  W\"urzburg, Germany\\
         \and
             Dr.\ Karl Remeis-Sternwarte, Universit\"at Erlangen-N\"urnberg, and Erlangen Centre for Astroparticle Physics, Sternwartstr.~7, 96049 Bamberg, Germany\\
       	 \and
       		 Centre for Space Research, North-West University, Private Bag X6001, Potchefstroom 2520, South Africa\\
         \and
             NASA, Goddard Space Flight Center, Astrophysics Science Division, Greenbelt, MD 20771, USA\\         
       	 \and
             NASA, Marshall Space Flight Center, Science Research Office, Huntsville, AL 35812, USA\\
             }

   \date{Received 9 January 2020 / Accepted 21 March 2020}

  \abstract
  % context heading (optional), {} leave it empty if necessary
   {Hard X-ray properties of beamed AGN have been published in the 105-month \textit{Swift}/BAT catalog, but there have not been any studies carried out so far on a well-defined, radio-selected sample of low-peaked blazars in the hard X-ray band.}
  % aims heading (mandatory)
   {Using the statistically complete MOJAVE-1 sample, we aim to determine the hard X-ray properties of radio-selected blazars, including the enigmatic group of gamma-ray-faint blazars. Additionally, we aim to determine the contribution of radio-selected low-peaked blazars to the diffuse CXB.}
  % methods heading (mandatory)
   {We determined photon indices, fluxes, and luminosities in the range of 20\,keV -- 100\,keV of the X-ray spectra of blazars and other extragalactic jets from the MOJAVE-1 sample, derived from the 105-month \textit{Swift}/BAT survey. We calculated $\log N$-$\log S$ distributions and determined the luminosity functions.}
  % results heading (mandatory)
   {The majority of the MOJAVE-1 blazars are found to be hard X-ray emitters albeit many at low count rates. The $\log N$-$\log S$ distribution for the hard X-ray emission of radio-selected blazars is clearly non-Euclidean, in contrast to the radio flux density distribution. Approximately 0.2\% of the CXB in the 20\,keV -- 100\,keV band can be resolved into MOJAVE-1 blazars.} 
   % conclusions
   {The peculiar $\log N$-$\log S$ distribution disparity might be attributed to different evolutionary paths in the X-ray and radio bands, as tested by luminosity-function modeling. X-ray variability can be ruled out as the dominant contributor. Low-peaked blazars constitute an intrinsically different source population in terms of CXB contribution compared to similar studies of X-ray-selected blazars. The hard X-ray flux and spectral index can serve as a good proxy for the gamma-ray detection probability of individual sources. Future observations combining deep X-ray survey, for example, with eROSITA, and targeted gamma-ray observations with CTA can benefit strongly from the tight connection between these high-energy bands for the different blazar sub-classes.}

   \keywords{Galaxies: active --
                Methods: statistical --
                X-rays: galaxies
               }
   \maketitle

\section{Introduction}

\blfootnote{$^\dagger$Deceased}
At the very low and high end of the electromagnetic spectrum, in the radio and gamma-ray regime, the extragalactic sky is dominated by highly beamed jets from active galactic nuclei \citep[AGN, see, e.g.,][]{Becker1995,Ackermann2015}. Blazars are a particularly variable and luminous AGN group and the subject of multiple monitoring campaigns at all accessible wavelengths \citep[e.g.,][]{Villata2008,Ojha2010,Jorstad2016}. Aside from their prominent occurrence at radio and gamma-ray energies, blazars, and especially low-peaked blazars, are notoriously hard to detect in the hard X-rays, just above 10\,keV to 20\,keV. This is largely due to the characteristic spectral minimum between the low-energy synchrotron bump and the high-energy (HE) emission bump at MeV to TeV energies.
Also, currently operating hard X-ray instruments such as \textit{INTEGRAL} and \textit{Swift}/BAT are usually background-dominated and do not reach the flux sensitivity of soft X-ray observatories by several orders of magnitude \citep[see, e.g.,][]{Bottacini2012, Wang2016}. 

Because of the proximity of the keV to the gamma-ray range, multiwavelength studies of blazars rely on this sampling point in order to model the start of the HE bump of a blazar's spectral energy distribution (SED). Particularly, the modeling of SEDs of the enigmatic group of gamma-ray-faint blazars requires good coverage of the neighboring spectral range.
Past studies \citep[e.g.,][]{Giommi2006,Ajello2009,Draper2009} suggest that between soft X-ray and MeV energies, blazars become the dominating contributor to the diffuse cosmic X-ray background.  

Previous all-sky surveys and studies at hard X-rays with the \textit{Swift}/BAT instrument featured a moderately high significance cutoff at approximately $5\,\sigma$, effectively hiding a large number of blazars, and consequently leading to small sample sizes \citep[see, e.g.,][]{baumgartner2013,Krivonos2015}.
The 105-month BAT survey catalog \citep{Oh2018} counts 158 ``beamed AGN'', including blazars and flat-spectrum radio quasars (FSRQ), out of 1632 sources, including 114 unclassified AGN and an additional 129 unidentified sources. The recent 4FGL catalog of the gamma-ray all-sky survey from Fermi/LAT \citep{4FGL2019} alone contains 1102 BL Lac type and 681 FSRQ type blazars out of a total of 2940 associated blazars and blazar candidates.

In this work, we perform a detailed analysis of X-ray data from the most recent \textit{Swift}/BAT survey maps for the radio-selected MOJAVE-1 (Monitoring Of Jets in Active galactic nuclei with VLBA Experiments) beamed AGN sample. We report the hard X-ray characteristics in the 20\,keV -- 100\,keV band of this statistically complete extragalactic jet sample, which is mostly composed of blazars. 

This paper is structured as follows: Sect.~\ref{sec:sam} gives a short introduction to the MOJAVE program as well as the \textit{Swift} mission and describes the object samples and data basis we used. The analysis, including the applied methods and assumptions, is presented in Sect.~\ref{sec:ana}. The results, including the statistical properties of the entire sample and distributions for flux and luminosity, are presented in Sect.~\ref{sec:res}. Several key aspects of this study are discussed in Sect.~\ref{sec:disc}, such as the peculiar hard X-ray properties of individual sources, the potential \textit{Fermi}/LAT detection of gamma-faint blazars, and the $\log N$-$\log S$ distribution of the MOJAVE-1 blazars. A short summary of the main results and conclusions can be found in Sect.~\ref{sec:conc}. The derived data set for all spectral fits for all sources is presented in Appendix \ref{tab:app}.

\section{Object samples}
\label{sec:sam}

\subsection{The MOJAVE-1 sample and Swift}

The MOJAVE program\footnote{http://www.physics.purdue.edu/astro/MOJAVE/} provides continuous interferometric measurements of the radio-brightest AGN in the northern hemisphere with VLBI \citep{Lister2009_137}. We concentrate on the flux-limited MOJAVE-1 sample, counting 135 sources \citep{Lister2005}. The selection criteria for this sample were: $\delta\ge -20^\circ$, Galactic latitude $\textbar b\textbar\,\geq\,2\fdg5$, a total 2\,cm (15\,GHz) flux density greater than 1.5\,Jy at any epoch between 1994.0 and 2004.0, and more than 2\,Jy for sources below the celestial equator.
This northern AGN sample can be considered statistically complete in terms of 15\,GHz radio flux density due to strict selection criteria applied over a large amount of time, as well as the extensive flux density database that was used \citep[][and references therein]{Lister2009_137}.
Following the optical classification scheme of \cite{Veron2003}, the sources of this sample can be divided into 104 FSRQs, eight radio galaxies, 21 BL Lacs, and two unidentified objects with no known optical counterpart. 
The source 1219+044 has been re-classified as a NLSY1 galaxy by \citet{Yao2015}. We keep the original FSRQ classification \citep{Hovatta2014} since the intrinsic X-ray luminosity and photon index are notably atypical for the MOJAVE-1 sources classified radio galaxy, and very similar to the FSRQ classification. In general, for the naming of all sources the IAU B1950 coordinate format is applied.

In November 2004, the \textit{Neil Gehrels Swift Observatory} \citep{Swift} was launched by NASA, bringing a multiwavelength observatory in low-Earth orbit that is dedicated to the detection and study of gamma-ray bursts (GRBs). The \textit{Swift} satellite is equipped with the optical/UV telescope UVOT \citep[][170\,nm $-$ 600\,nm]{Roming2005}, the narrow-field X-ray telescope XRT \citep[][0.2\,keV -- 10\,keV]{Burrows2005}, and the wide-field coded-mask system, the burst alert telescope \citep[BAT,][14\,keV -- 195\,keV]{BAT}. The purpose of BAT, aside from GRB detection, is the monitoring of hard X-ray light curves for a list of transient and variable sources, as well as a continuous blind all-sky survey. The most recent catalog of galactic and extragalactic hard X-ray sources from \textit{Swift}/BAT for the first 105 months of operation reaches a flux level of $7.24 \cdot 10^{-12}$ $\mathrm{erg s^{-1} cm^{-2}}$ over 50\% of the sky and $8.40 \cdot 10^{-12}$ $\mathrm{erg s^{-1} cm^{-2}}$ over 90\% of the sky at very uniformly distributed exposure times \citep{Oh2018}. 

In previous versions of the \textit{Swift}/BAT all-sky survey catalog, only sources above a Crab-weighted signal-to-noise ratio (S/N) of 4.8\,$\sigma$ found in a blind search were included due to the dominance of local noise below this threshold \citep{Tueller2008,baumgartner2013,Oh2018}. Consequently, sources that are expected to have low emission in the hard X-ray range of BAT are likely to be under-represented in the catalogs.

We performed the extraction of spectra and Crab-weighted S/Ns using the \textit{Swift}/BAT 8-band all-sky mosaic images of 105 months of survey data. 
Following the procedure described by \cite{baumgartner2013}, the extraction of fluxes of each sample source is performed with an exclusion zone of $40.^\prime 5$ (15 pixels) around every target using the \textit{ftools} software \textit{batcelldetect}. The associated errors are determined by the rms value in the surrounding map area with the radius of $4\fdg5$ (100 pixels). Crab-weighted S/N values are computed for all source coordinates using Crab-weighted versions of the all-sky images.

\subsection{Comparison with other catalogs}

We compare the MOJAVE-1 sample with other catalogs in the hard X-ray and gamma-ray regime in terms of detection statistics in order to assert the nature of spectral properties in the hard X-rays, as well as selection biases of other surveys. For this purpose we use the \textit{Swift}/BAT 105-month source catalog \citep{Oh2018}, the \textit{INTEGRAL}/IBIS AGN survey catalog  \citep{Malizia2012}, the \textit{INTEGRAL}/IBIS 11-year survey catalog \citep{Krivonos2015}, the \textit{Fermi}/LAT 4-year AGN catalog \citep[3LAC]{Ackermann2015}, and the \textit{Fermi}/LAT 8-year source catalog \citep[4FGL]{4FGL2019}.

In order to determine the number of common sources in the BAT 105-month source catalog and the MOJAVE-1 sample, the empirical relation of 90\% error radius and signal strength is applied as a guideline \citep{Oh2018},
\begin{linenomath*}\begin{equation}
	r_{\mathrm{error}}(\mathrm{'})=\left(\left(\frac{30.5}{\mathrm{S/N}} \right)^2+0.1^2 \right)^{\frac{1}{2}}.
	\label{eq.r_error}
\end{equation}\end{linenomath*} 
Within the source-specific X-ray error radius we find 36 sources that are common between the two samples. We estimate false and missing associations as very unlikely, since there is a significant difference (factor of about 100) in angular separation between the coordinates of BAT counterpart coordinates and MOJAVE source positions, for sources marked associated and unassociated, respectively. The largest separation of an associated source is $0.^\prime03$ (2201+315), while the mean separation of unassociated MOJAVE-1 sources to the nearest BAT catalog source is $3\fdg11$, with a standard deviation of $1\fdg65$. Based on the angular separation of 2201+315 we calculate the probability of $4 \cdot 10^{-8}$ of a pure chance association between both sub-samples.
Objects with moderately high BAT S/N values, larger than approximately 4\,$\sigma$, are registered in both catalogs\footnote{The FSRQ 1502+106 is an exception because of source confusion with the bright Seyfert galaxy Mrk 841}.

The 36 MOJAVE-1 sources in the 105-month BAT catalog consist of 28 FSRQs, five radio galaxies, and three BL Lacs. The vast majority of the 36 common sources between both catalogs is classified as ``beamed AGN'' in the BAT 105-month catalog. Only two sources, both radio galaxies, are listed as Seyfert galaxies: 0415+379 (3C 111) and 1957+405 (Cygnus A) according to the classification in the catalog. 

We compare the positional information of the MOJAVE-1 sources with another deep hard X-ray survey, the \textit{INTEGRAL}/IBIS 11-year survey \citep{Krivonos2015}, consisting of all-sky measurements above 100\,keV. 
From 35 AGN that have been detected by \textit{INTEGRAL} above 4\,$\sigma$ in the 100\,keV to 150\,keV band, six sources are shared with the MOJAVE-1 sample: three FSRQs and two radio galaxies with BAT S/N values all larger than 37\,$\sigma$, and the FSRQ 1219+044 with a BAT S/N of 13\,$\sigma$. In general, common sources in the MOJAVE-1 sample and the \textit{INTEGRAL} survey have the highest BAT S/N values in the MOJAVE-1 source list. The distribution of BAT S/N values of the MOJAVE-1 sample is described in Sect.~\ref{subsec:res:SNR} in more detail.

The \textit{INTEGRAL}/IBIS AGN catalog \citep{Malizia2012} lists 272 AGN that were observed in the X-ray bands of 2\,keV -- 10\,keV and 20\,keV -- 100\,keV, of which 57\% are classified as Type 1 AGN, including blazars, Seyferts 1 to 1.5 as well as several sources of mixed classification such as 3C 273 (classified as Sy1/QSO). Fourteen sources are present in both the MOJAVE-1 and the \textit{INTEGRAL}/IBIS AGN catalog, five of them categorized as radio galaxies in the MOJAVE-1 sample and Seyfert or QSO Type 2 in the \textit{INTEGRAL} sample. The nine remaining sources are classified as blazars. 

The most comprehensive all-sky survey at gamma-ray energies has been performed by the \textit{Fermi}/LAT instrument, resulting in the 4FGL catalog (50\,MeV -- 1\,TeV), which includes 5098 sources above a significance level of 4\,$\sigma$, 2940 of them blazars or blazar candidates. The 4FGL catalog shares 112 sources with the MOJAVE-1 sample, including 86 FSRQs, 21 BL Lacs, three radio galaxies, and two unidentified types at a broad range of BAT S/N values. 

Whereas the fraction of sources in the MOJAVE-1 sample with a counterpart in the 4FGL catalog is 83\%, the fraction in the beamed AGN sub-sample of the BAT 105-month catalog is only 65\%. 
A significant correlation between radio flux densities and gamma-ray flux measured by \textit{Fermi}/LAT has been found since its operation \citep[e.g.,][]{Ackermann2011,Mufakharov2015}. The relation of the detection statistic in the LAT survey of X-ray-bright blazars is discussed further in Sect.~\ref{subsec:res:gamma}. For the sake of comparing gamma-ray detections, we also include data from the previous 3LAC catalog, which shares 100 sources with the MOJAVE-1 sample but also features an integration time frame that is much more compatible with the BAT data (2008 -- 2012 and 2004 -- 2013, respectively). 
All common sources between the MOJAVE-1 sample and the aforementioned catalogs are marked in Table~\ref{tab:app}.

\section{Analysis}
\label{sec:ana}

\subsection{Spectral fitting}
\label{subsec:spectralfitting}

The full spectral range of the BAT survey comprises 14\,keV -- 195\,keV. Due to the background-dominated character of the BAT instrument, especially for sources with low S/N values around 2\,$\sigma$ -- 3\,$\sigma$, the first spectral bin (14\,keV -- 20\,keV) is characterized by very low or even negative count rates. Likewise, the last two bins (100\,keV -- 150\,keV, 150\,keV -- 195\,keV) feature the very low count rates at high energies. Consequently, we only fit the reduced energy range of 20\,keV to 100\,keV for all sample sources.
We model the \textit{Swift}/BAT spectra, extracted from the 105-month survey maps, with the XSPEC model \textit{pegpwlw} that is a simple power law with norm $K$ in the aforementioned fixed energy band, expressed as
\begin{linenomath*}\begin{equation}
f(E)=K E^{-\Gamma},
\label{eqn:power}
\end{equation}\end{linenomath*}
with the photon index $\Gamma$. Spectral fitting of the BAT spectra in the previous object catalogs \citep{Tueller2008,baumgartner2013,Oh2018} made use of the $\chi^2$ fitting statistic. Since this fitting statistic assumes Gaussian-distributed data with at least about 30 counts per energy bin, fitting of sources with lower numbers of counts is not feasible. Many of the BAT spectra that are analyzed in this work are low count spectra and, therefore, need to be treated in a different way.
Fitting spectra with very low count rates is typically done using the maximum likelihood-based statistic for Poisson-distributed data according to \citet{cash1979}.
However, this only accounts for counts that can be described by Poisson statistics with no background, which can usually be modeled separately. BAT spectra are count rate spectra already subtracted by the dominant instrument background. In order to model Poisson-distributed counts of the source itself and a Gaussian-distributed background we use the recommended profile likelihood statistic PGSTAT from the XSPEC statistics appendix\footnote{https://heasarc.gsfc.nasa.gov/xanadu/xspec/manual/XSappendix Statistics.html} with zero background counts,
\begin{linenomath*}\begin{equation}
PG = 2 \sum_{i=0}^N t_s (m_i+f_i) - S_i \ln(t_s m_i + t_s f_i) + \frac{1}{2 \sigma_i^2} t_b^2 f_i^2 - S_i (1 - \ln S_i) \, ,
\end{equation}\end{linenomath*} \label{eq:PG}
where
\begin{linenomath*}\begin{equation}
f_i = \frac{-t_s \sigma_i^2 - t_b^2 m_i + d_i}{2 t_b^2},
\end{equation}\end{linenomath*}
and
\begin{linenomath*}\begin{equation}
d_i = \sqrt{(t_s \sigma_i^2 + t_b^2 m_i)^2 - 4 t_b^2(t_s \sigma_i^2 m_i - S_i \sigma_i^2)} \, ,
\end{equation}\end{linenomath*}
with the observed rates $S_i$, the corresponding uncertainties $\sigma_i$, and the predicted rates $m_i$ per channel $i$. The exposure times for source and background are $t_s$ and $t_b$, respectively.

A spectral fit is performed for 77 of all 135 sources in the energy range of 20\,keV -- 100\,keV (five energy channels). The spectra of all remaining sources possess at least one energy channel with a negative number of counts, which is the result of the background subtraction of these already faint sources. The second logarithmic term in Eq.~\ref{eq:PG} does not permit a spectral fit in these cases. Instead, the flux is determined using a simulated spectrum with the equal number of counts of the real spectrum and a photon index that has been frozen. The applied photon index is derived as a weighted mean of all photon indices of the fitted spectra in the sample of the same object class (FSRQ, BL Lac, or radio galaxy). For the group of unidentified sources the index of FSRQs is applied, which is the largest sub-group in the sample. Since sources of very high significance and low uncertainty heavily dominate the weighted mean only sources below a significance of 20\,$\sigma$ were used in calculating the mean. The template indices are listed in Table \ref{tab:number_sources1}.

\begin{table}
\caption{Optical classifications of MOJAVE-1 sources after \cite{Veron2003}.}             % title of Table
\label{tab:number_sources1}
\centering
\begin{tabular}{ccccc}
\hline\hline
		AGN Class 	& Total			& Fit\tablefootmark{a} & $\Gamma_{\mathrm{template}}$\tablefootmark{b}	& $\langle z\rangle$\tablefootmark{c}\\	
\hline                        
        FSRQ      			& 104		& 60     		& $1.51 \pm 0.05$		& 1.19 $\pm$ 0.08\\ 
        BL Lac      		& 21	 	& 11            & $1.57 \pm 0.14$		& 0.308 $\pm$ 0.053\\ 
        Galaxy      		& 8			& 6               & $1.86 \pm 0.17$		& 0.0417 $\pm$ 0.0123\\
        Unidentified 		& 2			& 0				  & $1.51 \pm 0.05$		&\\
\hline                                   
\end{tabular}
\tablefoot{
\tablefoottext{a}{Number of sources that are fit by a power-law.}
\tablefoottext{b}{Derived template photon indices of the fitted source spectra (see text).}
\tablefoottext{c}{Mean redshift of sources in the fitted sub-sample.}
}
\end{table}

Error bars corresponding to 90\% uncertainty ranges for flux and photon indices are calculated for each source via a Monte Carlo approach: the derived values for X-ray flux, photon index, and the \textit{Swift}/BAT survey response file\footnote{https://swift.gsfc.nasa.gov/results/bs105mon/inc/data/\\swiftbat\_survey\_full.rsp} are used to simulate a new ideal BAT spectrum. A new and randomized spectrum is then created using the \textit{ftools} program \textit{batphasimerr}. Both the flux and photon index of a new power-law fit are then determined. This procedure is repeated 2000 times for every source. The resulting distributions of flux and photon index are Gaussian-shaped. The 90\% error ranges are then determined using a Gaussian fit function.

The uncertainties of the spectral parameters for sources with negative spectral counts are derived from the count rate error bars of every energy channel as well as the uncertainty of each template photon index. If the flux of a source in any case is less than the error bar of the flux corresponding to 3\,$\sigma$, it is considered an upper limit.

Spectral contamination of three nearby X-ray sources also necessitates the calculation of an upper limit flux value. The ratio of contamination rate (estimated count rate of all nearby sources from \textit{batcelldetect}) and count rate is defined as the contamination ratio. We adopt the convention by \citet{baumgartner2013}, where a source is considered confused or contaminated if the contamination fraction surpasses a value of 0.02.

\subsection{Source counts and CXB contribution}
\label{sec:lognlogs}

The number-count diagram, or $\log N$-$\log S$, conveys a number of statistical properties of the flux emission of any given sample as well as possible selection effects.
The cumulative $\log N$-$\log S$ distribution with the number of sources per square degree with fluxes greater than $F_j$ of source $j$ is defined as:
\begin{linenomath*}\begin{equation}
N(>F_j) = \frac{\Omega_{\mathrm{survey}}}{\Omega_{\mathrm{sky}}} \sum_{i=0}^N \frac{1}{\Omega_j},
\label{eqn:logn_logs}
\end{equation}\end{linenomath*}
with the sky area $\Omega_j$ corresponding to the flux $F_j$, the total sky area $\Omega_{\mathrm{sky}}$, and the total sky area covered by the MOJAVE-1 survey $\Omega_{\mathrm{survey}}$\footnote{We continue to use ``$F$'' for the source flux, although many other studies use ``$S$'' for flux in $\log N$-$\log S$ diagrams}. Based on Poisson statistics the error bars are calculated via:
\begin{linenomath*}\begin{equation}
N_{\mathrm{err}}(>F_j) = \frac{N(>F_j)}{\sqrt{N}}.
\label{eqn:logn_logs_err}
\end{equation}\end{linenomath*}
We use the $\log N$-$\log S$ statistic to compare the sample source density distribution in space to the Euclidean distribution of flat space, that is described by $N=A F^{-\alpha}$ with $\alpha=1.5$. 
Any divergence from this shape indicates an evolution of the source flux in this sample and/or a source detection bias \citep[see, e.g,][]{Longair1966,Mateos2008}.

From the fitted $\log N$-$\log S$ diagram we also derive the contribution of blazars to the cosmic X-ray background (CXB) that has been studied in a number of cases in the past \citep[e.g.][]{Gilli2007,Beckmann2006,Ajello2009}. Depending on the specific energy range, the CXB can be resolved into different AGN sub-classes. Studies of the blazar contribution above 10\,keV suffer from low sample sizes and are often limited by the low flux of these sources \citep[see, e.g.,][]{Ajello2008_bat_III, Ajello2009}. Other studies rely on the extrapolation of the contribution to the CXB from lower energies \citep[see, e.g.,][]{Giommi2006}. Results therefore vary considerably with the used blazar sample and analytic approach. A contribution of approximately 1.5\% in the 14\,keV -- 170\,keV range and 10\% -- 20\% for the 15\,keV -- 55\,keV range has been reported by \citet{Ajello2008_bat_III} and \citet{Ajello2009}, respectively.
The role of blazars in the creation of the CXB above its maximum flux at approximately 30\,keV \citep{Gruber1999} remains subject of discussion. Previous studies have addressed samples of X-ray-bright blazars usually detected with S/N values larger than $5\,\sigma$ \citep[e.g.,][]{Sazonov2007,Ajello2009,Bottacini2012}. 
Here, we incorporate a well studied blazar sample, defined by long-term radio surveillance and a broad X-ray S/N distribution. 
Based on the known blazar positions of the MOJAVE-1 sample we measure their flux distribution to the CXB in the energy range of 20\,keV -- 100\,keV.

In order to determine the flux $F_{\text{CXB}}^{\mathrm{deg^2}}$ per sky area of the X-ray background itself in this range we integrate the CXB spectrum given by Eq.~5 of \citet[][]{Ajello2008_cxb_albedo} and obtain:
\begin{linenomath*}\begin{equation}
F_{\mathrm{CXB,\,20 - 100\,keV}}^{\mathrm{deg^2}} = 3.14 \cdot 10^{-11} \mathrm{erg \, s^{-1} cm^{-2} deg^{-2}}.
\label{eqn:intensity_cxb}
\end{equation}\end{linenomath*}
To compare the results of this analysis with the flux data of the beamed AGN class in the most recent 105-month BAT source catalog we also calculate the X-ray background flux in the full BAT range of 14\,keV -- 195\,keV, and obtain:
\begin{linenomath*}\begin{equation}
F_{\mathrm{CXB,\,14 - 195\,keV}}^{\mathrm{deg^2}} = 4.61 \cdot 10^{-11} \mathrm{erg \, s^{-1} cm^{-2} deg^{-2}}.
\label{eqn:intensity_cxb2}
\end{equation}\end{linenomath*}
The contributing flux of a given source sample can be estimated by integrating the differential $\log N$-$\log S$ distribution, that is, the derivative of the function $N(>F)$, multiplied with the source flux \citep[see, e.g.,][]{Giommi2006},
\begin{linenomath*}\begin{equation}
F_{\mathrm{contrib}}^{\mathrm{deg^2}} = \int_{F_{\mathrm{min}}}^{F_{\mathrm{max}}} \frac{\mathrm{d}N}{\mathrm{d}F} F dF
\label{eqn:contrib_cxb}
\end{equation}\end{linenomath*}
with $F_{\text{min}}$ and $F_{\text{max}}$ usually describing the lower limit and upper limits of source fluxes in the sample. The integral represents the contribution of flux of all sources (of this population in the sky) brighter than the minimum flux $F_{\text{min}}$. This integration limit is usually equal to the smallest source flux in the sample, or can be extrapolated to lower fluxes, albeit under the assumption of a constant slope of the $\log N$-$\log S$ distribution. 

\begin{figure}
\centering
\includegraphics[scale=0.55]{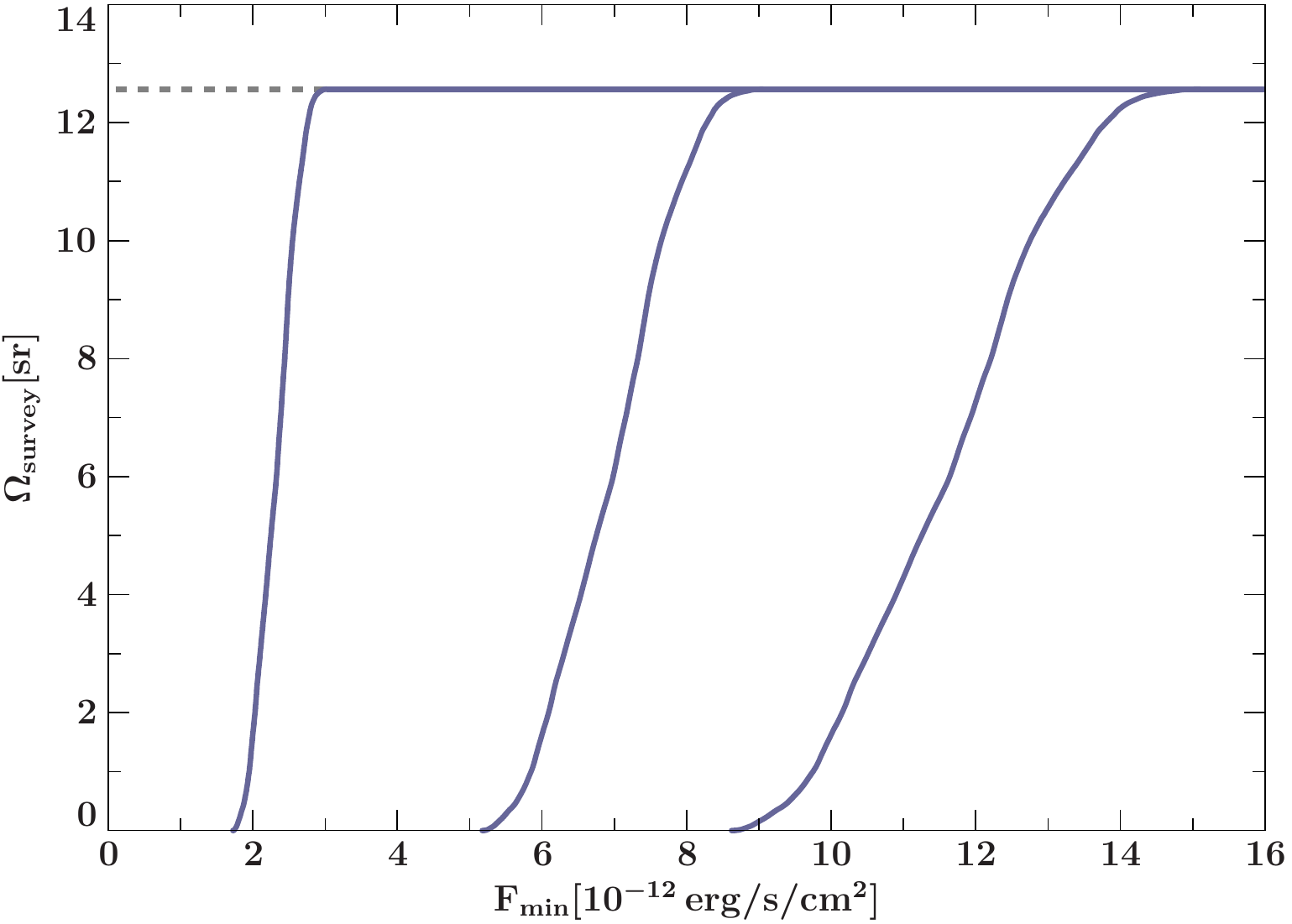}
\caption{\textit{Swift}/BAT 105 month survey sky coverage as the function of minimal flux corresponding to a detection at a level of 1, 3, and 5\,$\sigma$ in the band 20\,keV -- 100\,keV, from the left to the right. The dashed line indicates the full sky.}
         \label{fig:sky_cov_curve}
\end{figure}

In order to determine the $\log N$-$\log S$ distribution, we compute the empirical sky survey area as a function of the minimal flux that corresponds to a detection. We extract the function from the BAT 105-month survey maps for a minimal flux corresponding to a detection at 1\,$\sigma$, 3\,$\sigma$, and 5\,$\sigma$ in the 20\,keV -- 100\,keV band (graphed in Fig.~\ref{fig:sky_cov_curve}). For the calculation of the $\log N$-$\log S$ distribution of the MOJAVE-1 sample in the BAT band we choose the 1\,$\sigma$ curve that has been extracted from the survey maps. Although a measured signal equivalent to 1\,$\sigma$ above the background can hardly be called a detection this criterion only applies to blind surveys, unlike the procedure described here: the position of every source in the sample is already known and, therefore, the effective survey sky area at a given flux must be higher in any case.

\subsection{X-ray and radio luminosity function}
\label{sec:xlf_analy}

The shape of the $\log N$-$\log S$ diagram of any given source sample is related to the distribution of the sources in space. However, the data are possibly influenced by a number of factors such as selection effects or an intrinsic evolution of the emission in the observed band. In order to understand the shape of the $\log N$-$\log S$ distribution of the MOJAVE-1 blazar sample and possibly responsible biases and intrinsic flux evolution we calculate the hard X-ray and radio luminosity functions (LF). 
The differential luminosity function reveals the density of the sample sources per unit redshift and luminosity. Our approach closely follows the methods implemented by \citet{Ajello2009}, \citet{Ebrero2009}, and \citet{Miyaji2015}. The goal of this part of the analysis is to compare models of positive evolution of luminosity with a no-evolution scenario for the X-ray and radio bands.

The binned LF is determined by the number of sources $N$ in a logarithmic luminosity bin $L_{\text{bin,min}}-L_{\text{bin,max}}$ divided by the co-moving volume $\mathrm{d}V/\mathrm{d}z$ \citep[see, e.g.,][]{Hogg1999}, integrated over the total luminosity range of the sample $L_{\text{min}}-L_{\text{max}}$ and the redshift range $z_{\text{min}}-z_{\text{max}}$:
\begin{linenomath*}\begin{equation}
\frac{\mathrm{d}\Phi}{\mathrm{d}\log L} = \frac{N}{\int^{L_{\text{max}}}_{L_{\text{min}}} \int^{z_{\text{max}}}_{z_{\text{min}}} \frac{\mathrm{d}V}{\mathrm{d}z} ~ \mathrm{d}z ~ \mathrm{d}\log L}.
\label{eqn:fl_binned}
\end{equation}\end{linenomath*}
The double integral in Eq.~\ref{eqn:fl_binned} is solved using a numerical approach: the co-moving volume is calculated on a discrete $\log L$-$z$ grid with the size of 50 by 50 steps. Each volume element is then multiplied by the step size in both $\log L$ and $z$ and summed up.

We also fit an analytic model of the LF directly to the data using a maximum likelihood method. A single and a double power-law model of the differential LF are fitted to the data. The present day LFs ($z=0$) are expressed by:
\begin{linenomath*}\begin{equation}
\frac{\mathrm{d}\Phi}{\mathrm{d}\log L} = A \left[  \left(\frac{L}{L_*}\right)^{\gamma_1} + \left(\frac{L}{L_*}\right)^{\gamma_2}  \right]^{-1},
\label{eqn:fl_double_pw}
\end{equation}\end{linenomath*}
with the normalization $A$, along with power-law indices $\gamma_1$ and $\gamma_2$ for a double power law. The parameter $L_*$ functions as a break luminosity for the double power-law model, that is also fitted. For single power-law models, $L_*$ is fixed and the second term in Eq.~\ref{eqn:fl_double_pw} is 0. The evolution of the LF with $z$ is given by the term $e$ with the two parameters $k$ and $g$:
\begin{linenomath*}\begin{equation}
e = (1+z)^{(k+g z)}.
\label{eqn:fl_e}
\end{equation}\end{linenomath*}
It is applied for two different scenarios: pure luminosity evolution (PLE) and pure density evolution (PDE):
\begin{linenomath*}\begin{equation}
\frac{\mathrm{d}\Phi}{\mathrm{d}\log L}(L,z) \biggr\rvert_{\mathrm{PLE}} = \frac{\mathrm{d}\Phi}{\mathrm{d}\log L}(L/e,z=0) \, , 
\label{eqn:fl_ple}
\end{equation}\end{linenomath*}
\begin{linenomath*}\begin{equation}
\frac{\mathrm{d}\Phi}{\mathrm{d}\log L}(L,z) \biggr\rvert_{\mathrm{PDE}} = \frac{\mathrm{d}\Phi}{\mathrm{d}\log L}(L,z=0) \cdot e.
\label{eqn:fl_pde}
\end{equation}\end{linenomath*}
We can further divide the models by setting the evolutionary parameter $g=0$ for one set of PLE and PDE models and allow the parameter to vary for others, named PLEg and PDEg.
The maximum likelihood algorithm that is utilized, following, for instance, \citet{Miyaji2015}, minimizes the expression of: 
\begin{linenomath*}\begin{equation}
S = -2 \sum_{i=0}^N \log \frac{\mathrm{d}\Phi}{\mathrm{d}log L}\biggr\rvert_{i} - \log \int_L \int_z \frac{\mathrm{d}\Phi}{\mathrm{d}\log L} \frac{\mathrm{d}V}{\mathrm{d}z} dz ~ d\log L \, ,
\label{eqn:s_minimization}
\end{equation}\end{linenomath*}
for all sample sources $i$. For the minimization of the function and the calculation of the best fit parameters and errors we use the MINUIT software package \citep{James1994}. 
In order to determine a best-fit model, we apply the Akaike information criterion \citep[$\mathrm{AIC}$,][]{Akaike1973}:
\begin{linenomath*}\begin{equation}
\mathrm{AIC} = 2k - 2log \mathcal{L} \, ,
\label{eqn:aic}
\end{equation}\end{linenomath*}
with the number of free parameters $k$ and the MINUIT return value $\mathcal{L}$ (maximum likelihood value). A minimal $\mathrm{AIC}$ indicates the best fit LF model. As outlined in, for example, \citet{Burnham2004} the difference of $\mathrm{AIC}$ values $\Delta_j = \mathrm{AIC}_j - \mathrm{AIC}_{\mathrm{min}}$ of model $j$ to the model with the lowest $\mathrm{AIC}$ value determines if the model is equally or less probable. 
For practical purposes, the likelihood, $p_j$, of model, $j$, can be calculated. It expresses the relative probability, compared to $\mathrm{AIC}_{\mathrm{min}}$ / the best fit, that model $j$ minimizes the $\mathrm{AIC}$:
\begin{linenomath*}\begin{equation}
p_j = e^{-\Delta_j / 2}.
\label{eqn:p_j}
\end{equation}\end{linenomath*}

\section{Results}
\label{sec:res}

\subsection{Signal-to-noise ratio and BAT source detection}
\label{subsec:res:SNR}

\begin{figure}
   \centering
   \includegraphics[width=\hsize]{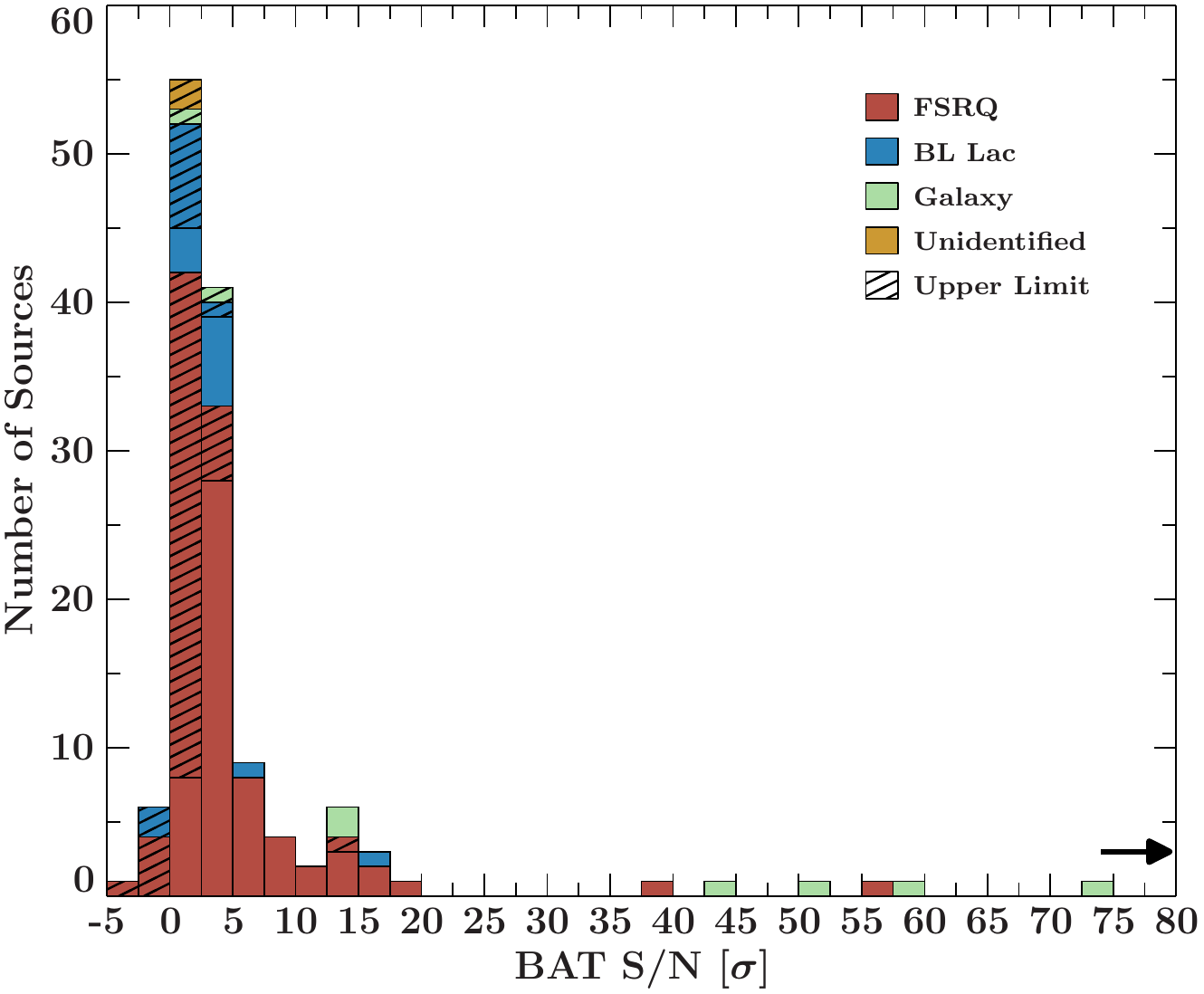}
      \caption{BAT S/N distribution of MOJAVE-1 sample for different source classes. The brightest source 3C 273 at 192\,$\sigma$ has been omitted for better readability.
              }
         \label{fig:histo_SNR}
\end{figure}

Other than the past BAT blind survey catalogs with a significance threshold of 4.8\,$\sigma$, we derive the significance values and spectra at known source positions, independent of signal strength. For each source in the MOJAVE-1 sample, the S/N value has been extracted from the Crab-weighted 105-month survey maps in the full energy range of 14\,keV -- 195\,keV. In Fig.~\ref{fig:histo_SNR} we present the distribution of the BAT S/N values for all source types in the sample. Hatched areas represent sources for which only upper limits for the X-ray fluxes could be determined. The highest value at about 192\,$\sigma$ belongs to the FSRQ 1226+023 (3C 273) and has been omitted in the histogram. 
27 sources have S/N values between 4.8\,$\sigma$ and 20\,$\sigma$, and can be considered clear detections, according to the blind survey conventions of \cite{Oh2018}, \citet{baumgartner2013}, and \citet{Tueller2008}. 
A tail of very X-ray bright and highly significant sources (three FSRQs and four radio galaxies) shows higher values still. The brightest source in the BL Lac category is BL Lac itself (2200+420) with a S/N of 16.8\,$\sigma$. All remaining BL Lac objects in the sample have S/N values under 5.4\,$\sigma$.
The sub-set of radio galaxies appears relatively bright compared to the other source types, which is due to the close proximity of these sources. The average redshift of the radio galaxies in the sample is 0.06, whereas BL Lacs and FSRQs are characterized by noticeably higher redshifts of 0.39 and 1.16, respectively.

\begin{figure}
\centering
\includegraphics[scale=0.55]{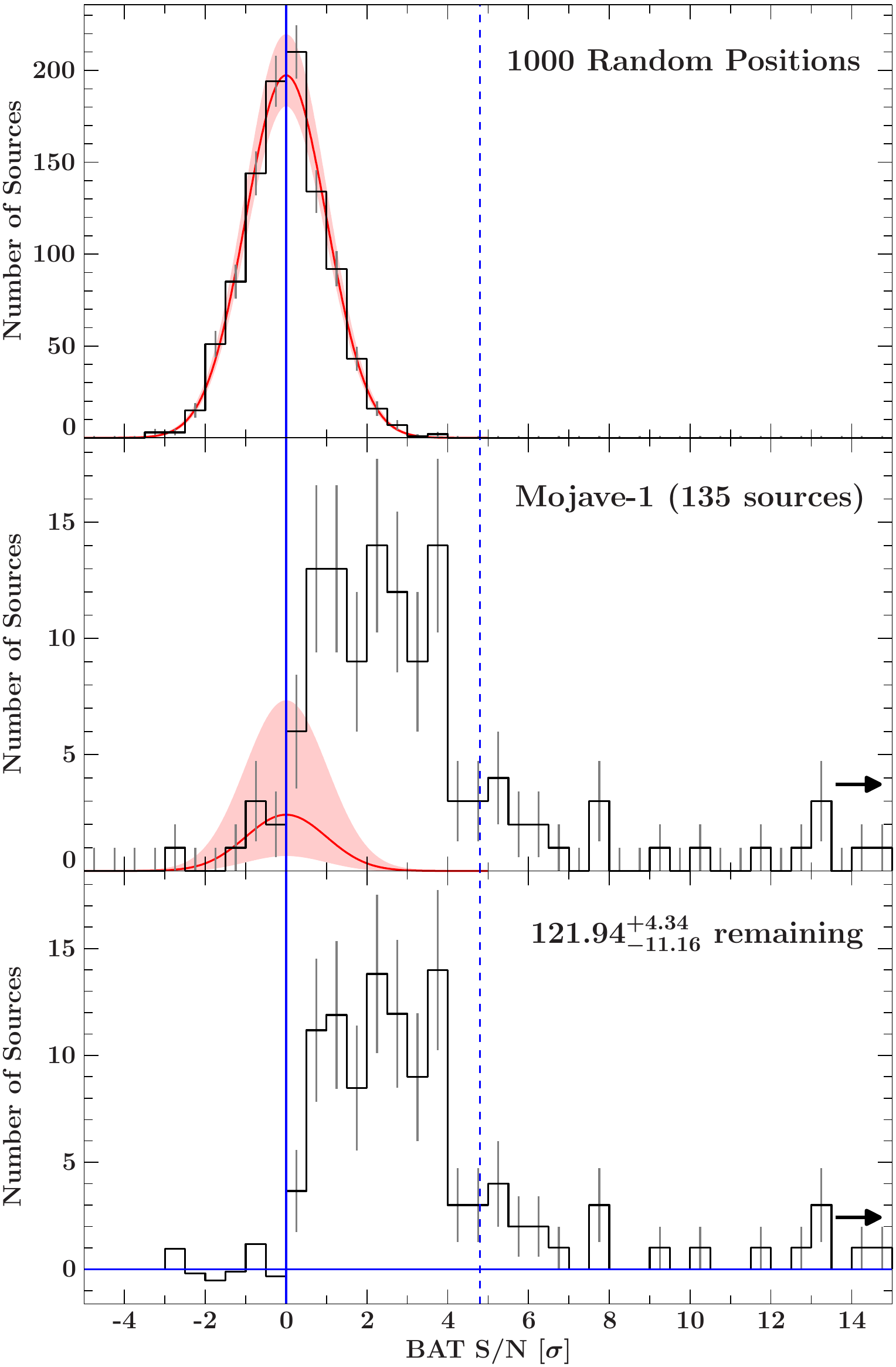}
\caption{Top panel: Distribution of \textit{Swift}/BAT S/N values of 1000 random blank sky positions. The red line indicates a Gaussian fit with center of 0\,$\sigma$ and width 1\,$\sigma$. The red shaded region indicates 3\,$\sigma$ uncertainty ranges for the norm of the fitted function. Middle panel: Distribution of \textit{Swift}/BAT S/N values of the MOJAVE-1 sample. The red line indicates a Gaussian fit with center of 0\,$\sigma$ and width 1\,$\sigma$ to the distribution with values smaller than 0\,$\sigma$. The red shaded region indicates 3\,$\sigma$ error ranges. The 4.8\,$\sigma$ threshold of the 105-month BAT survey is shown by the dashed line. Bottom panel: resulting histogram after subtracting the MOJAVE-1 S/N distribution by the distribution described by the Gaussian fit. Arrows indicate the cutoff at 15\,$\sigma$ (for better readability).
              }
         \label{fig:subtr_moj}
\end{figure}

The majority of 101 sources of the sample lies below the 105-month catalog threshold of 4.8\,$\sigma$, although with a clear offset from 0\,$\sigma$ towards positive S/N values. Negative S/N values are viable since the survey maps that we used are already subtracted with a dominant instrument background count rate value. Hence, areas with no bright source emission can be over-subtracted, leading to negative S/N values. 
Since BAT is a background-dominated coded mask instrument, low signal measurements from source positions might also be of random nature. In order to estimate the number of signals from true sources we compare a pure noise distribution with the S/N distribution of our sample. Figure~\ref{fig:subtr_moj}, top panel, shows the S/N characteristic of 1000 pointings over the entire sky which have at least a 100 pixel ($4\fdg5$) distance from any known source brighter than 4.8\,$\sigma$ in the 105-month survey maps. The resulting distribution is centered around 0\,$\sigma$ and ranges from about $-3.3\,\sigma$ to 3.5\,$\sigma$. It can be described very well with a Gaussian function  that is centered at 0\,$\sigma$ with a width of 1\,$\sigma$, as indicated by the red line (reduced Cash statistic of 0.86 for 19 degrees of freedom). The 99.73\% confidence intervals (3\,$\sigma$) of the fit function are shown by the red shaded area. 

We want to stress the difference in the idea of ``noise'' in terms of signal strength and noise contribution to the signal distribution of a source sample. The former relates to the S/N of the count rate of a central pixel of a source's coordinates in relation to the mean instrumental background. The latter describes the contribution of low-significance sources (low or even negative S/N values) to the overall S/N distribution of a sample.

A direct comparison of the top and middle panel of Fig.~\ref{fig:subtr_moj} clearly demonstrates that the MOJAVE-1 S/N characteristic is significantly different from random noise\footnote{Formally, using a 2-sample KS test the null hypothesis of a common distribution can be rejected at a level of $\alpha = 7 \cdot 10^{-12}$.}.
In order to determine the fraction of a source sample whose hard X-ray emission is not compatible with random map noise we subtract a Gaussian model fit from the S/N distribution of the sample. The Gaussian, with a center at 0\,$\sigma$ and a variance of 1\,$\sigma$, is modeled using a fit to all negative S/N values in the histogram. The model function and its set parameters for center and width follow the measured pixel significance distribution by \citet{Tueller2008}. We apply this method to the measured MOJAVE-1 sample S/N distribution.

The S/N distribution of the MOJAVE-1 sample is shown in the middle panel of Fig.\,\ref{fig:subtr_moj}. The errors are purely statistical. We fit a Gaussian function with a center of 0 and width of 1\,$\sigma$ to the negative part of the distribution, which describes the map noise contribution. The source distribution is then subtracted by the values indicated by the Gaussian fit function. In this way we obtain the number of sources not compatible with random map noise. It remains a number of $121.94^{+4.39}_{-11.16}$ sources, representing 82.06\% to 93.58\% of the MOJAVE-1 sources, which is the percentage of the sample that is emitting hard X-rays with respect to the achievable flux limit of the BAT instrument after 105 months of observations. 

We note that this global consideration does not yield an immediate statement regarding which individual sources of the low-S/N sources are real detections (i.e., belonging to the subtracted sample within the error margins). However, within a given significance bin, we can determine the probability that a given source is a real detection by comparing the number of expected signals through background fluctuations to the actual number of counts in this bin. For example, in the (0.5\,$\sigma$ to 1\,$\sigma$) bin, we expect up to approximately five counts from the fitted background distribution (within the 3\,$\sigma$ uncertainty range of the fitted Gaussian; see Fig.~\ref{fig:subtr_moj}, middle panel). We find thirteen sources in this bin. Each individual of these thirteen sources thus has a probability of less than 5/13 of being associated with a background fluctuation. Within the higher significance bins, this chance-fluctuation probability drops quickly. For example, in the (2\,$\sigma$ to 2.5\,$\sigma$) bin, we expect only up to 0.5 counts but detect fourteen sources. Each of those thus has a probability of larger than 96\% of not being associated with a background fluctuation.

The most recent \textit{Swift}/BAT catalog \citep{Oh2018} was only able to identify 36 out of 135 radio-selected MOJAVE-1 AGN, which are for the vast majority blazars. We show that a lower S/N threshold for BAT detections leads to a large increase in number of detected hard X-ray emitting blazars.

\subsection{Spectral shape -- photon index $\Gamma$}
\label{subsec:res:gamma}

\subsubsection{General description}
\label{subsec:res:gamma_general}

The BAT spectra of 77 out of all 135 sources were modeled with power law continua. This includes 46 sources below the 105-month catalog 4.8\,$\sigma$ threshold. Spectra with at least one negative count rate bin (58 sources) due to low signal strength have been processed using template spectra (see Sect.~\ref{subsec:spectralfitting}). No photon index has been derived for the three sources 0917+624, 1502+106, and 1928--179, which have sufficient count rates for fitting, but are spectrally contaminated by other nearby sources. 

In a canonical spectral energy distribution (SED) of a blazar in a $\nu F_\nu$ plot the photon index of $\Gamma=2$ indicates a flat spectrum. Smaller values correspond to a rising spectrum, typically the left flank of the HE bump at hard X-rays. The majority of all measured blazar photon indices in the sample are smaller than 2, which locates the BAT band at the rising part of the HE bump. Radio galaxies show a different behavior and concentrate around $\Gamma=2$ with the largest value at 3.2 (0316+414 / 3C 84).

\begin{figure}
   \centering
   \includegraphics[width=\hsize]{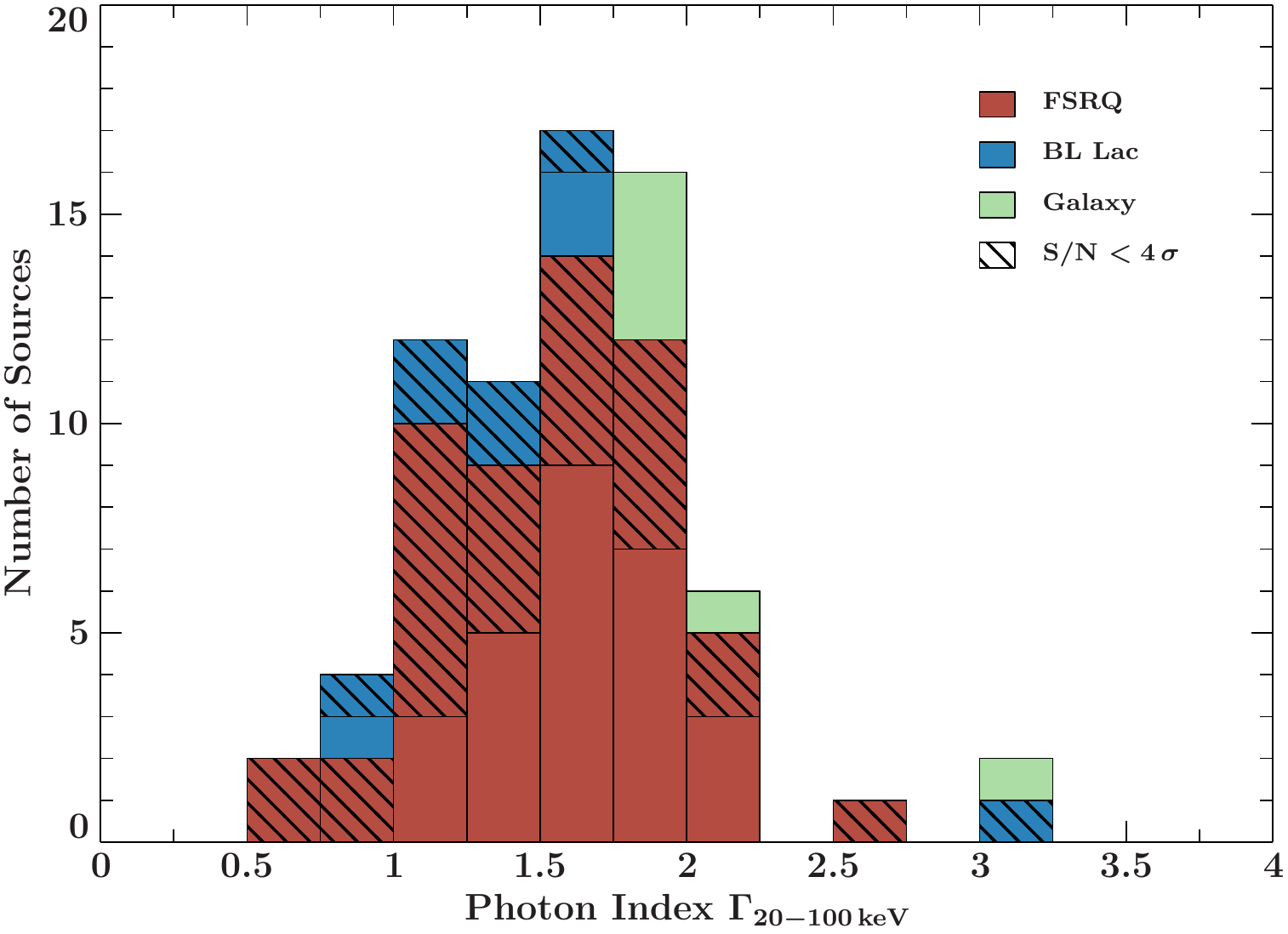}
      \caption{Distribution of photon index $\Gamma$ for all fitted MOJAVE-1 sources. Hatched boxes indicate sources with S/N values below 4\,$\sigma$ (fainter 50\% of the sample).
              }
         \label{fig:histo_gamma}
\end{figure}

The distribution of derived photon indices $\Gamma$ is shown in Fig.~\ref{fig:histo_gamma}. FSRQs form a quasi-Gaussian distribution around 1.6, with values ranging from about 0.5 to 2.7. The distribution for the low number of BL Lacs shows a broader and less peaked shape, ranging from 0.8 to 3. In the plot the fainter half of the sources (S/N $< 4\,\sigma$) is also marked. The brighter half of the shown source sample (omitting 3C 84) is characterized by a smaller spread compared to the fainter sources, approximately 0.8 -- 2.1 vs.~0.5 -- 3.0.
Sources that are less significant show large uncertainties of the photon index, approximately 0.4 for brighter sources and up to 1.7 for the faintest of the fitted spectra. Consequently, the distribution of the photon index becomes noticeably broader.

\subsubsection{\textit{Fermi}/LAT detections}
\label{subsec:res:gamma_3lac}

\begin{figure}
   \centering
   \includegraphics[width=\hsize]{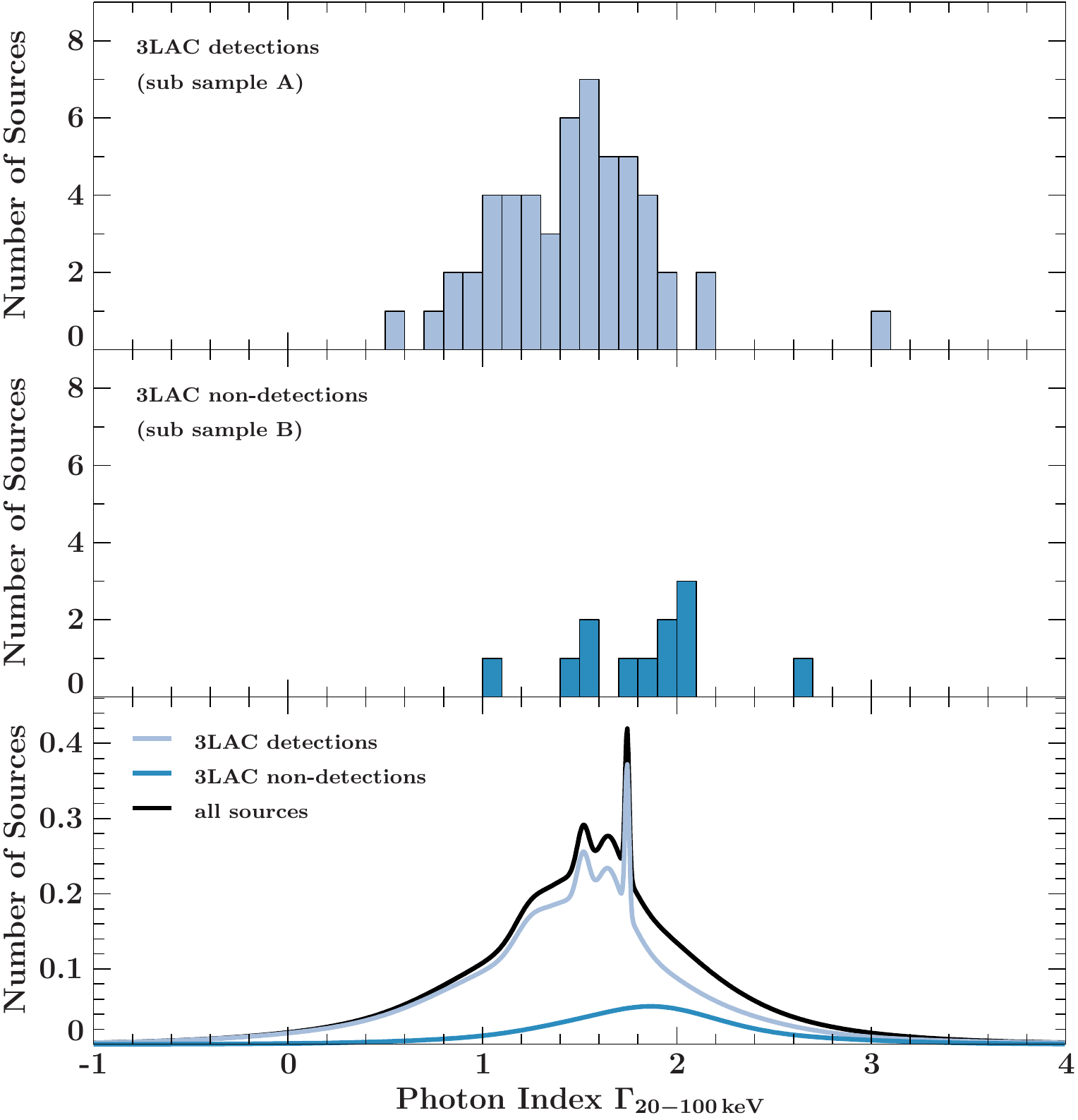}
      \caption{Distribution of BAT photon index $\Gamma$ for all fitted MOJAVE-1 blazars, detected and not detected by \textit{Fermi}/LAT, respectively. The bottom panel shows the added Gaussian distributions for each photon index. The sharp peak at 1.75 stems from the brightest blazar in the sample, 3C 273.
              }
         \label{fig:histo_gamma_3lac}
\end{figure}

The slope of the hard X-ray spectrum pinpoints in most cases the rising part of the HE emission bump and therefore holds information about the behavior of the SED at higher energies. Figure~\ref{fig:histo_gamma_3lac} displays the distribution of the photon index for all fitted MOJAVE-1 blazars, and compares gamma-bright sources which are listed in the 3LAC catalog (top panel) and gamma-faint sources (middle), respectively. For this part we choose to incorporate the \textit{Fermi}/LAT detection statistics from the 4 year catalog (3FGL / 3LAC), since the observation time slot of 2008 to 2012 is close to the BAT 105-month survey time range. We compare this with the later 4FGL \textit{Fermi}/LAT source catalog at the end of this section.
The MOJAVE-1 blazars in the 3LAC catalog (sub-sample $A$) have noticeably harder photon indices than gamma-faint sources (sub-sample $B$). We perform a two-sample KS-test for the two distributions. The null hypothesis of both data sets drawn from a common distribution can be rejected at a level $\alpha=0.008$ with a test statistic of $D_{\mathrm{sample}}=0.497$. Additionally, we apply a one-sided Z-test, comparing the difference of means of $A$ and $B$, depending on the sample size. The resulting $p$-value of 0.0028 also strongly implies a distinct difference of distribution of values.

Even so, the tests ignore the partially substantial uncertainties of the photon indices. In order to take into account the (Gaussian-distributed) uncertainties for every source we model the probability distribution of each index as a Gaussian curve with a width corresponding to its uncertainty range. 
Then, the distributions of probability for every source in each data set $A$ and $B$ are added up. The resulting distributions are shown in the bottom panel of Fig.~\ref{fig:histo_gamma_3lac}. Since the area of every individual Gaussian curve is equal, bright sources with small uncertainties translate into sharp peaks, whereas sources with large uncertainties equal low and broadened curves. The distributions are plotted as fine histograms on a grid of 1000 bins, normalized to their respective sample size.
From the summed distribution of all sources (black line) $10^5$ times two test samples $A_{\mathrm{simul}}$ and $B_{\mathrm{simul}}$ with the respective sample size of $A$ and $B$ are randomly drawn and each time a KS-test is performed. 
The probability of a certain photon index value to be drawn is determined by the height of the probability function / summed distribution at that value.
The goal of the analysis is to find the number of randomly chosen sample pairs, which give a KS-test statistic $D$ that is equal or higher than the test statistic $D_{\mathrm{sample}}$ of the original samples $A$ and $B$. This percentage indicates how likely it is that the photon index distributions of $A$ and $B$ are distinct form each other.

The fraction of $A_{\mathrm{simul}}$ and $B_{\mathrm{simul}}$ with a KS-test statistic $D$ equal or greater than the test statistic $D_{\mathrm{sample}}$ of $A$ and $B$ is 0.9\%. This is equivalent to a significance of 2.61\,$\sigma$ for a Gaussian distribution of test statistic values. Thus, sub-samples of \textit{Fermi}/LAT-detected and non-detected blazars in the MOJAVE-1 sample depict two significantly different groups of sources in terms of spectral shape in the BAT energy range: \textit{Fermi}/LAT-detected sources are harder, while non-detected sources tend to be softer.

Including the remaining two \textit{Fermi}/LAT-detected and four non-detected radio galaxies in the MOJAVE-1 sample gives an even more distinctive result. We obtain a sample test statistic of $D_{\mathrm{sample}}=0.55$, equal to a rejection level of the null hypothesis of $\alpha=0.002$. The fraction of KS-tests with $D > D_{\mathrm{sample}}$ is 0.05\%.
This effect however is largely due to the photon index of the radio galaxies near $\Gamma=2$ in sub-sample $B$, emphasizing the difference of both distributions.

\begin{figure}
   \centering
   \includegraphics[width=\hsize]{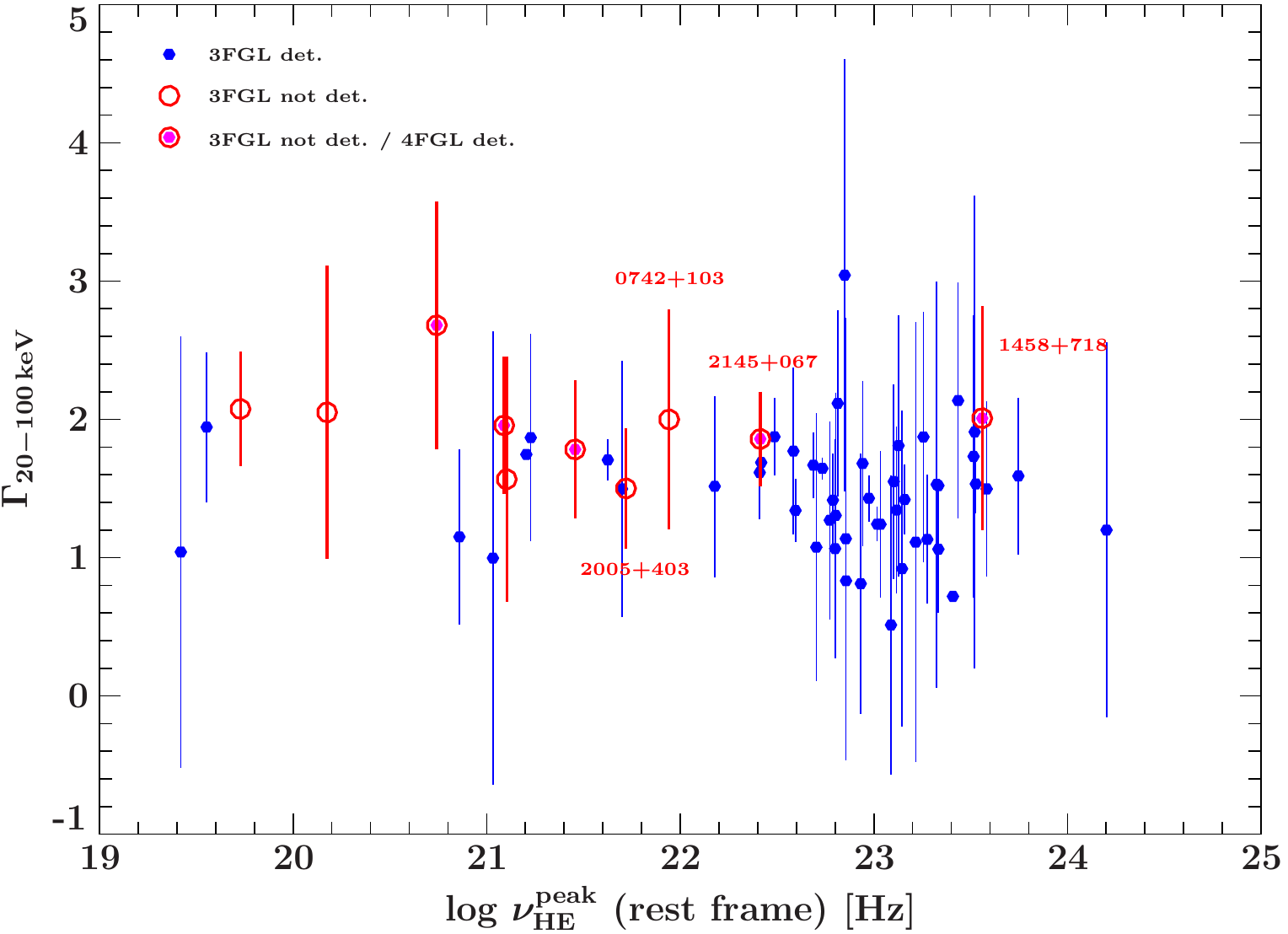}
      \caption{SED HE peak frequency of MOJAVE-1 blazars (rest frame) against BAT photon index $\Gamma$.
              }
         \label{fig:graph_nu_peak-gamma}
\end{figure}

A difference in the photon index or spectral slope in an energy band at the rising part of the SED's HE bump would correlate with the position of the SED along the frequency axis.
In Fig.~\ref{fig:graph_nu_peak-gamma} we show the photon indices of both \textit{Fermi}/LAT-detected and non-detected sub-samples against the HE emission peak frequencies, taken from \citet{Chang2010} and transformed to the rest frame. Whereas \textit{Fermi}/LAT-detected MOJAVE-1 blazars tend to concentrate around HE bump peak frequencies $\log \nu_{\mathrm{HE}}^{\mathrm{peak}}$ of 22.5 to 23.5 non-detected blazars span a range of approximately 19 to 24. 
Taking into account the previous result of 3LAC non-detected sources having noticeably harder BAT photon indices than non-detected sources (see testing procedure above), this indicates a possible correlation of spectral shape in the BAT 20\,keV -- 100\,keV band and the probability for a detection at GeV energies. Additionally, the sources which have a 4FGL catalog entry are plotted using filled circles in Fig.~\ref{fig:graph_nu_peak-gamma}. The two sources 1458+718 (3C 309.1) and 2145+067, previously undetected by \textit{Fermi}/LAT, are now sharing the same space in the diagram with the majority of detected sources at high HE peak frequencies. 

\subsection{Hard X-ray flux}
\label{subsec:res:xflux}

\begin{figure}
	\centering
	\includegraphics[width=\hsize]{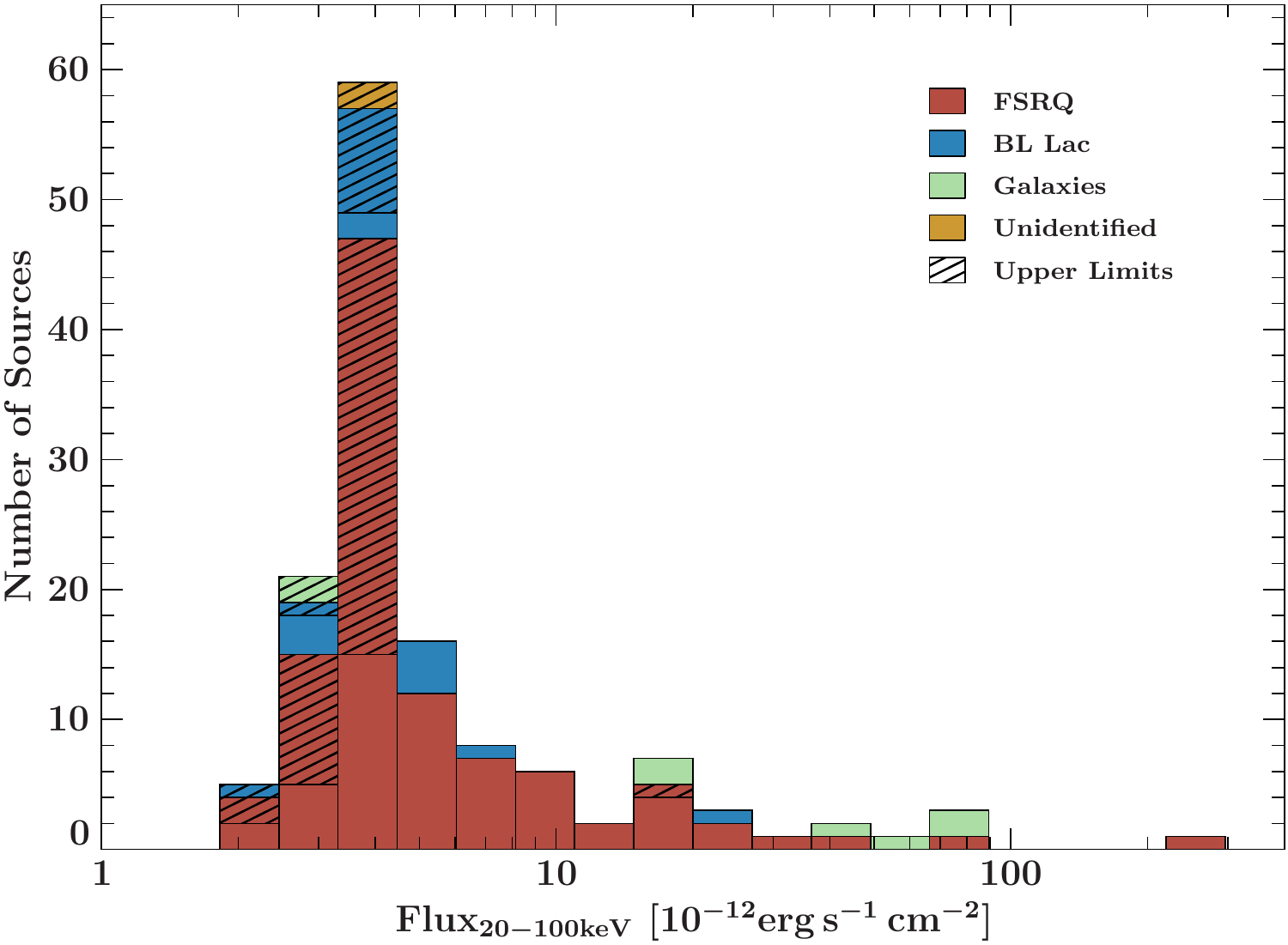}
		\caption{Distribution of hard X-ray flux of the MOJAVE-1 sample.} 
	\label{fig:histo_xflux}
\end{figure}

The histogram of hard X-ray flux measurements is presented in Fig.~\ref{fig:histo_xflux}. We show both the 77 fitted sources and the 58 remaining sources. The vast majority of 127 of 135 sources lies below $30\cdot10^{-12}\,\mathrm{erg\,s^{-1}\,cm^{-2}}$  with four FSRQs and four radio galaxies having higher significance values. Most of the sample, especially BL Lacs and unidentified sources, is characterized by low hard X-ray flux, as shown by the median of $4\cdot10^{-12}\,\mathrm{erg\,s^{-1}\,cm^{-2}}$, also taking into account upper limits.
Following the flux calculation described in Sect.\,\ref{sec:ana}, we obtain a number of 59 upper limits, including three sources that have been fitted, but are also designated upper limits because of spectral contamination of nearby X-ray sources.

\begin{figure}
   \centering
   \includegraphics[width=\hsize]{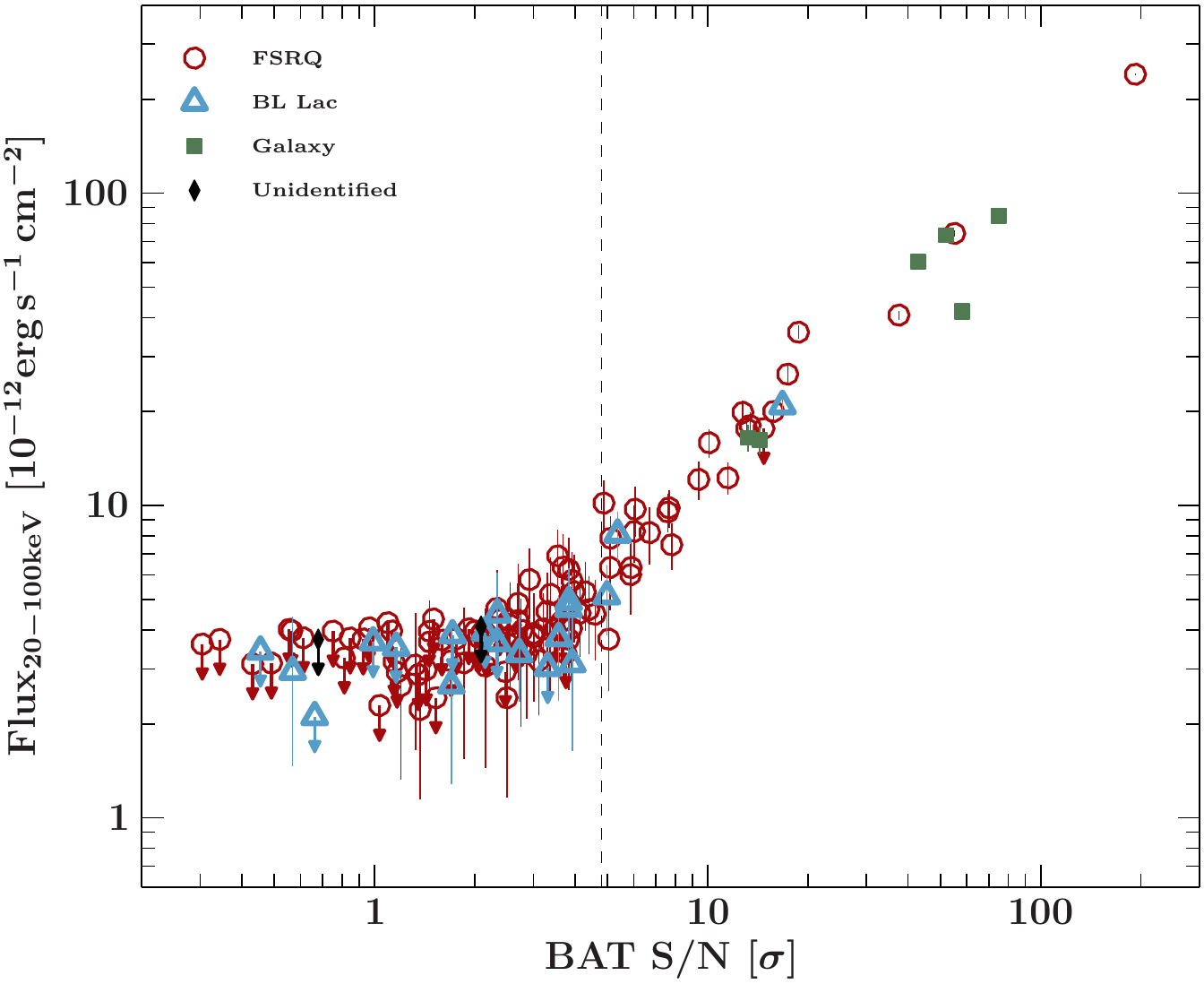}
      \caption{Relation of hard X-ray flux to S/N. The dashed line represents the S/N value of $4.8\sigma$. The source 1504$-$166 at 0.01$\sigma$ has been omitted for better readability.}
         \label{fig:graph_flux-snr}
\end{figure}

Figure~\ref{fig:graph_flux-snr} shows the relation between X-ray flux and S/N. Seven data points at negative S/N values as well as source 1504$-$166 (PKS 1504--167) at 0.01\,$\sigma$ have been omitted. The 105-month catalog threshold value of $4.8\,\sigma$ is indicated by the dashed line. Beginning at about 2\,$\sigma$, the slope of the narrow distribution in Fig.~\ref{fig:graph_flux-snr} is about unity, making the relation of flux and S/N linear, as expected. Below about 2\,$\sigma$ the flux value saturates around $(3-4) \cdot 10^{-12} \mathrm{erg\,s^{-1}\,cm^{-2}}$ at a point where almost all flux values are described as upper limits.

\subsection{Hard X-ray luminosity}
\label{subsec:res:xlum}

\begin{figure}
\centering
\includegraphics[width=\hsize]{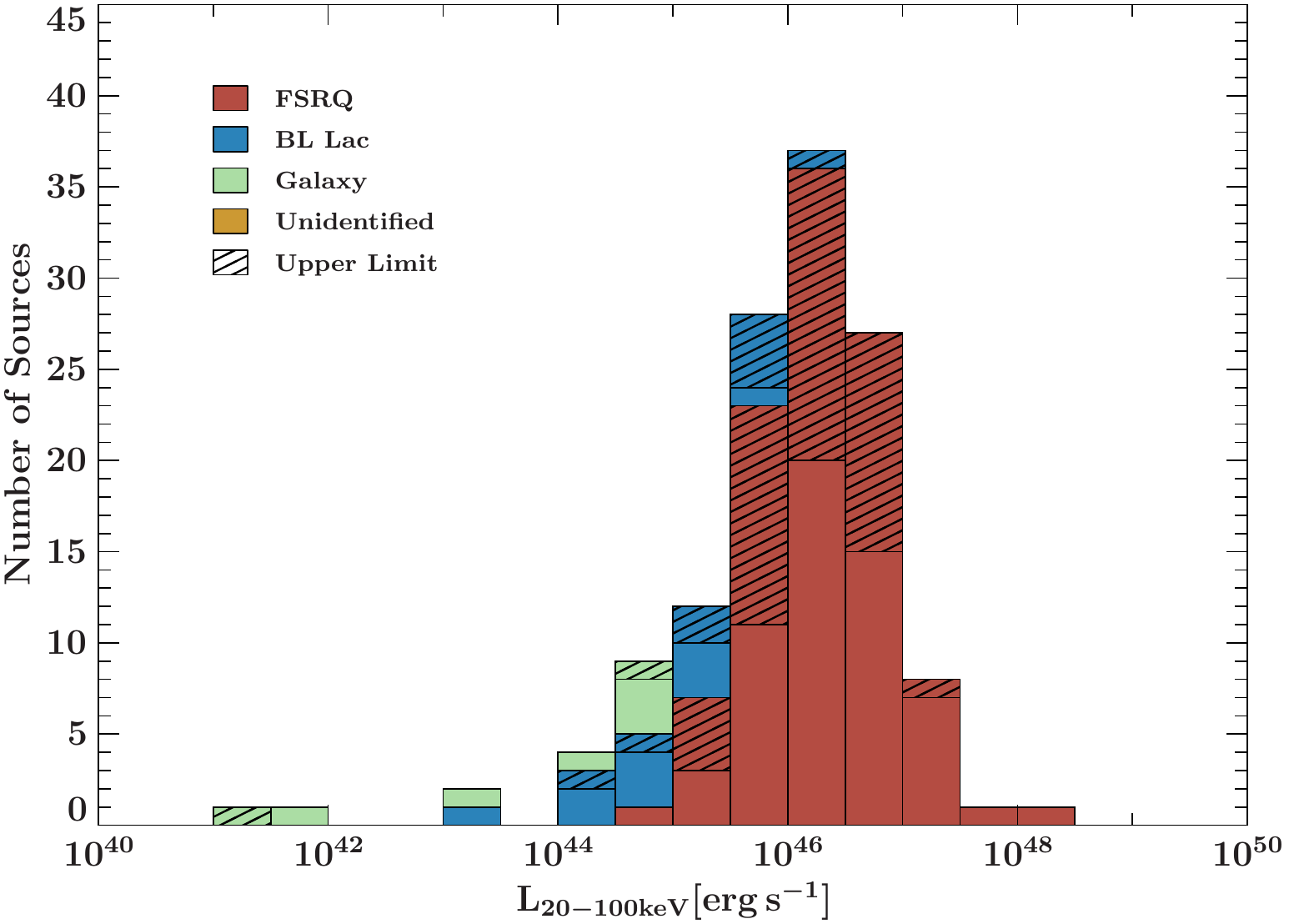}
\caption{K-corrected X-ray luminosity distribution for the energy range of (20 -- 100) keV of the MOJAVE-1 sample.}
\label{fig:histo_xlum} 
\end{figure}

We calculate the hard X-ray luminosities including the K-correction,
\begin{linenomath*}\begin{equation}
L=\frac{1}{(1+z)^{2-\Gamma}}4 \pi d_L^2 F,
\label{eqn:lum}
\end{equation}\end{linenomath*}
where $d_L$ is the luminosity distance, and assuming the cosmological parameters:  $H\mathrm{_0=70.0\,km\,s^{-1}\,Mpc^{-1}}$, $\Omega\mathrm{_M=0.30}$ and $\Omega_{\mathrm{\lambda}}=0.70$.
The resulting distribution of the hard X-ray luminosity is presented in Fig.~\ref{fig:histo_xlum}. Four sources with missing redshift information (BL Lacs and unidentified objects) were not included. 
This quasi-Gaussian distribution centers around $(10^{46}-10^{47})\,\mathrm{erg\,s^{-1}}$, with all BL Lacs and radio galaxies below $10^{46}\mathrm{erg\,s^{-1}}$. BL Lacs have minimum values of about $10^{43}\mathrm{erg\,s^{-1}}$. The brightest sources are all FSRQs with luminosities up to $10^{48}\mathrm{erg\,s^{-1}}$ (0836+710) and upper limits up to $5 \cdot 10^{47}\mathrm{erg\,s^{-1}}$. Radio galaxies possess lower luminosities around  $10^{41}\,\mathrm{erg\,s^{-1}}$ -- $10^{45}\,\mathrm{erg\,s^{-1}}$.  

The two different blazar classifications, FSRQ and BL Lac, clearly do not follow a common distribution. FSRQs exhibit a pronounced maximum, which is more than an order of a magnitude higher compared to BL Lacs. A KS-test for both data sets reveals a test statistic of $D=0.88$. We can reject the null hypothesis that the distributions are equal at a level of $\alpha=3.2 \cdot 10^{-6}$.

In Sect.~\ref{subsec:res:gamma} we have presented data that suggests a correlation of spectral shape in the hard X-ray regime and the probability of a detection at gamma-ray energies. In order to test for a difference of the intrinsic luminosity of both \textit{Fermi}/LAT-detected and non-detected sub-samples, we apply a KS-test to both data sets.
We obtain a KS test statistic of $D=0.22$. The null hypothesis that both luminosity distributions of \textit{Fermi}/LAT-detected and non-detected MOJAVE-1 blazars are the same can be rejected at a level of $\alpha=0.35$. Both distributions are not significantly different from each other in the BAT band.

\begin{figure*}
   \centering
   \includegraphics[width=\hsize]{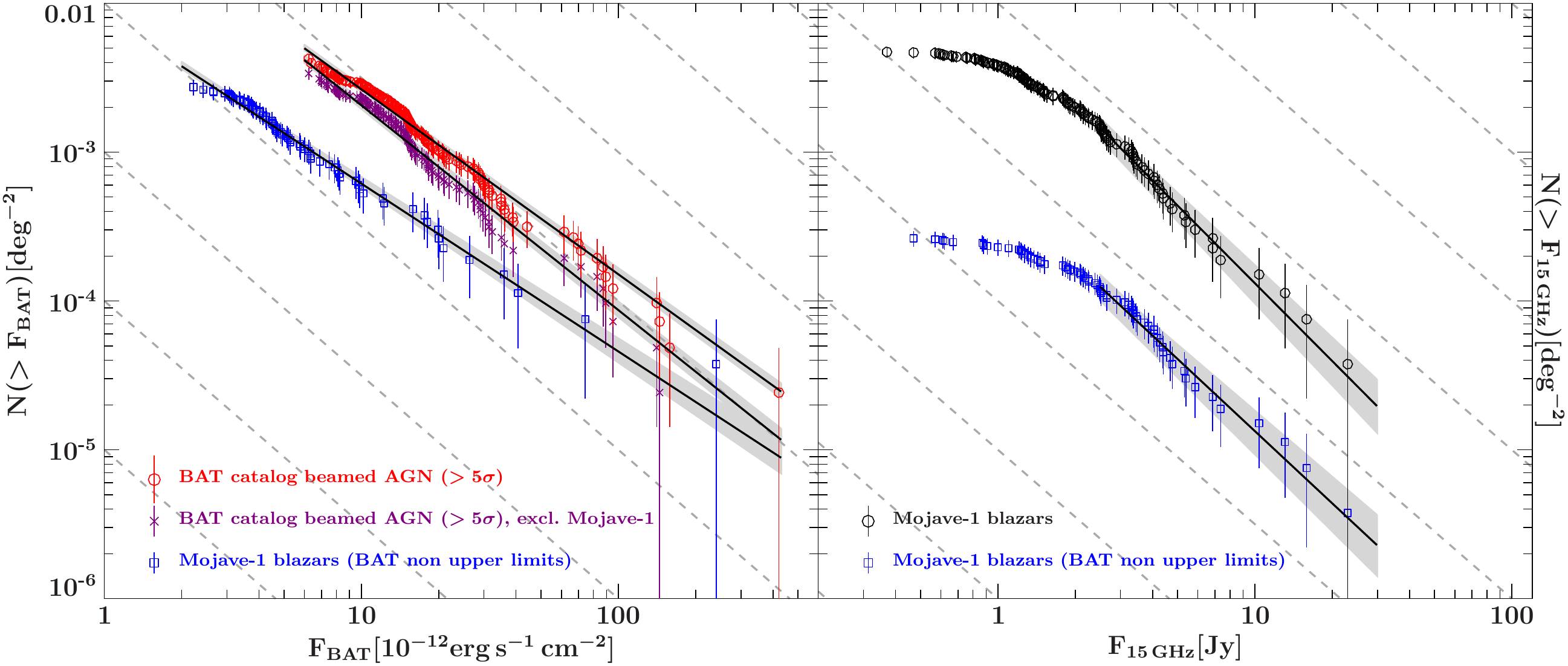}
      \caption{Left: Cumulative $\log N$-$\log S$ distribution of the MOJAVE-1 sample excluding radio galaxies, omitting all upper limits (flux for 20 -- 100\,keV, blue squares) and $\log N$-$\log S$ for the beamed AGN sub-sample of the BAT 105-month source catalog (flux for 14\,keV -- 195\,keV, red circles) and the same sub-sample excluding all MOJAVE-1 sources (14\,keV -- 195\,keV, purple x). Right: $\log N$-$\log S$ for 15\,GHz flux density VLBI measurements from \citet{Lister2015} of all blazars in the MOJAVE-1 sample (black circles) and MOJAVE-1 blazars from left graph (blue squares, norm with a factor of 0.1 for better readability). The black lines represent power-law fits, the gray area shows the corresponding error range. The radio data have been fit for all fluxes higher than 2.5\,Jy. The dashed lines indicate a slope of -1.5.}
         \label{fig:graph_logn-logs}
\end{figure*}

\subsection{Source count statistics}

\subsubsection{$\log N$-$\log S$ distribution}
\label{subsubsec:logn-logs}

Figure~\ref{fig:graph_logn-logs} (left panel) shows the $\log N$-$\log S$ distribution of the BAT flux values of the MOJAVE-1 blazars (open squares), excluding all upper limit sources, leaving 70 sources.
The distribution follows a power law, which is significantly flatter than -1.5 for all fluxes $F \gtrsim 2.2 \cdot 10^{-12} \mathrm{erg s^{-1} cm^{-2}}$. 
We fit the data using a least squares approach and obtain the parameters for the best fit with $A_{\mathrm{Moj,BAT}}=(8.25 \pm 0.57) \cdot 10^{-3}\,\mathrm{deg}^{-2}$ and $\alpha_{\text{Moj,BAT}}=1.13 \pm 0.04$ for the fitting function $A F^{-\alpha}$, where $F$ is the flux in $10^{-12} \mathrm{erg \, s^{-1} cm^{-2}}$. 
Below a flux of approximately $3 \cdot 10^{-12} \mathrm{erg s^{-1} cm^{-2}}$, the distribution of the data points is saturated.

The faintest source in the plot (1417+385) is not compatible with the power-law fit within the error margins. The fitted power law at the lowest flux in the sample amounts to a cumulative source number of $(3.33 \pm 0.22) \cdot 10^{-3} \mathrm{deg}^{-2}$, or $88 \pm 6$ sources in the survey part of the sky. The difference to the 70 sources of the fitted sample likely stems from the missing contribution of at least some of the upper limits which have not been included in the plot.

\begin{table*}[t]
\caption{Results of power-law fits of the blazar $\log N$-$\log S$ distributions. The first two rows are fits to the BAT flux data (20\,keV -- 100\,keV), the three rows corresponding to the BAT catalog are fitted to the BAT flux catalog data set (14\,keV -- 195\,keV), and the last group of three rows are fits to the 15\,GHz flux density data.}             % title of Table
\label{tab:logn-logs-parameters}
\centering
\begin{tabular}{ccccc}
\hline\hline
		Sample 	& Fitted Sources			& normalization $A$ [$\mathrm{10^{-3}\,deg^{-2}}$] & slope $a$ & Original Sample\\	
\hline                        
        MOJAVE-1, BAT      & 70		& $8.25 \pm 0.57$ & $1.13 \pm 0.04$ & \citet{Lister2009_137}\\ 
        MOJAVE-1.5, BAT      & 92		& $9.22 \pm 0.45$ & $1.13 \pm 0.03$ & \citet{Lister2013}\\ 
        & & & & \\
        BAT cat.~(beamed AGN)      & 143	& $46.20 \pm 2.50$ & $1.24 \pm 0.02$ & \citet{Oh2018}\\ 
        BAT cat.~(beamed AGN, excl.~MOJAVE-1)      & 115	& $48.96^{+3.57}_{-3.09}$ & $1.38 \pm 0.03$ & \citet{Oh2018}\\  
        BAT cat.~(AGN)      & 955	& $338.99^{+2.72}_{-2.70}$ & $1.27 \pm 0.01$ & \citet{Oh2018}\\ 
        & & & & \\       
        MOJAVE-1\tablefootmark{a}  & 40 & $7.17^{+1.19}_{-0.99}$ & $1.74^{+0.13}_{-0.12}$ & \citet{Lister2009_137}\\
        MOJAVE-1\tablefootmark{b}  & 32 & $5.46^{+1.1}_{-0.89}$ & $1.61^{+0.15}_{-0.14}$ & \citet{Lister2009_137}\\
        MOJAVE-1.5 & 170 & $7.20^{+0.98}_{-0.84}$ & $1.72^{+0.11}_{-0.10}$ & \citet{Lister2013}\\        
\hline                                   
\end{tabular}
\tablefoot{
\tablefoottext{a}{Source selection based on all 125 MOJAVE-1 blazars. Sources with a VLBI flux density less than 2.5 Jy are excluded.}
\tablefoottext{b}{Source selection based on all 70 MOJAVE-1 blazars that provide flux non upper limits in the BAT band. Sources with a VLBI flux density less than 2.5 Jy are excluded from the fit.}
}
\end{table*}

Within the uncertainty range of $\alpha_{\text{Moj,BAT}}$ the $\log N$-$\log S$ distribution for the hard X-ray data is not compatible with a regular Euclidean distribution. For comparison we also show the $\log N$-$\log S$ distribution of the beamed AGN sub-sample of the 105-month \textit{Swift}/BAT source catalog \citep{Oh2018}, indicated by open circles, and for flux values in the complete BAT range of 14\,keV -- 195\,keV. The BAT catalog is compiled purely on the basis of BAT significance and does not suffer from any selection effects that might be introduced by a catalog at a different wavelength. The original sub-sample of objects designated beamed AGN includes 158 sources. We exclude the brightest source, Centaurus A, which is more commonly classified as a radio galaxy \citep[see, e.g.,][for a review]{Steinle2006}. Also, all sources with a significance smaller than $5\,\sigma$ are excluded, leaving 143 beamed AGN. A power-law fit of the distribution yields a slope of $\alpha_{\text{cat,BAT}}=1.24 \pm 0.02$. A fit to the $\log N$-$\log S$ distribution of the entire BAT 105-month AGN sample (955 sources fitted, not shown) gives a very similar power-law index of $\alpha_{\text{BAT AGN}}=1.27 \pm 0.01$. 
The slope of the $\log N$-$\log S$ distribution of the MOJAVE-1 hard X-ray emission is thus not compatible with the most recent BAT source catalog and the beamed AGN sub-sample (within a 2\,$\sigma$ uncertainty range of both results). However, the close proximity of the determined power-law indices ($1\,\sigma$ errors) and the significantly greater distance of all indices to the expected Euclidean slope of $\alpha = 1.5$ suggests a similar underlying reason for this behavior of blazar samples in the hard X-ray domain. In terms of sky area density the BAT catalog sample shows approximately 1.6 times the density of beamed AGN compared to the MOJAVE-1 blazars (non upper limits) regarding the X-ray-faintest source in both distributions. Fitting the $\log N$-$\log S$ distribution of the BAT catalog beamed AGN sample excluding all MOJAVE-1 sources results in $\alpha_{\text{cat,BAT,noMoj}}=1.38 \pm 0.03$, bringing the slope closer to the Euclidean case.

Additionally, we fit the MOJAVE-1 $\log N$-$\log S$ distribution of available VLBI radio flux density data from \citet{Lister2015}, using median flux density values. The observation time of the measurements spans 2008 -- 2012, reasonably close to the BAT survey time. The right panel of Fig.~\ref{fig:graph_logn-logs} displays the $\log N$-$\log S$ distribution of the MOJAVE-1 blazars in the 15 GHz band from \citet{Lister2015}, where the black circles denote the full sample of MOJAVE-1 blazars (122 sources, all with known redshift). The blue squares indicate the same sub-set of X-ray bright blazars as in the the left graph. Both distributions are plotted and fitted in order to determine possible selection effects in the $\log N$-$\log S$ distribution of the BAT data. Fitting the complete flux range can introduce a new data bias, however. Because the MOJAVE-1 sample is just a sub-sample of the compiled flux-limited compact object radio samples of \citet{Lister2015} we choose a cutoff at 2.5 Jy, below which all data are ignored. The vast majority of additional sources to the MOJAVE-1 sample in both larger samples are located below this cutoff. By ignoring all data below the cutoff we avoid using a sample with a flux distribution that is not representative of a flux-limited survey. The normalizations and power-law slopes of all samples that we used are listed in Table~\ref{tab:logn-logs-parameters}. 

The power-law fits of the radio flux densities larger than 2.5\,Jy (122 source and 70 source sub-sample) are described by $a_{\mathrm{Moj,15GHz}}^{122} = 1.74^{+0.13}_{-0.12}$, and $a_{\mathrm{Moj,15GHz}}^{70} = 1.61^{+0.15}_{-0.14}$, respectively. The smaller sample of 70 sources lies well within the Euclidean distribution, and is not compatible with the slope of the BAT data within their 1 $\sigma$ uncertainty range. The fit of the larger sample of 122 sources in the radio band produces a slope that is even larger than for a Euclidean distribution. 
Since the difference in slopes from radio to X-ray band is likely not directly due to a selection effect, we investigate the influence of variability and intrinsic evolution of the emitted X-ray flux in the following sections.
The influence of different contributions to the slopes from FRSQs and BL Lacs individually cannot be examined 
in detail because of the low number of only four BL Lacs in the reduced sample above 2.5 Jy. 

To test for possible biases from selection effects of the MOJAVE-1 sample, we analyze the extended MOJAVE-1.5 sample \citep{Lister2013} and the corresponding $\log N$-$\log S$ distribution for radio and BAT fluxes. Contrary to the main sample of this study all sources with a minimum 15\,GHz flux density of 1.5\,Jy and with $\delta\ge -30^\circ$ and the Galactic plane are included. However, the source confusion problem is not trivial in the hard X-ray band because of the large instrument PSF and high source density, especially within the Galactic plane.

The larger fitted sample of blazars in the radio band reveals a steep slope of $a_{\mathrm{Moj 1.5,15GHz}}^{170} = 1.72^{+0.11}_{-0.10}$, which is compatible with the results of the MOJAVE-1 / Lister data set. As expected, the normalization of the fitted power law for the BAT fluxes is somewhat higher because of the small number of additional sources. Only 22 newly added blazars from the MOJAVE-1.5 sample were fitted because of the high number of upper limits around $F=3\cdot10^{-12}\mathrm{erg\,s^{-1}\,cm^{-2}}$.

The derived slope in the hard X-ray band is practically identical to the MOJAVE-1 data set with $\alpha_{\text{Moj 1.5,BAT}}=1.13 \pm 0.03$. It can thus be concluded that the slightly different selection criteria of the earlier MOJAVE-1 sample are not a significant influence on the relatively flat slope of the X-ray $\log N$-$\log S$ distribution. Both samples share the behavior of flux distribution in space. It is therefore highly suggested that the samples depict the same blazar population.

\begin{figure}
\centering
\includegraphics[width=\hsize]{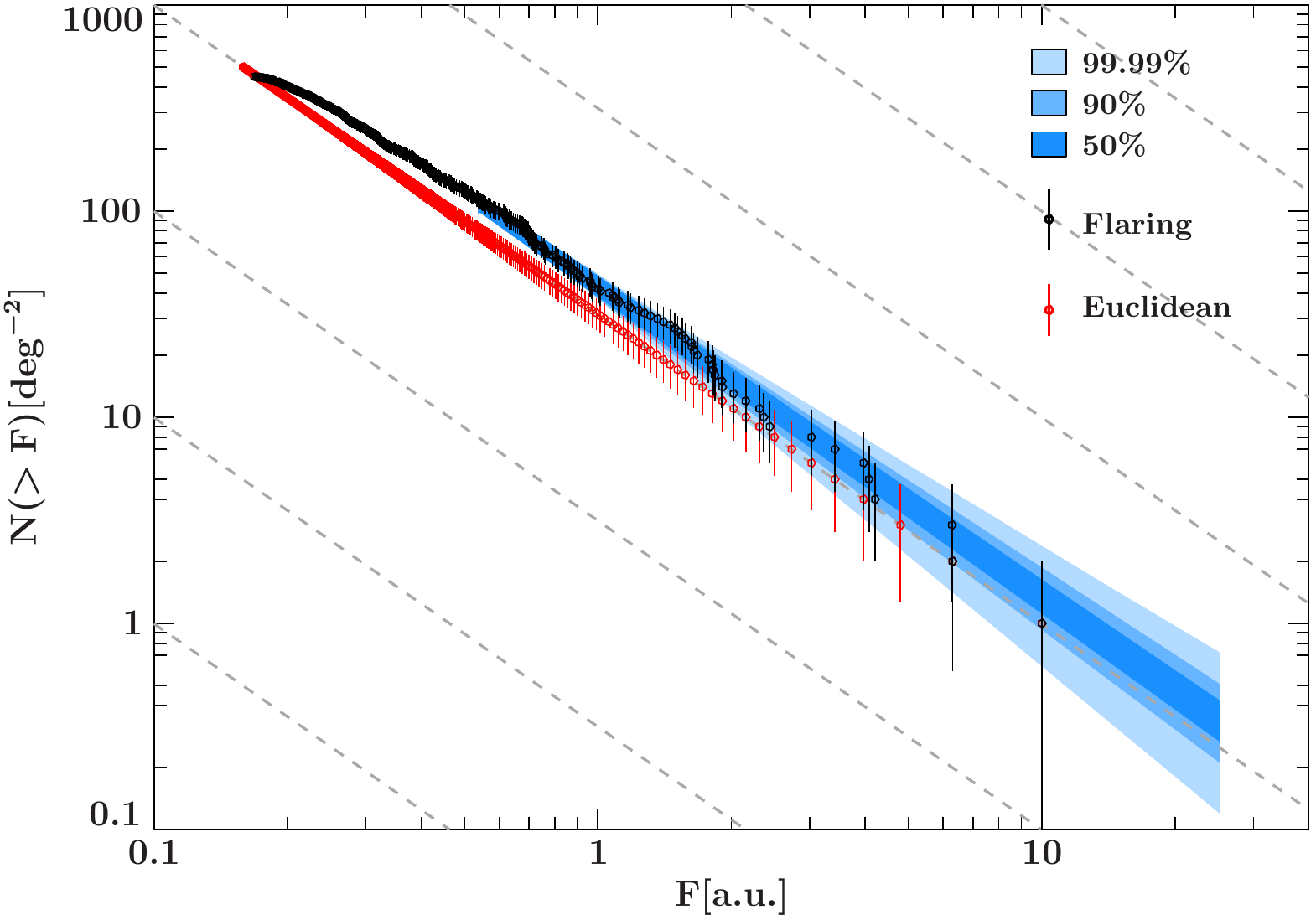}
\caption{Simulated sample of sources with uniform luminosity and ideal Euclidean $\log N$-$\log S$ flux distribution (red) and one instance of the same sample with random flaring (black, see text). The shaded areas indicate the error ranges derived by the Monte-Carlo approach, with 99.99\% of all fitted power laws lying inside the outer shaded area.}
\label{fig:simul_lognlogs}
\end{figure}

\begin{figure}
\centering
\includegraphics[width=\hsize]{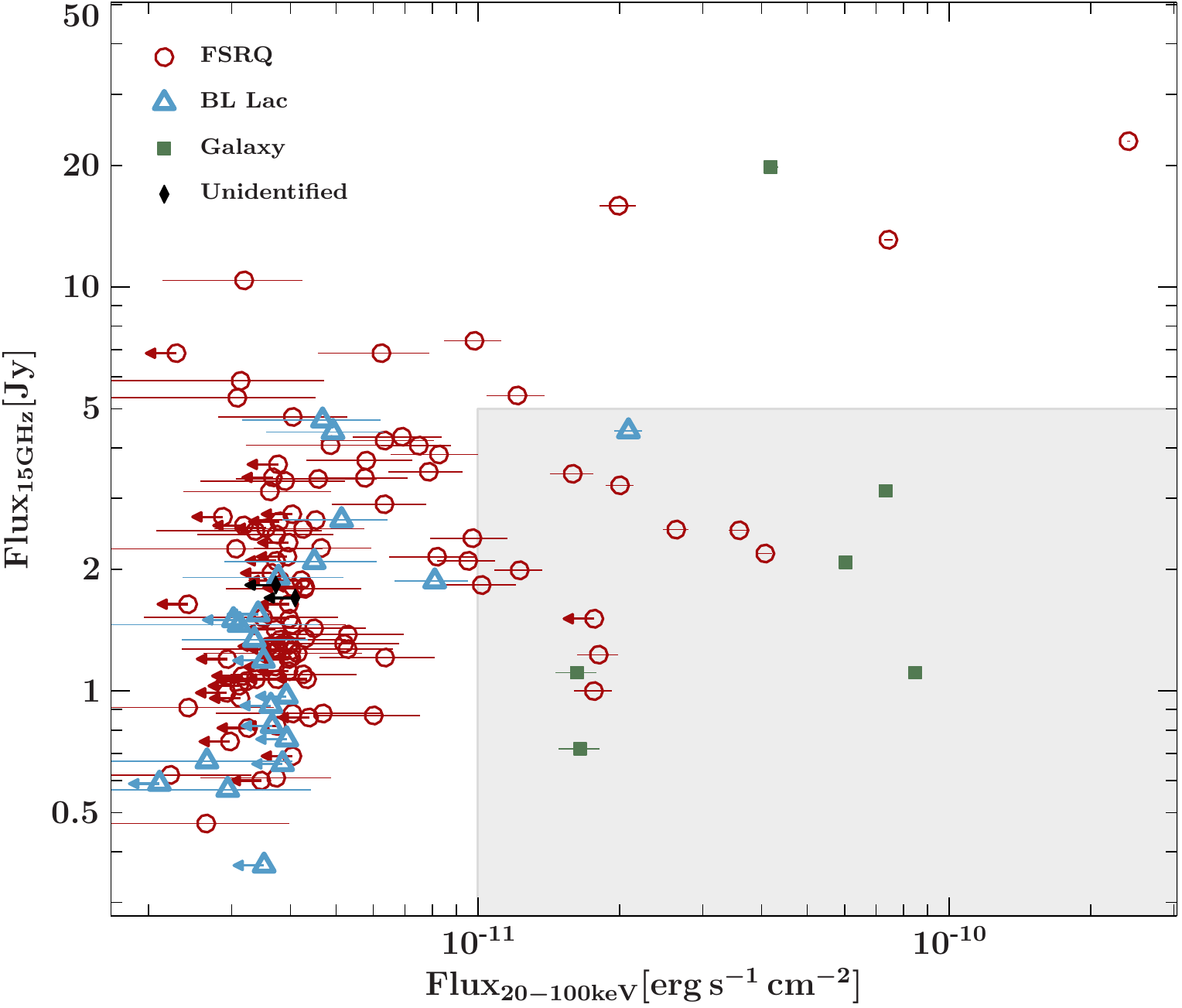}
\caption{MOJAVE-1 sample: BAT 20\,keV -- 100\,keV flux plotted against 15\,GHz flux density. The symbols with arrows to the left represent upper limit values. The shared area indicates a selected number of X-ray-bright while radio-faint sources (see text).}
\label{fig:graph_xflux-rflux} 
\end{figure}

\subsubsection{Influence of flaring sources}
\label{subsubsec:flaring_simul}

Since blazars are known to exhibit strong flaring behavior on a variety of time scales and because the BAT fluxes are derived from an integration over 105 months, it is reasonable to assume that at least a part of the blazar sample is characterized by fluxes that are distinctively higher compared to their quiet state. These increased fluxes may distort the $\log N$-$\log S$ distribution, moving data points toward higher fluxes, and, consequently, change the distribution's slope. 
A long-term variability study of AGN at hard X-rays, conducted by \citet{Soldi2014}, revealed significant variation of source flux in the 14\,keV -- 195\,keV band of the first 66 months of light curve data collected by \textit{Swift}/BAT. 
The authors calculated the variability estimator $S_v$, which expresses the intrinsic variability of a light curve of a source. The estimator is based on the maximum-likelihood estimate of the variability parameter $\sigma_\mathrm{Q}$, which depends on the individual light curve flux values and measurement errors \citep[see][Eq.~1]{Soldi2014}. This parameter is then normalized by the average flux of the specific source. An average value $\langle S_v \rangle$ in per cent for a sample of sources can then be calculated.
The blazar sub-sample in their study showed the highest variability ($\langle S_v \rangle = 33\% \pm 2 \%$) as well as an exceptionally high variability of two gamma-ray bright blazars ($\langle S_v \rangle = 90 \%$). About 80\% of the surveyed AGN are found to have variability on the time scale of months to years. 

In order to estimate the influence of randomly increased flux within our sample we assume that 80\% of a sample are variable with the maximum flux increase factor equal to the variability estimator $S_v$, for instance, a factor of 1.33 for $S_v = 33\%$. 
From our sample of 70 MOJAVE-1 blazars in the $\log N$-$\log S$ distribution we calculate $S_v$ for the available monthly binned \textit{Swift}/BAT light curves in the 105-month time span, and additionally the 70-month time span\footnote{https://swift.gsfc.nasa.gov/results/bs105mon/, https://swift.gsfc.nasa.gov/results/bs70mon/}. The results are presented in Table~\ref{tab:sv_results}.

\begin{table}
\caption{Mean variability amplitude estimator $\langle S_v \rangle$ derived from \textit{Swift}/BAT light curves of the 70- and 105-month data sets for the 70 X-ray-bright MOJAVE-1 blazars. The numbers in parentheses are the number of analyzed light curves in that category.}
\label{tab:sv_results}
\centering
\begin{tabular}{ccc}
\hline\hline
		Data Set 	& non-3FGL			& 3FGL\\	
\hline                        
        70-month      & $45 \pm 12$ (3)		& $39 \pm 6$ (12)\\ 
        105-month     & $69 \pm 9$ (3)			& $56 \pm 9$ (17)\\ 
\hline                                   
\end{tabular}
\end{table}

Based on the number of available light curves, it can be stated that the variability amplitude estimator $S_v$ for \textit{Fermi}/LAT-detected sources (on the basis of the 3FGL catalog) is approximately 33\% to 55\%. The results for \textit{Fermi}/LAT-non-detected sources differ significantly from approximately 30\% in the 70-month data set to almost 80\% in the 105-month data set. This, however, is likely due to the very low number of only three light curves per data set. Furthermore, the longer light curves also tend to have larger variability estimators.
As a conservative estimate we assume that the variability amplitude does not exceed 100\% for all light curves, equal to a maximum flux increase of a factor of two.

We start from a perfectly Euclidean flux distribution of 500 sources. The brightest source is set to 10 arbitrary flux units, the next source at a flux that is determined by following a slope in the $\log N$-$\log S$ diagram of $-1.5$. 
In this simulation we produce two separate effects. 
First, the flattening of the $\log N$-$\log S$ distribution at the lower end, which is caused by graphing the median radio flux whereas the selection of the sample was purely based on a possibly one time high state of a source, pushing the flux reading over the registration threshold. Second, a random flaring of 80\% of all sources in the simulated sample, as outlined for the X-ray band. 

A simulated flaring is induced in 400 randomly picked sources with a maximum random factor for the flux increase in the range of [1 -- 2] and a uniform probability distribution. We simulate a large number of instances of the first and the second flaring behaviors and fit a power law to each of the resulting distributions. All data points with fluxes larger than 0.5 of the arbitrary flux units are included in the fit (see Fig.~\ref{fig:simul_lognlogs}). A highly significant majority of 99.99\% of all fitted instances have power-law slopes larger than 1.28, which is thus not compatible with the fit of the BAT data ($1.13 \pm 0.04$), within $4\,\sigma$ of the slope's uncertainty range. This result makes a simple flaring scenario of the X-ray emission unlikely to be solely responsible for the small slope of the X-ray $\log N$-$\log S$ distribution.

\subsubsection{High X-ray and low radio flux sources}
\label{subsubsec:high_xflux_sources}

For the $\log N$-$\log S$ distribution of the BAT data to be steeper (more Euclidean), either faint sources are missing or bright sources are too numerous. Flux values that are only upper limits, and are consequently not in the $\log N$-$\log S$ diagram, cannot be responsible for the missing faint sources since the vast majority of all upper limits are located below $4 \cdot 10^{-12} \mathrm{erg \, s^{-1} cm^{-2}}$, a region that is still very well described by the power-law fit. 

The potential over-abundance of X-ray-bright sources compared to the Euclidean radio measurements is examined in the following.
As indicated by Fig.~\ref{fig:graph_xflux-rflux}, no noticeable correlation of radio against X-ray flux is present in the whole MOJAVE-1 source sample. Also, a Spearman's rank correlation coefficient of $\rho = 0.31$ does not indicate any significant link between the flux values in both bands. The figure shows a scatter plot with the flux data of both bands covering little more than two decades. At the same time, the distribution along the flux axes in both bands is dissimilar. Compared to the radio data set, the BAT fluxes (excluding upper limits) are concentrated at low values, below $10^{-11} \mathrm{erg \, s^{-1} cm^{-2}}$.  

The gray area in Fig.~\ref{fig:graph_xflux-rflux} marks X-ray fluxes greater than $10^{-11} \mathrm{erg \, s^{-1} cm^{-2}}$ and radio flux densities smaller than 5\,Jy, comprising 10 sources (9 FSRQs and BL Lac itself). 
If this set of X-ray bright and radio-faint blazars is removed, the resulting slope of the $\log N$-$\log S$ distribution is $1.54 \pm 0.06$, well compatible with the Euclidean case.

The scenario of a number of (bright) X-ray sources that govern the behavior of the blazar sample in the $\log N$-$\log S$ plot is equal to a population that shows intrinsic evolution in luminosity. The following section analyzes the sample in terms of evolution of luminosity per unit co-moving volume.

\subsubsection{LF data fit}
\label{subsec:lf_res}

We can exclude a dominant influence on the difference of the $\log N$-$\log S$ shapes through variability. 
In the following, we present the results of the test for intrinsic emission evolution. 
In fitting the hard X-ray (XLF) and radio (RLF) luminosity function, we have two goals: first, study the evolution of the respective emission in the MOJAVE-1 blazar sample, and second, answer the question whether the difference in slope in the $\log N$-$\log S$ distributions can be attributed to a different evolution in flux emission in both bands. 

We analyze the data sets for the hard X-ray and 15\,GHz bands for the MOJAVE-1 blazars. $\mathrm{AIC}$ values are calculated for all described models as well as the probability $p_j$ that model $j$ also maximizes the $\mathrm{AIC}$ value compared to the best fit model.
The fitted models are as described in Sect.~\ref{sec:xlf_analy} for single power laws: PDE, PLE, PDEg, PLEg, and the PLE case with no evolution ($k=g=0$). The double power-law models are generally less likely to produce an optimal fit, but are often not significantly less likely than the best fit ($p_i \approx 0.05 - 0.70$). However, we choose to neglect these models because of the additional number of free parameters which do not introduce significant improvements in the fits. The fitting parameters and all of the following results are also listed in Table~\ref{tab:LF_pj_results}.

The best fit for the X-ray data set (69 sources\footnote{The reduction from 70 to 69 sources stems from the missing redshift information of source 0300+470}) is achieved with the single power-law PDEg and PLEg models. Double power-law models resulted of PLE and PLEg are considerably less probable ($p_{\mathrm{PLE}} = 0.247$, $p_{\mathrm{PLEg}} = 0.320$). The model corresponding to no evolution can be excluded ($p_{\mathrm{no~evol}} = 0.0269$). 

The same sub-sample at radio frequencies gives a somewhat different picture. A simple evolution scenario (PLE, PDE) is preferred, with the second most likely cases of PLEg and PDEg considerably close ($p_{\mathrm{PLEg}} = p_{\mathrm{PDEg}} = 0.737$). The no evolution scenario can also not be fully excluded ($p_{\mathrm{no\,evol}} = 0.12$).

The analysis of the larger sample of 123 MOJAVE-1 blazars\footnote{The reduction from 125 to 123 sources stems from the missing redshift information of sources 0300+470 and 0814+425} in the radio band shows that the evolutionary models of PLEg and PDEg are the preferred case against no evolution ($p_{\mathrm{no~evol}} = 0.003$). Also, the largest available sample from \citet{Lister2015}, counting 170 blazars with redshift information, shows that PLEg is the preferred model against the no evolution case $p_{\mathrm{no~evol}} = 0.0049$). 

\begin{table*}
\caption{Results of the maximum likelihood fits of the analytic luminosity functions for X-ray and radio data sets. The relative probability of a model to describe the data compared to the best fit is expressed by $p_j$.}             % title of Table
\label{tab:LF_pj_results}
\centering
\begin{tabular}{ccccccccc}
\hline\hline
		Sample 	& LF Model & $A \, [\mathrm{Mpc^{-3}}]$ & $L_* \, [\mathrm{erg \, s^{-1}}]$ & $\gamma_1$ & $\gamma_2$ & $k$ & $g$ & $p_j$\\	
\hline                        
	69$_\mathrm{X}$	& PLEg & $8.99 \cdot 10^{-9}$ & $10^{44}$ & $1.05 \pm 0.12$ & & $2.20 \pm 0.65$ & $-0.56 \pm 0.19$ & 1.000 \\ 
				& PDEg & $3.57 \cdot 10^{-9}$ & $10^{44}$ & $1.05 \pm 0.12$ & & $4.51 \pm 1.54$ & $-1.14 \pm 0.41$ & 1.000 \\
				& PLEg (dpw) & $2.11 \cdot 10^{-10}$ & $(0.52 \pm 1.56) \cdot 10^{46}$ & $1.35 \pm 0.31$ & $0.68 \pm 0.32$ & $1.57 \pm 0.86$ & $-0.35 \pm 0.25$  & 0.320 \\
				& PLE (dpw)	& $1.97 \cdot 10^{-10}$ & $(1.62 \pm 2.05) \cdot 10^{46}$ & $0.54 \pm 0.19$ & $1.52 \pm 0.29$ & $0.38 \pm 0.30$ & & 0.247 \\
				& PLE		& $7.44 \cdot 10^{-9}$ & $10^{44}$ & $0.89 \pm 0.09$ &  & $0.29 \pm 0.31$ &  & 0.015\\ 
				& PDE		& $7.44 \cdot 10^{-9}$ & $10^{44}$ & $0.89 \pm 0.09$&  & $0.54 \pm 0.61$ &  & 0.015\\ 		
				& no evol.	& $7.41 \cdot 10^{-9}$  & $10^{44}$ & $0.82 \pm 0.05$ &  & $0$ &  & 0.027\\ 
				& 		&  &  &  &  &  &  & \\ 
	69$_\mathrm{R}$	& PLE	& $5.06 \cdot 10^{-10}$ & $10^{34}$ & $0.50 \pm 0.08$ &  & $-0.83 \pm 0.39$ &  & 1.000\\ 
				& PDE	& $5.07 \cdot 10^{-10}$ & $10^{34}$ & $0.50 \pm 0.08$ &  & $-1.25 \pm 0.54$ &  & 1.000\\ 
				& PLEg	& $2.61 \cdot 10^{-10}$ & $10^{34}$ & $0.53 \pm 0.09$ &  & $0.03 \pm 0.86$ & $-0.26 \pm 0.25$ & 0.737\\ 
				& PDEg	& $3.33 \cdot 10^{-10}$ & $10^{34}$ & $0.53 \pm 0.09$ &  & $0.05 \pm 1.32$ & $-0.41 \pm 0.38$ & 0.737\\ 
				& no evol.		& $2.92 \cdot 10^{-10}$ & $10^{34}$ & $0.65 \pm 0.05$ &  & $0$ &  & 0.120\\ 
				& 		&  &  &  &  &  &  & \\ 
	123$_\mathrm{R}$ & PLEg	& $2.00 \cdot 10^{-10}$ & $10^{34}$ & $0.53 \pm 0.07$ &  & $0.64 \pm 0.65$ & $-0.45 \pm 0.19$ & 1.000\\
				& PDEg	& $2.01 \cdot 10^{-10}$ & $10^{34}$ & $0.53 \pm 0.07$ &  & $0.98 \pm 1.01$ & $-0.69 \pm 0.30$ & 1.000\\
				& PLE	& $4.17 \cdot 10^{-10}$ & $10^{34}$ & $0.48 \pm 0.06$ &  & $-0.80 \pm 0.29$ &  & 0.105\\ 
				& PDE	& $4.17 \cdot 10^{-10}$ & $10^{34}$ & $0.48 \pm 0.06$ &  & $-1.19 \pm 0.39$ &  & 0.105\\ 
				& no evol.		& $2.38 \cdot 10^{-10}$ & $10^{34}$ & $0.63 \pm 0.04$ &  & 0 &  & 0.003\\ 
				& 		&  &  &  &  &  &  & \\ 
	$170_\mathrm{R}$ & PLEg & $2.17 \cdot 10^{-10}$ & $10^{34}$ & $0.57 \pm 0.06$ &  & $0.48 \pm 0.53$ & $-0.35 \pm 0.15$ & 1.000\\ 
				& PDEg	& $2.21 \cdot 10^{-10}$ & $10^{34}$ & $0.57 \pm 0.06$ &  & $0.76 \pm 0.83$ & $-0.56 \pm 0.24$ & 0.368\\  	
				& PLE	& $3.97 \cdot 10^{-10}$ & $10^{34}$ & $0.53 \pm 0.05$ &  & $-0.66 \pm 0.24$ &  & 0.118\\  	
				& PDE	& $3.97 \cdot 10^{-10}$ & $10^{34}$ & $0.53 \pm 0.05$ &  & $-1.01 \pm 0.34$ &  & 0.118\\ 
				& no evol.	& $2.45 \cdot 10^{-10}$ & $10^{34}$ & $0.65 \pm 0.03$ &  & 0 &  & 0.004\\  	
\hline                                   
\end{tabular}
\end{table*}

\begin{figure}
\centering
\includegraphics[width=\hsize]{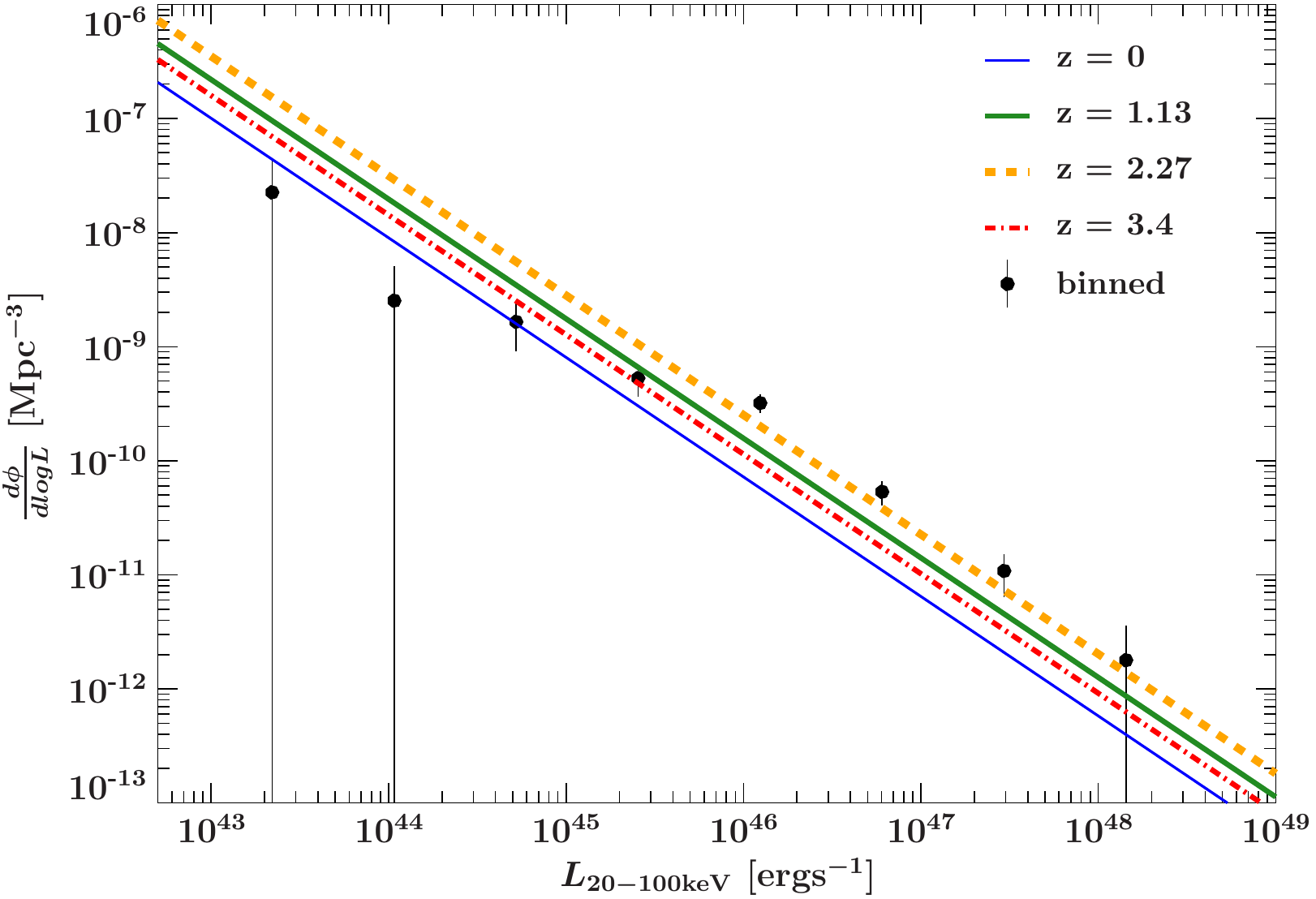}
\caption{MOJAVE-1 XLF model PLEg, one of the two best fit models. Binned LF and analytic form are fitted to the 70 source sub-sample for which BAT fluxes are non upper limits.}
\label{fig:graph_xlf} 
\end{figure}

Figure~\ref{fig:graph_xlf} shows the binned XLF and best fit model of the XLF for a number of redshifts. Both the binned XLF and the model fit show the general falling trend typical for luminosity functions of this type. Luminosity and density evolution fit the data equally well. In any case, it can be stated that evolution of the emission is favored by the analytic models compared to the static one. 

\begin{figure}
\centering
\includegraphics[width=\hsize]{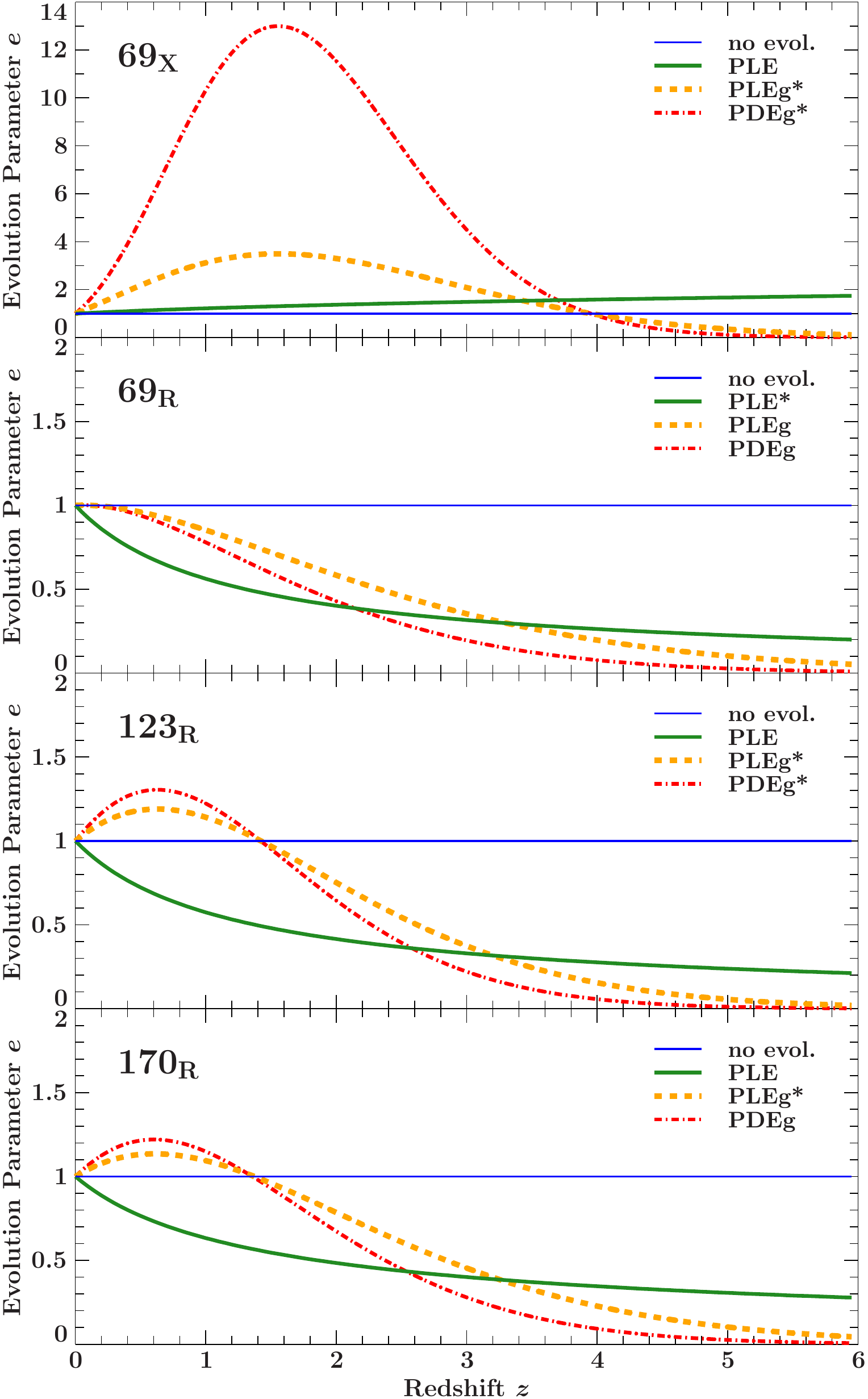}
\caption{Evolution parameter $e$ as a function of redshift for some of the fitted luminosity function models. The identification in the upper left corner corresponds to the samples in Table~\ref{tab:LF_pj_results}. The models that describe the data best are marked with an * in the legend.}
\label{fig:graph_evol_parameter} 
\end{figure}

In Fig.~\ref{fig:graph_evol_parameter}, we graph the evolutionary factor $e$ against redshift for all analyzed samples and for most of the luminosity function models in Table~\ref{tab:LF_pj_results}.
For the 69 source X-ray data set both best-fit models show a trend of positive evolution, that is, an increase in X-ray luminosity or density up to $z\approx 1.5$ and a negative trend after this point towards higher redshifts where sources tend to be less dense or luminous. Previous studies \citep[e.g.,][]{Ueda2003, Miyaji2015, Ranalli2016} regarding the XLF of AGN samples also have applied a model of luminosity-dependent density evolution. However, because of the relatively low samples size, and additional number of parameters in such models, we choose to only use the simplest available models for the XLF. 

Generally, all analyzed samples prefer an evolutionary scenario against no evolution at all. The preferred RLF models of the data sets of the 123 and 170 source samples show a peak in the at $z \approx 0.6$. The likeliest evolution models for both radio samples and the 69 source X-ray data set show that the luminosity output per volume increased up to a point in time for the given population and decreased again. However, the same 69 sample of the hard X-ray bright sources in the radio band shows no such peak. The evolutionary parameter $e$ only decreases with redshift. In any case, it is suggested that in terms of emission, that is, intrinsic luminosity, the MOJAVE-1 blazar sample behaves differently in the radio and hard X-ray regime. 

However, with the available data, we cannot fully exclude the contribution of a selection effect that is given by the reduction of the full radio sample to the 69 sources that are analyzed in the X-ray band. In any way, the conclusion of a wavelength-dependent evolution through the analysis of the luminosity functions and the fact that the X-ray sample includes over-proportionally many bright sources, leads us to believe that this explanation is consistent and viable.  

Assuming that a truly different evolutionary path for the radio and hard X-ray emission is present, that is, a steady decrease in the radio band with redshift or a small maximum at low redshifts but a large increase in luminosity of the population per volume for X-rays up to $z \approx 1.5$, would imply a relatively low proportion of low flux sources in the X-ray data. 
This relation can thus be seen to be at least partially responsible for the lack of low flux sources and therefore flat slope of the $\log N$-$\log S$ distribution.
  
We also test for evolution of the 69 X-ray bright MOJAVE-1 blazars using a standard $V/V_{\mathrm{max}}$ test \citep{Schmidt1968}. The result of the test value and its mean absolute error $\langle V/V_{\mathrm{max}}\rangle = 0.451 \pm 0.184$ suggest, although not significantly, a qualitative agreement with our previous results of (negative) evolution of the hard X-ray emission.

\subsection{CXB contribution}
\label{subsec:cxb}

We calculate the contribution to the hard X-ray background from blazars of the MOJAVE-1 type and the BAT catalog beamed AGN sample for comparison. The MOJAVE-1 blazar sub-sample yields the following results: Using the fit of the hard X-ray $\log N$-$\log S$ data set (see Fig.~\ref{fig:graph_logn-logs}), we extrapolate the number of sources brighter than $2.2 \cdot 10^{-12} \mathrm{erg s^{-1} cm^{-2}}$ (flux of the faintest source) for the entire sky and obtain:
\begin{linenomath*}\begin{equation}
N_{\mathrm{all sky}}^{\mathrm{MOJAVE}}(> 2.2 \cdot 10^{-12}) = N(> 2.2 \cdot 10^{-12}) \cdot \Omega_{\mathrm{sky}} \approx 139 \pm 9.
\label{eqn:result_xray_sources_allsky}
\end{equation}\end{linenomath*}
This number exceeds the actual total number of MOJAVE-1 sources for which a hard X-ray flux could be calculated (70 sources; see Sect.~\ref{subsubsec:logn-logs}) but is consistent with the larger total number of sources contributing to the BAT signal (Sect.~\ref{subsec:res:SNR}).
The integrated flux of radio-selected blazars of the MOJAVE-1 type above $2.2 \cdot 10^{-12} \mathrm{erg s^{-1} cm^{-2}}$ over the whole sky can thus be calculated from Eq.~\ref{eqn:contrib_cxb} using the parameters in Table \ref{tab:logn-logs-parameters}:
\begin{linenomath*}\begin{equation}
\begin{split}
F_{\mathrm{contrib}}^{\mathrm{deg^2,\,MOJAVE}} & = \int_{2.2 \cdot 10^{-12}}^{\infty} \alpha \cdot A \cdot F^{-\alpha} dF \\ & = (6.47^{+2.83}_{-1.51}) \cdot 10^{-14} \mathrm{erg s^{-1} cm^{-2} deg ^{-2}}
\label{eqn:result_xray_contribution_flux}
\end{split}
\end{equation}\end{linenomath*}
which is equivalent to 0.21\% of $F_{\mathrm{CXB,\,20 - 100\,keV}}^{\mathrm{deg^2}}$, with the uncertainty range of 0.16\% to 0.30\%. The contribution of MOJAVE-1-type blazars to the CXB increases only slowly with lower integration thresholds.

In order to reach a total blazar contribution to the CXB of 20\%, as determined by \citet{Ajello2009} based on X-ray selected blazars in the 15\,keV -- 55\,keV band, the lower flux integration limit needs to be $F=1.15 \cdot 10^{-27} \mathrm{erg s^{-1} cm^{-2}}$ for the measured slope of -1.13 of the MOJAVE-1 blazars. A number of $10^{14}$ sources per $\mathrm{deg^2}$ would be needed to contribute to the signal, which is orders of magnitude larger than the estimated total number of $2 \cdot 10^{12}$ galaxies up to a redshift of $z=8$ \citep{Conselice2016}. In contrast, a $\log N$-$\log S$ power law, normalized on the brightest source and with a slope of $-1.5$ would correspond to a more realistic minimum integration flux of $F=5 \cdot 10^{-15} \mathrm{erg s^{-1} cm^{-2}}$, or about 421 sources per $\mathrm{deg^2}$. 

In the study by \citet{Ajello2009}, the blazar $\log N$-$\log S$ distribution was likewise characterized by very steep slopes ($\alpha = $1.7 -- 2.0) and substantial contribution percentages. In the same study a contribution of Seyfert galaxies in the same energy band (15\,keV -- 55\,keV) was determined to be approximately 22\% to 55\%, depending on the specific evolutionary model, already putting another restriction on the maximum value for blazars. The noticeably different results of blazar CXB contribution of our study and that of \citet{Ajello2009} do not contradict each other because of the selection criteria for each blazar sample, that is, radio- vs.~X-ray-bright sources.
Finally, we compare the results of the MOJAVE-1 blazar sample to the $\log N$-$\log S$ distribution of the BAT 105-month source catalog beamed AGN sub-sample using the catalog flux values in the 14\,keV -- 195\,keV band. The analysis shows a contribution to the CXB of 0.31\% (0.29\% -- 0.35\%) when integrating starting at the faintest source flux in the sample ($6.3 \cdot 10^{-12} \mathrm{erg s^{-1} cm^{-2}}$).
Removing all MOJAVE-1 sources from that sub-sample reveals a contribution of 0.20\% (0.18\% -- 0.23\%). Both the MOJAVE-1 and BAT beamed AGN samples have a very similar flux contribution to the CXB. For reaching a total blazar contribution of 20\% a lower integration limit of $F = 2.5 \cdot 10^{-19} \mathrm{erg s^{-1} cm^{-2}}$ is required, corresponding to a source count of $7.3 \cdot 10^6$ per $\mathrm{deg^2}$. Both the radio-selected MOJAVE-1 sample and the BAT catalog beamed AGN show two distinct groups of sources that, through their different $\log N$-$\log S$ slopes, contribute in a significantly different way to the CXB in terms of necessary numbers.

\section{Discussion}
\label{sec:disc}

\subsection{Individual sources with peculiar hard X-ray properties}
\label{subsec:disc_data}

The distribution of photon indices for sources shown in Fig.~\ref{fig:histo_gamma} is very similar to the distribution for the soft X-ray regime (2\,keV -- 10 keV) as obtained by \textit{Swift}/XRT \citep{Chang2010}. A notable difference between both distributions	is a softer spectrum for the radio galaxy sub-set at hard X-rays whereas FSRQs and BL Lacs are in general broader distributed with no significant offset to the soft X-ray data. 
A peculiar case is set by the X-ray-bright radio galaxy 0316+413 (3C 84) at S/N = 58.2\,$\sigma$ with $\Gamma = 3.17 \pm 0.12$. The poor fit statistic of the power-law fit ($PG$ = 14.8), however, indicates that a more complex fit model and measurements of higher spectral resolution is needed. Similar fit results are presented in the 105-month BAT source catalog with $\Gamma = 3.82 \pm 0.09$ and the reduced fit statistic $\chi^2 = 3.8$ \citep{Oh2018}. Since the X-ray emission of this source in particular is not dominated by synchrotron or inverse-Compton (IC) processes, but by thermal emission \citep{Churazov2003} we do not expect a typical blazar spectrum.

Another source with relatively soft photon index is the BL Lac 0754+100 with $\Gamma = 3.04 \pm 1.56$ at S/N = 3.9\,$\sigma$. The fitted SED from \citet{Chang2010} suggests a HE emission bump smaller and flatter than the synchrotron bump in $\nu F_\nu$ scaling with the BAT energy band at the rising part of the HE bump. Considering the large error a BAT photon index $\Gamma < 2$ seems reasonable and more realistic. 

The BL Lac 0716+714 has the hardest spectrum of all sources in this class with $\Gamma = 0.83 \pm 0.46$ while being still reasonably bright at 5.4\,$\sigma$. Additionally, the very strong variability of this source \citep[e.g.,][]{Wagner1996} makes meaningful spectral measurements difficult. A number of studies in the hard X-ray band in the past have not been able to derive a spectral index above 10\,keV due to variability issues and low significance. The second \textit{INTEGRAL} AGN catalog \citep{Beckmann2009} lists 0716+714 as a source of strong variability ($> 0.5$\,mag) in the optical photometric V-band. The authors analyzed the data and derived a flux value under the assumption of a photon index of $\Gamma = 2$ because of low significance in the observed 18\,keV -- 60\,keV band of less than 5\,$\sigma$. 
In another study \citet{Pian2005} derived the flux of the source in the range of 30\,keV -- 60\,keV using high-state-triggered observations. This was also done by freezing the photon index to the known value of the Crab pulsar $\Gamma = 2.1$, because of low significance. 
The very hard photon index of our measurement can be attributed to the position of the BAT band at a very steep rising part of the HE bump. 
Recent observations of the source with \textit{Swift}/XRT (0.3\,keV -- 10\,keV) and \textit{NuSTAR} (3\,keV -- 50\,keV) in a flaring state \citep{Wierzcholska2016} have clearly shown the spectral minimum between both emission bumps in the covered X-ray range at $E_{\mathrm{br}} = (8.01 \pm 0.56)$\,keV. The spectrum has been modeled with a broken power law with the resulting photon indices of $\Gamma_1 = 2.40 \pm 0.01$ and $\Gamma_2 = 1.61 \pm 0.05$. The found break energy $E_{\mathrm{br}}$ has been the highest value ever measured.
A previous study by \citet{Wierzcholska2015} found break energy values between 2\,keV and 5\,keV during another flare state. Also, a direct correlation of elevated flux levels and break energy was found. The position of the very steep rising part of the HE bump can be placed within the BAT band (20 keV – 100 keV) with high confidence. The detection of this source by \textit{Fermi}/LAT \citep{Lister2009_fermi} supports this assumption, since the HE bump is comparable to the majority of the sample, shifted to higher energies, allowing \textit{Fermi}/LAT to detect more high energy photons.

A relatively broad distribution of spectral shapes for low-peaked sources was also reported in a recent study by \citet{Marchesini2019}. The authors analyzed a large sample of high- and low-peaked BL Lac type objects in the soft X-ray band and found an especially broad distribution of hardness ratios for the log-peaked sub-set, which has been attributed to the intersecting region of synchrotron and HE bump in the SED.

\subsection{Fermi detections of gamma-faint blazars}
\label{subsec:disc_fermi_detections}

Despite the substantial integration time of the LAT instrument during the four years (3FGL catalog), some of the most luminous blazars at radio wavelengths are not detected by \textit{Fermi}/LAT (Sect.~\ref{sec:sam}). \citet{Lister2015}, using the extended MOJAVE-1.5 sample, found that the synchrotron bump of the \textit{Fermi}/LAT-non-detected radio-loud AGN tend to be characterized by a lower peak frequency than the \textit{Fermi}/LAT-detected part of the sample. This was partially explained by the SED position and consequently lower flux values above 0.1\,GeV besides low relativistic boosting of the jet and its velocity. Also, low radio variability levels were associated with a low probability of detection in the gamma-ray range.

In Sect.~\ref{subsec:res:gamma_3lac}, we have presented the distribution of the measured BAT photon indices of all fitted MOJAVE-1 blazars, \textit{Fermi}/LAT-detected and non-detected. The analysis, using the extended 2-sample KS testing method, showed that both \textit{Fermi}/LAT-detected and non-detected sub-samples exhibit significantly different distributions, suggesting that detected sources (harder) are associated with higher peaked and non-detected sources (softer) with lower peaked SEDs.
The shown distribution of HE peak frequencies against BAT photon index in Fig.~\ref{fig:graph_nu_peak-gamma} supports this trend. 
A very similar relation has been found by \citet{Abdo2010_sed_fermi_blazars}. The authors report a strong correlation of gamma-ray photon index and the synchrotron as well as HE bump peak frequency. The higher (softer) the index, the lower the peak frequency. Furthermore, the gamma-ray photon index was found to be strongly correlated with the X-ray photon index.

Previous to the publication of the forth \textit{Fermi}/LAT source catalog \citep{4FGL2019}, \citet{Lister2015} identified three blazars in the extended 1.5\,Jy MOJAVE sample which were thought to be strong candidates for a future detection by \textit{Fermi}/LAT on the basis of their observed properties, that is, high radio modulation index and high apparent jet speed. 
None of the three mentioned blazars (III Zw 2, PKS 0119+11, and 4C +69.21) are listed in the 4FGL catalog, which includes approximately double the number of blazars compared to the the 3FGL/3LAC catalog.

We propose a different approach for predicting which blazars in the MOJAVE-1 sample will most likely be detected in future \textit{Fermi} catalogs. We compare the HE peak frequencies and BAT photon indices of non-detected with detected sample sources.
The two previously non-detected FSRQs (3LAC) 1458+718 and 2145+067 have HE peak frequencies comparable with the majority of detected sources, as seen in Fig.~\ref{fig:graph_nu_peak-gamma}. However, when compared with the rest of all MOJAVE-1 blazars, both sources have relatively low apparent maximum jet speeds of $6.6c$ and $2.8c$, respectively. As indicated in the plot both sources have since been detected, reinforcing the hypothesis that HE peak position is likely the dominant factor for a significant gamma-ray signal. 

Based on our findings, we can predict likely gamma-ray detections of not yet detected blazars in the MOJAVE-1 sample.  
The mean of the maximum jet speeds of all \textit{Fermi}/LAT-detected MOJAVE-1 blazars is $12.8c$ and a maximum value of $41.8c$ \citep{Lister2013}. A source with a hard BAT spectrum ($\Gamma = 1.50 \pm 0.44$) and reasonably high apparent jet speed of $9.9c$ is the FSRQ 2005+403. Its HE peak frequency is on the order of a magnitude lower than the bulk of all detected MOJAVE-1 blazars, although higher than the majority of all non-detected ones.
Also, the source 0742+103 can be considered a detection candidate because of its relatively high peak frequency. This case is of particular interest, since the previously proposed criteria of high jet speed and variability index ($v = 2.8c$ and $m = 0.016$) are far from ideal considering the mean values of these quantities in the mentioned study ($v_{\mathrm{mean}} = 10.1c$ and $m_{\mathrm{mean}} = 0.2$). A future gamma-ray detection would certainly emphasize the importance and impact of soft and hard X-ray coverage in order to estimate the HE peak position and, thus, the gamma-ray detection probability for a blazar.

\subsection{log N-log S distribution and LF}
\label{subsec:disc_logn-logs}

The $\log N$-$\log S$ distribution of the MOJAVE-1 blazars is characterized by a flat and non-Euclidean slope in the 20\,keV -- 100\,keV band ($\alpha_{\text{Moj,BAT}}=1.13 \pm 0.04$). Most X-ray AGN surveys, which are usually dominated by Seyferts (e.g., surveys with \textit{INTEGRAL} \citep{Beckmann2006} and \textit{Swift}/BAT \citep{Ajello2012}), show a slope that is compatible with a Euclidean distribution in space. However, the beamed AGN sub-sample of the hard X-ray BAT source catalog \citep{Oh2018} is also described by a relatively flat fit function to the $\log N$-$\log S$ distribution ($\alpha_{\text{cat,BAT}}=1.24 \pm 0.02$), reinforcing this peculiar trend of blazars in this energy range. It could be argued that a substantial number of the BAT catalog's unidentified 243 sources are indeed blazars, which would likely influence the distribution's slope. The fraction of unidentified sources in the MOJAVE-1 sample, on the other hand, is with the two sources 0648--165 and 1213--172 relatively small.

The radio data set incorporates the full set of all 125 MOJAVE-1 blazars, whereas the BAT data that we used comprise the 70 brightest blazars in the 20\,keV -- 100\,keV energy band. One likely effect responsible for the different slopes is the measurement of two different frequency bands. The sample is statistically complete in the radio band by definition. The X-ray fluxes, on the other hand, do not correlate strongly with the radio emission. 
The plot of BAT flux vs.~VLBI MOJAVE flux at 15\,GHz (Fig.~\ref{fig:graph_xflux-rflux}) shows no significant positive correlation of both data sets for any of the sub types. However, a majority of all the sample is concentrated at low X-ray flux (Sect.~\ref{subsec:res:xflux}) whereas these sources scatter over a wide range of radio fluxes (approximately 0.6\,Jy -- 5\,Jy). The position of data points in the 15\,GHz $\log N$-$\log S$ diagram are consequently shifted towards low X-ray flux, and in this case even below the set flux limit of the X-ray $\log N$-$\log S$ plot, lowering the number of sources at the bright end, and thus reducing the distribution's slope. 
This on its own could lead to the conclusion that the difference between $\alpha_{\text{Moj,BAT}}$ and $\alpha_{\text{Moj,15GHz}}$ is due to source selection bias of a sample in a different wavelength regime. However, the very similar slope of the $\log N$-$\log S$ distribution for the MOJAVE-1 sample and the beamed AGN sub-sample of BAT 105-month source catalog make a pure selection bias unlikely.
The influence of flaring sources can be neglected, as well. A likely conclusion is a different evolutionary path for X-ray and radio emission in the source sample. Whereas the preferred XLF model indicates a maximum of the evolution parameter $e$ at around $z=1.5$ the RLFs of the blazars of the full MOJAVE-1 and the extended MOJAVE-1.5 samples have maxima at approximately $z<1$.

In a recent study, \citet{Ighina2019} have found a significant difference of the evolutionary trend of soft X-ray to radio luminosity ratio in a high-redshift blazar sample ($z>4$) compared to a sample at low redshifts. The ratio was found to be approximately 2 -- 3 times larger for the high-redshift sample. This trend, although of the soft X-ray band (0.2\,keV -- 10\,keV), generally supports our results, which indicate an earlier X-ray emission maximum compared to the radio data.
The authors tentatively attribute the decreasing fraction of X-ray to radio emission with the interaction of extended jet regions with the CMB by the IC process, which is strongly dependent on the redshift. 
However, \citet{Ighina2019} also make the argument that, in order to reconcile previous results from radio- and X-ray-selected source samples \citep{Ajello2009,Caccianiga2019}, the ratio of extended to compact emission has to be non-uniform, likely depending on the specific source and the source selection.

In general, the $\log N$-$\log S$ distribution stays consistent to very low fluxes in the BAT band. Thus, the behavior of the fainter sources in the plotted sample is not significantly different to the rest. This consistency reinforces our approach of subtracting the fitted noise component of the S/N distribution from the measured S/N distribution of the blazar sample (Sect.~\ref{subsec:res:SNR}). Also, the analysis of the extended MOJAVE-1.5 sample shows the same slope of the distribution, excluding any influence that might have been introduced by non-uniform source selection criteria in the original MOJAVE-1 sample.

The contribution of AGN, blazars, and other source types to the CXB has primarily been examined using source samples that are also X-ray or gamma-ray bright \citep[see, e.g.,][]{Mateos2008, Ajello2009, Bottacini2012}.
The results for the blazar contribution to the hard CXB vary greatly in previous works, largely due to different source samples and analytic approaches. In this work we use a sample of radio-bright low-peaked blazars and obtain up to 0.30\% of the contribution in the 20\,keV -- 100\,keV band above $2.2 \cdot 10^{-12} \mathrm{erg s^{-1} cm^{-2}}$, which again is compatible with the beamed AGN sub-sample of the 105-month BAT survey catalog.

While the contribution to the approximately 20\% of the CXB from blazars \citep{Ajello2009} can be reached with a low enough X-ray flux integration limit of $F=5 \cdot 10^{-15} \mathrm{erg s^{-1} cm^{-2}}$ for a Euclidean slope, the measured slope of $-1.13$ from the MOJAVE-1 sources is too flat for a physically reasonable result (Sect.~\ref{subsec:cxb}). 
Our results thus suggest that X-ray and radio-selected blazars behave like two two different populations of which the radio-selected population makes only a negligible contribution to the CXB.
The parameters of the fits of the luminosity functions to the MOJAVE-1 BAT data show a key difference to comparable blazar studies of in soft and hard X-ray bands. Studies by, for example, \citet{Ajello2009}, \citet{Sazonov2007}, and \citet{Hasinger2005} revealed very similar parameter values for the evolutionary term of the luminosity functions of larger AGN and blazar samples. However, the slopes of the fitted power law are usually relatively steep, that is, around $\gamma=2-3$. If a double power law was applied the index of the lower luminosity part was flat at around $\gamma=0.5$. In contrast, the most likely models of the fitted MOJAVE-1 BAT data show indices of approximately $\gamma=1$. The analyzed radio data sets reveal even flatter indices of $\gamma=0.5-0.65$. The fitted blazar luminosity functions from \citet{Ajello2009} for the energy range of 15\,keV -- 55\,keV show very similar source densities at low luminosities around $10^{44}\,\mathrm{erg\,s^{-1}}$. For higher luminosities of $10^{48}\,\mathrm{erg\,s^{-1}}$ the source density is approximately $10^3$ times lower compared to our studies. The relative scarcity of low luminosity sources in the MOJAVE-1 sample could also be attributed to the 52 blazars that have been omitted in the analysis because of low signal strength (upper limits). The radio luminosity function of the complete MOJAVE-1 blazar sample of 122 sources shows a similar behavior without the possible influence of missing contributing sources, however. It is therefore highly suggested that the X-ray-selected blazar samples of previous studies are of a different blazar population and evolutionary behavior than the radio-selected MOJAVE-1 sample.

Compared to the MOJAVE-1 blazar sample, the 38 hard X-ray selected sources in the study by \citet{Ajello2009} show a large amount of very low (18 with $z<0.5$) and high redshift sources (10 with $z>2$). The HE emission bump of high-redshift sources is shifted down to the BAT detection band. Also, the included BL Lac type sources are almost all HBLs. This creates a fundamentally different selection of sources and spectral characteristics in the hard X-ray regime relative to the radio-selected sample with a much more even redshift distribution in this study. Again, this shows that the different results of positive blazar evolution \citep{Ajello2009} against the negative evolution (MOJAVE-1, hard X-rays) do not directly contradict each other.

\section{Summary and conclusions}
\label{sec:conc}

In the following, we list the major results of our study of the hard X-ray properties of the radio-selected and statistically complete MOJAVE-1 AGN sample, consisting mainly of low-peaked blazars. We present the spectral characteristics of this under-represented and rare AGN type at hard X-rays in the 20\,keV -- 100\,keV band, based on the \textit{Swift}/BAT 105-month survey maps.

\begin{itemize}
	\item[$\bullet$] The X-ray flux, luminosities, and photon indices for the MOJAVE-1 AGN sample are now available for the first time in this spectral band. Out of the 135 sources in the sample, 121 are characterized by a hard X-ray signal that is not compatible with background noise, although most sources have low S/N values. 
	\item[$\bullet$] The derived distributions of hard X-ray flux, luminosity, and photon index clearly distinguish FSRQs, radio galaxies, and BL Lacs. Furthermore, the distribution of photon indices is significantly different for \textit{Fermi}/LAT-detected and non-detected sources, indicating that the position of the HE peak in the SED is one likely contributor to the question why certain radio-bright AGN are not also gamma-bright.
	\item[$\bullet$] The source count distribution / $\log N$-$\log S$ for the 70 X-ray-brightest blazars in the MOJAVE-1 sample shows a relatively flat slope of  $\alpha = 1.13 \pm 0.04$, which is clearly not Euclidean. One possible main contributor responsible for the difference in slope, and consequently the scarcity of hard X-ray detected blazars, is the different evolution of hard X-ray emission, peaking at around $z=1.5$. 
	\item[$\bullet$] We derive the contribution of the hard X-ray flux for the MOJAVE-1 and beamed AGN BAT catalog samples to the CXB and find that approximately 0.2\% -- 0.3\% of the CXB can be resolved into the sample sources for fluxes greater that their respective survey limits. Compared to earlier studies with X-ray-selected blazar samples this contribution is significantly lower, which suggests a different blazar population due to the selection criteria of the MOJAVE-1 survey, that is, low-peaked and radio-loud compact and beamed sources. 
\end{itemize}

The obtained properties for the various AGN types will become important for further research in areas such as, for example, the blazar sequence and broadband SED modeling, especially near the spectral minimum between synchrotron and HE emission peak. 
Large and deep future X-ray surveys, using, for example, the recently launched eROSITA X-ray telescope, will further complete the understanding the source types that are responsible for the creation of the CXB. Since the soft and hard X-ray regimes are instrumental for the estimation of future gamma-ray detections of beamed AGN in terms of SED position, complementary studies of both bands are essential for finding interesting new targets for gamma-ray studies using GeV and TeV observatories, such as the currently constructed Cherenkov Telescope Array (CTA).  

\begin{acknowledgements}

We want to thank Moritz B\"ock and Takamitsu Miyaji for useful discussion of the analysis methods. We also thank the anonymous referee for many useful comments and suggestions that helped improving the paper. This work was partially funded by Deutsche Forschungsgemeinschaft grant 50\,OR\,1709.
This research has made use of data from the MOJAVE database that is maintained by the MOJAVE team \citep{Lister2018} and the ISIS functions (ISISscripts), provided by the ECAP/Remeis observatory and MIT (http://www.sternwarte.uni-erlangen.de/isis/).
The \textit{Swift}/BAT survey maps have been provided by the \textit{Swift} team.
Furthermore, this research has made use of the NASA/IPAC Extragalactic Database NED (Jet Propulsion Laboratory, California
Institute of Technology / NASA) and of the SIMBAD database (CDS, Strasbourg, France). 
The work of M. Kreter is supported by the South African Research Chairs Initiative (grant no. 64789) of the Department of Science and Innovation and the National Research Foundation\footnote{Any opinion, finding and conclusion or recommendation expressed in this material is that of the authors and the NRF does not accept any liability in this regard.} of South Africa.

\end{acknowledgements}

\newpage

\bibliographystyle{aa}
{\small
\bibliography{mybibliography}

\begin{thebibliography}{65}
\expandafter\ifx\csname natexlab\endcsname\relax\def\natexlab#1{#1}\fi

\bibitem[{{Abdo} {et~al.}(2010){Abdo}, {Ackermann}, {Agudo}, {Ajello}, {Aller},
  {Aller}, {Angelakis}, {Arkharov}, {Axelsson}, {Bach}, \&
  et~al.}]{Abdo2010_sed_fermi_blazars}
{Abdo}, A.~A., {Ackermann}, M., {Agudo}, I., {et~al.} 2010, \apj, 716, 30

\bibitem[{{Ackermann} {et~al.}(2011){Ackermann}, {Ajello}, {Allafort},
  {Angelakis}, {Axelsson}, {Baldini}, {Ballet}, {Barbiellini}, {Bastieri},
  {Bellazzini}, {Berenji}, {Blandford}, {Bloom}, {Bonamente}, {Borgland},
  {Bouvier}, {Bregeon}, {Brez}, {Brigida}, {Bruel}, {Buehler}, {Buson},
  {Caliandro}, {Cameron}, {Cannon}, {Caraveo}, {Casand jian}, {Cavazzuti},
  {Cecchi}, {Charles}, {Chekhtman}, {Cheung}, {Ciprini}, {Claus},
  {Cohen-Tanugi}, {Cutini}, {de Palma}, {Dermer}, {Silva}, {Drell}, {Dubois},
  {Dumora}, {Escande}, {Favuzzi}, {Fegan}, {Focke}, {Fortin}, {Frailis},
  {Fuhrmann}, {Fukazawa}, {Fusco}, {Gargano}, {Gasparrini}, {Gehrels},
  {Giglietto}, {Giommi}, {Giordano}, {Giroletti}, {Glanzman}, {Godfrey},
  {Grandi}, {Grenier}, {Guiriec}, {Hadasch}, {Hayashida}, {Hays}, {Healey},
  {J{\'o}hannesson}, {Johnson}, {Kamae}, {Katagiri}, {Kataoka},
  {Kn{\"o}dlseder}, {Kuss}, {Lande}, {Lee}, {Longo}, {Loparco}, {Lott},
  {Lovellette}, {Lubrano}, {Makeev}, {Max-Moerbeck}, {Mazziotta}, {McEnery},
  {Mehault}, {Michelson}, {Mizuno}, {Monte}, {Monzani}, {Morselli},
  {Moskalenko}, {Murgia}, {Naumann-Godo}, {Nishino}, {Nolan}, {Norris}, {Nuss},
  {Ohsugi}, {Okumura}, {Omodei}, {Orlando}, {Ormes}, {Ozaki}, {Paneque},
  {Pavlidou}, {Pelassa}, {Pepe}, {Pesce-Rollins}, {Pierbattista}, {Piron},
  {Porter}, {Rain{\`o}}, {Razzano}, {Readhead}, {Reimer}, {Reimer}, {Richards},
  {Romani}, {Sadrozinski}, {Scargle}, {Sgr{\`o}}, {Siskind}, {Smith},
  {Spandre}, {Spinelli}, {Strickman}, {Suson}, {Takahashi}, {Tanaka}, {Taylor},
  {Thayer}, {Thayer}, {Thompson}, {Torres}, {Tosti}, {Tramacere}, {Troja},
  {Vandenbroucke}, {Vianello}, {Vitale}, {Waite}, {Wang}, {Winer}, {Wood},
  {Yang}, \& {Ziegler}}]{Ackermann2011}
{Ackermann}, M., {Ajello}, M., {Allafort}, A., {et~al.} 2011, \apj, 741, 30

\bibitem[{{Ackermann} {et~al.}(2015){Ackermann}, {Ajello}, {Atwood}, {Baldini},
  {Ballet}, {Barbiellini}, {Bastieri}, {Becerra Gonzalez}, {Bellazzini},
  {Bissaldi}, {Blandford}, {Bloom}, {Bonino}, {Bottacini}, {Brandt}, {Bregeon},
  {Britto}, {Bruel}, {Buehler}, {Buson}, {Caliandro}, {Cameron}, {Caragiulo},
  {Caraveo}, {Carpenter}, {Casandjian}, {Cavazzuti}, {Cecchi}, {Charles},
  {Chekhtman}, {Cheung}, {Chiang}, {Chiaro}, {Ciprini}, {Claus},
  {Cohen-Tanugi}, {Cominsky}, {Conrad}, {Cutini}, {D'Abrusco}, {D'Ammando}, {de
  Angelis}, {Desiante}, {Digel}, {Di Venere}, {Drell}, {Favuzzi}, {Fegan},
  {Ferrara}, {Finke}, {Focke}, {Franckowiak}, {Fuhrmann}, {Fukazawa},
  {Furniss}, {Fusco}, {Gargano}, {Gasparrini}, {Giglietto}, {Giommi},
  {Giordano}, {Giroletti}, {Glanzman}, {Godfrey}, {Grenier}, {Grove},
  {Guiriec}, {Hewitt}, {Hill}, {Horan}, {Itoh}, {J{\'o}hannesson}, {Johnson},
  {Johnson}, {Kataoka}, {Kawano}, {Krauss}, {Kuss}, {La Mura}, {Larsson},
  {Latronico}, {Leto}, {Li}, {Li}, {Longo}, {Loparco}, {Lott}, {Lovellette},
  {Lubrano}, {Madejski}, {Mayer}, {Mazziotta}, {McEnery}, {Michelson},
  {Mizuno}, {Moiseev}, {Monzani}, {Morselli}, {Moskalenko}, {Murgia}, {Nuss},
  {Ohno}, {Ohsugi}, {Ojha}, {Omodei}, {Orienti}, {Orlando}, {Paggi}, {Paneque},
  {Perkins}, {Pesce-Rollins}, {Piron}, {Pivato}, {Porter}, {Rain{\`o}},
  {Rando}, {Razzano}, {Razzaque}, {Reimer}, {Reimer}, {Romani}, {Salvetti},
  {Schaal}, {Schinzel}, {Schulz}, {Sgr{\`o}}, {Siskind}, {Sokolovsky}, {Spada},
  {Spandre}, {Spinelli}, {Stawarz}, {Suson}, {Takahashi}, {Takahashi},
  {Tanaka}, {Thayer}, {Thayer}, {Tibaldo}, {Torres}, {Torresi}, {Tosti},
  {Troja}, {Uchiyama}, {Vianello}, {Winer}, {Wood}, \&
  {Zimmer}}]{Ackermann2015}
{Ackermann}, M., {Ajello}, M., {Atwood}, W.~B., {et~al.} 2015, \apj, 810, 14

\bibitem[{{Ajello} {et~al.}(2012){Ajello}, {Alexander}, {Greiner}, {Madejski},
  {Gehrels}, \& {Burlon}}]{Ajello2012}
{Ajello}, M., {Alexander}, D.~M., {Greiner}, J., {et~al.} 2012, \apj, 749, 21

\bibitem[{{Ajello} {et~al.}(2009){Ajello}, {Costamante}, {Sambruna}, {Gehrels},
  {Chiang}, {Rau}, {Escala}, {Greiner}, {Tueller}, {Wall}, \&
  {Mushotzky}}]{Ajello2009}
{Ajello}, M., {Costamante}, L., {Sambruna}, R.~M., {et~al.} 2009, \apj, 699,
  603

\bibitem[{{Ajello} {et~al.}(2008{\natexlab{a}}){Ajello}, {Greiner}, {Sato},
  {Willis}, {Kanbach}, {Strong}, {Diehl}, {Hasinger}, {Gehrels}, {Markwardt},
  \& {Tueller}}]{Ajello2008_cxb_albedo}
{Ajello}, M., {Greiner}, J., {Sato}, G., {et~al.} 2008{\natexlab{a}}, \apj,
  689, 666

\bibitem[{{Ajello} {et~al.}(2008{\natexlab{b}}){Ajello}, {Rau}, {Greiner},
  {Kanbach}, {Salvato}, {Strong}, {Barthelmy}, {Gehrels}, {Markwardt}, \&
  {Tueller}}]{Ajello2008_bat_III}
{Ajello}, M., {Rau}, A., {Greiner}, J., {et~al.} 2008{\natexlab{b}}, \apj, 673,
  96

\bibitem[{{Akaike}(1973)}]{Akaike1973}
{Akaike}, H. 1973, Second International Symposium on Information Theory
  (Tsahkadsor, 1971), Akadémiai Kiadó, Budapest, 267–281

\bibitem[{{Barthelmy} {et~al.}(2005){Barthelmy}, {Barbier}, {Cummings},
  {Fenimore}, {Gehrels}, {Hullinger}, {Krimm}, {Markwardt}, {Palmer},
  {Parsons}, {Sato}, {Suzuki}, {Takahashi}, {Tashiro}, \& {Tueller}}]{BAT}
{Barthelmy}, S.~D., {Barbier}, L.~M., {Cummings}, J.~R., {et~al.} 2005, SSR,
  120, 143

\bibitem[{{Baumgartner} {et~al.}(2013){Baumgartner}, {Tueller}, {Markwardt},
  {Skinner}, {Barthelmy}, {Mushotzky}, {Evans}, \& {Gehrels}}]{baumgartner2013}
{Baumgartner}, W.~H., {Tueller}, J., {Markwardt}, C.~B., {et~al.} 2013, \apjs,
  207, 19

\bibitem[{{Becker} {et~al.}(1995){Becker}, {White}, \& {Helfand}}]{Becker1995}
{Becker}, R.~H., {White}, R.~L., \& {Helfand}, D.~J. 1995, \apj, 450, 559

\bibitem[{{Beckmann} {et~al.}(2009){Beckmann}, {Soldi}, {Ricci},
  {Alfonso-Garz{\'o}n}, {Courvoisier}, {Domingo}, {Gehrels}, {Lubi{\'n}ski},
  {Mas-Hesse}, \& {Zdziarski}}]{Beckmann2009}
{Beckmann}, V., {Soldi}, S., {Ricci}, C., {et~al.} 2009, \aap, 505, 417

\bibitem[{{Beckmann} {et~al.}(2006){Beckmann}, {Soldi}, {Shrader}, {Gehrels},
  \& {Produit}}]{Beckmann2006}
{Beckmann}, V., {Soldi}, S., {Shrader}, C.~R., {Gehrels}, N., \& {Produit}, N.
  2006, \apj, 652, 126

\bibitem[{{Bottacini} {et~al.}(2012){Bottacini}, {Ajello}, \&
  {Greiner}}]{Bottacini2012}
{Bottacini}, E., {Ajello}, M., \& {Greiner}, J. 2012, \apjs, 201, 34

\bibitem[{{Burnham} \& {Anderson}(2004)}]{Burnham2004}
{Burnham}, K.~P. \& {Anderson}, D.~R. 2004, Sociological Methods \& Research,
  33, 261–304

\bibitem[{{Burrows} {et~al.}(2005){Burrows}, {Hill}, {Nousek}, {Kennea},
  {Wells}, {Osborne}, {Abbey}, {Beardmore}, {Mukerjee}, {Short}, {Chincarini},
  {Campana}, {Citterio}, {Moretti}, {Pagani}, {Tagliaferri}, {Giommi},
  {Capalbi}, {Tamburelli}, {Angelini}, {Cusumano}, {Br{\"a}uninger}, {Burkert},
  \& {Hartner}}]{Burrows2005}
{Burrows}, D.~N., {Hill}, J.~E., {Nousek}, J.~A., {et~al.} 2005, SSR, 120, 165

\bibitem[{{Caccianiga} {et~al.}(2019){Caccianiga}, {Moretti}, {Belladitta},
  {Della Ceca}, {Ant{\'o}n}, {Ballo}, {Cicone}, {Dallacasa}, {Gargiulo},
  {Ighina}, {March{\~a}}, \& {Severgnini}}]{Caccianiga2019}
{Caccianiga}, A., {Moretti}, A., {Belladitta}, S., {et~al.} 2019, \mnras, 484,
  204

\bibitem[{{Cash}(1979)}]{cash1979}
{Cash}, W. 1979, \apj, 228, 939

\bibitem[{{Chang}(2010)}]{Chang2010}
{Chang}, C.-S. 2010, PhD thesis, Max-Planck-Institut f{\"u}r Radioastronomie

\bibitem[{{Churazov} {et~al.}(2003){Churazov}, {Forman}, {Jones}, \&
  {B{\"o}hringer}}]{Churazov2003}
{Churazov}, E., {Forman}, W., {Jones}, C., \& {B{\"o}hringer}, H. 2003, \apj,
  590, 225

\bibitem[{{Conselice} {et~al.}(2016){Conselice}, {Wilkinson}, {Duncan}, \&
  {Mortlock}}]{Conselice2016}
{Conselice}, C.~J., {Wilkinson}, A., {Duncan}, K., \& {Mortlock}, A. 2016,
  \apj, 830, 83

\bibitem[{{Draper} \& {Ballantyne}(2009)}]{Draper2009}
{Draper}, A.~R. \& {Ballantyne}, D.~R. 2009, \apj, 707, 778

\bibitem[{{Ebrero} {et~al.}(2009){Ebrero}, {Carrera}, {Page}, {Silverman},
  {Barcons}, {Ceballos}, {Corral}, {Della Ceca}, \& {Watson}}]{Ebrero2009}
{Ebrero}, J., {Carrera}, F.~J., {Page}, M.~J., {et~al.} 2009, \aap, 493, 55

\bibitem[{{Gehrels} {et~al.}(2004){Gehrels}, {Chincarini}, {Giommi}, {Mason},
  {Nousek}, {Wells}, {White}, {Barthelmy}, {Burrows}, {Cominsky}, {Hurley},
  {Marshall}, {M{\'e}sz{\'a}ros}, {Roming}, {Angelini}, {Barbier}, {Belloni},
  {Campana}, {Caraveo}, {Chester}, {Citterio}, {Cline}, {Cropper}, {Cummings},
  {Dean}, {Feigelson}, {Fenimore}, {Frail}, {Fruchter}, {Garmire}, {Gendreau},
  {Ghisellini}, {Greiner}, {Hill}, {Hunsberger}, {Krimm}, {Kulkarni}, {Kumar},
  {Lebrun}, {Lloyd-Ronning}, {Markwardt}, {Mattson}, {Mushotzky}, {Norris},
  {Osborne}, {Paczynski}, {Palmer}, {Park}, {Parsons}, {Paul}, {Rees},
  {Reynolds}, {Rhoads}, {Sasseen}, {Schaefer}, {Short}, {Smale}, {Smith},
  {Stella}, {Tagliaferri}, {Takahashi}, {Tashiro}, {Townsley}, {Tueller},
  {Turner}, {Vietri}, {Voges}, {Ward}, {Willingale}, {Zerbi}, \&
  {Zhang}}]{Swift}
{Gehrels}, N., {Chincarini}, G., {Giommi}, P., {et~al.} 2004, ApJ, 611, 1005

\bibitem[{{Gilli} {et~al.}(2007){Gilli}, {Comastri}, \& {Hasinger}}]{Gilli2007}
{Gilli}, R., {Comastri}, A., \& {Hasinger}, G. 2007, \aap, 463, 79

\bibitem[{{Giommi} {et~al.}(2006){Giommi}, {Colafrancesco}, {Cavazzuti},
  {Perri}, \& {Pittori}}]{Giommi2006}
{Giommi}, P., {Colafrancesco}, S., {Cavazzuti}, E., {Perri}, M., \& {Pittori},
  C. 2006, \aap, 445, 843

\bibitem[{{Gruber} {et~al.}(1999){Gruber}, {Matteson}, {Peterson}, \&
  {Jung}}]{Gruber1999}
{Gruber}, D.~E., {Matteson}, J.~L., {Peterson}, L.~E., \& {Jung}, G.~V. 1999,
  \apj, 520, 124

\bibitem[{{Hasinger} {et~al.}(2005){Hasinger}, {Miyaji}, \&
  {Schmidt}}]{Hasinger2005}
{Hasinger}, G., {Miyaji}, T., \& {Schmidt}, M. 2005, \aap, 441, 417

\bibitem[{{Hogg}(1999)}]{Hogg1999}
{Hogg}, D.~W. 1999, arXiv e-prints, astro-ph/9905116, astro

\bibitem[{{Hovatta} {et~al.}(2014){Hovatta}, {Aller}, {Aller}, {Clausen-Brown},
  {Homan}, {Kovalev}, {Lister}, {Pushkarev}, \& {Savolainen}}]{Hovatta2014}
{Hovatta}, T., {Aller}, M.~F., {Aller}, H.~D., {et~al.} 2014, \aj, 147, 143

\bibitem[{{Ighina} {et~al.}(2019){Ighina}, {Caccianiga}, {Moretti},
  {Belladitta}, {Della Ceca}, {Ballo}, \& {Dallacasa}}]{Ighina2019}
{Ighina}, L., {Caccianiga}, A., {Moretti}, A., {et~al.} 2019, \mnras, 489, 2732

\bibitem[{{James}(1994)}]{James1994}
{James}, F. 1994, CERN Program Library Long Writeup D506 (Geneva: CERN)

\bibitem[{{Jorstad} \& {Marscher}(2016)}]{Jorstad2016}
{Jorstad}, S. \& {Marscher}, A. 2016, Galaxies, 4, 47

\bibitem[{{Krivonos} {et~al.}(2015){Krivonos}, {Tsygankov}, {Lutovinov},
  {Revnivtsev}, {Churazov}, \& {Sunyaev}}]{Krivonos2015}
{Krivonos}, R., {Tsygankov}, S., {Lutovinov}, A., {et~al.} 2015, \mnras, 448,
  3766

\bibitem[{{Lister} {et~al.}(2009{\natexlab{a}}){Lister}, {Aller}, {Aller},
  {Cohen}, {Homan}, {Kadler}, {Kellermann}, {Kovalev}, {Ros}, {Savolainen},
  {Zensus}, \& {Vermeulen}}]{Lister2009_137}
{Lister}, M.~L., {Aller}, H.~D., {Aller}, M.~F., {et~al.} 2009{\natexlab{a}},
  \aj, 137, 3718

\bibitem[{{Lister} {et~al.}(2018){Lister}, {Aller}, {Aller}, {Hodge}, {Homan},
  {Kovalev}, {Pushkarev}, \& {Savolainen}}]{Lister2018}
{Lister}, M.~L., {Aller}, M.~F., {Aller}, H.~D., {et~al.} 2018, apjs, 234, 12

\bibitem[{{Lister} {et~al.}(2013){Lister}, {Aller}, {Aller}, {Homan},
  {Kellermann}, {Kovalev}, {Pushkarev}, {Richards}, {Ros}, \&
  {Savolainen}}]{Lister2013}
{Lister}, M.~L., {Aller}, M.~F., {Aller}, H.~D., {et~al.} 2013, \aj, 146, 120

\bibitem[{{Lister} {et~al.}(2015){Lister}, {Aller}, {Aller}, {Hovatta},
  {Max-Moerbeck}, {Readhead}, {Richards}, \& {Ros}}]{Lister2015}
{Lister}, M.~L., {Aller}, M.~F., {Aller}, H.~D., {et~al.} 2015, \apjl, 810, L9

\bibitem[{Lister \& Homan(2005)}]{Lister2005}
Lister, M.~L. \& Homan, D.~C. 2005, AJ, 130, 1389

\bibitem[{{Lister} {et~al.}(2009{\natexlab{b}}){Lister}, {Homan}, {Kadler},
  {Kellermann}, {Kovalev}, {Ros}, {Savolainen}, \& {Zensus}}]{Lister2009_fermi}
{Lister}, M.~L., {Homan}, D.~C., {Kadler}, M., {et~al.} 2009{\natexlab{b}},
  \apjl, 696, L22

\bibitem[{{Longair}(1966)}]{Longair1966}
{Longair}, M.~S. 1966, \mnras, 133, 421

\bibitem[{{Malizia} {et~al.}(2012){Malizia}, {Bassani}, {Bazzano}, {Bird},
  {Masetti}, {Panessa}, {Stephen}, \& {Ubertini}}]{Malizia2012}
{Malizia}, A., {Bassani}, L., {Bazzano}, A., {et~al.} 2012, \mnras, 426, 1750

\bibitem[{{Marchesini} {et~al.}(2019){Marchesini}, {Paggi}, {Massaro},
  {Masetti}, {D'Abrusco}, {Andruchow}, \& {de Menezes}}]{Marchesini2019}
{Marchesini}, E.~J., {Paggi}, A., {Massaro}, F., {et~al.} 2019, \aap, 631, A150

\bibitem[{{Mateos} {et~al.}(2008){Mateos}, {Warwick}, {Carrera}, {Stewart},
  {Ebrero}, {Della Ceca}, {Caccianiga}, {Gilli}, {Page}, {Treister}, {Tedds},
  {Watson}, {Lamer}, {Saxton}, {Brunner}, \& {Page}}]{Mateos2008}
{Mateos}, S., {Warwick}, R.~S., {Carrera}, F.~J., {et~al.} 2008, \aap, 492, 51

\bibitem[{{Miyaji} {et~al.}(2015){Miyaji}, {Hasinger}, {Salvato}, {Brusa},
  {Cappelluti}, {Civano}, {Puccetti}, {Elvis}, {Brunner}, {Fotopoulou}, {Ueda},
  {Griffiths}, {Koekemoer}, {Akiyama}, {Comastri}, {Gilli}, {Lanzuisi},
  {Merloni}, \& {Vignali}}]{Miyaji2015}
{Miyaji}, T., {Hasinger}, G., {Salvato}, M., {et~al.} 2015, \apj, 804, 104

\bibitem[{{Mufakharov} {et~al.}(2015){Mufakharov}, {Mingaliev}, {Sotnikova},
  {Naiden}, \& {Erkenov}}]{Mufakharov2015}
{Mufakharov}, T., {Mingaliev}, M., {Sotnikova}, Y., {Naiden}, Y., \& {Erkenov},
  A. 2015, \mnras, 450, 2658

\bibitem[{{Oh} {et~al.}(2018){Oh}, {Koss}, {Markwardt}, {Schawinski},
  {Baumgartner}, {Barthelmy}, {Cenko}, {Gehrels}, {Mushotzky}, {Petulante},
  {Ricci}, {Lien}, \& {Trakhtenbrot}}]{Oh2018}
{Oh}, K., {Koss}, M., {Markwardt}, C.~B., {et~al.} 2018, \apjs, 235, 4

\bibitem[{{Ojha} {et~al.}(2010){Ojha}, {Kadler}, {B{\"o}ck}, {Booth}, {Dutka},
  {Edwards}, {Fey}, {Fuhrmann}, {Gaume}, {Hase}, {Horiuchi}, {Jauncey},
  {Johnston}, {Katz}, {Lister}, {Lovell}, {M{\"u}ller}, {Pl{\"o}tz}, {Quick},
  {Ros}, {Taylor}, {Thompson}, {Tingay}, {Tosti}, {Tzioumis}, {Wilms}, \&
  {Zensus}}]{Ojha2010}
{Ojha}, R., {Kadler}, M., {B{\"o}ck}, M., {et~al.} 2010, \aap, 519, A45

\bibitem[{{Pian} {et~al.}(2005){Pian}, {Foschini}, {Beckmann},
  {Sillanp{\"a}{\"a}}, {Soldi}, {Tagliaferri}, {Takalo}, {Barr}, {Ghisellini},
  {Malaguti}, {Maraschi}, {Palumbo}, {Treves}, {Courvoisier}, {Di Cocco},
  {Gehrels}, {Giommi}, {Hudec}, {Lindfors}, {Marcowith}, {Nilsson}, {Pasanen},
  {Pursimo}, {Raiteri}, {Savolainen}, {Sikora}, {Tornikoski}, {Tosti},
  {T{\"u}rler}, {Valtaoja}, {Villata}, \& {Walter}}]{Pian2005}
{Pian}, E., {Foschini}, L., {Beckmann}, V., {et~al.} 2005, \aap, 429, 427

\bibitem[{{Ranalli} {et~al.}(2016){Ranalli}, {Koulouridis}, {Georgantopoulos},
  {Fotopoulou}, {Hsu}, {Salvato}, {Comastri}, {Pierre}, {Cappelluti},
  {Carrera}, {Chiappetti}, {Clerc}, {Gilli}, {Iwasawa}, {Pacaud}, {Paltani},
  {Plionis}, \& {Vignali}}]{Ranalli2016}
{Ranalli}, P., {Koulouridis}, E., {Georgantopoulos}, I., {et~al.} 2016, \aap,
  590, A80

\bibitem[{{Roming} {et~al.}(2005){Roming}, {Kennedy}, {Mason}, {Nousek}, {Ahr},
  {Bingham}, {Broos}, {Carter}, {Hancock}, {Huckle}, {Hunsberger}, {Kawakami},
  {Killough}, {Koch}, {McLelland}, {Smith}, {Smith}, {Soto}, {Boyd},
  {Breeveld}, {Holland}, {Ivanushkina}, {Pryzby}, {Still}, \&
  {Stock}}]{Roming2005}
{Roming}, P.~W.~A., {Kennedy}, T.~E., {Mason}, K.~O., {et~al.} 2005, SSR, 120,
  95

\bibitem[{{Sazonov} {et~al.}(2007){Sazonov}, {Revnivtsev}, {Krivonos},
  {Churazov}, \& {Sunyaev}}]{Sazonov2007}
{Sazonov}, S., {Revnivtsev}, M., {Krivonos}, R., {Churazov}, E., \& {Sunyaev},
  R. 2007, \aap, 462, 57

\bibitem[{{Schmidt}(1968)}]{Schmidt1968}
{Schmidt}, M. 1968, \apj, 151, 393

\bibitem[{{Soldi} {et~al.}(2014){Soldi}, {Beckmann}, {Baumgartner}, {Ponti},
  {Shrader}, {Lubi{\'n}ski}, {Krimm}, {Mattana}, \& {Tueller}}]{Soldi2014}
{Soldi}, S., {Beckmann}, V., {Baumgartner}, W.~H., {et~al.} 2014, \aap, 563,
  A57

\bibitem[{{Steinle}(2006)}]{Steinle2006}
{Steinle}, H. 2006, ChJ Astron. Astrophys. Suppl., 6, 106

\bibitem[{{The Fermi-LAT collaboration}(2019)}]{4FGL2019}
{The Fermi-LAT collaboration}. 2019, arXiv e-prints, arXiv:1902.10045

\bibitem[{{Tueller} {et~al.}(2008){Tueller}, {Mushotzky}, {Barthelmy},
  {Cannizzo}, {Gehrels}, {Markwardt}, {Skinner}, \& {Winter}}]{Tueller2008}
{Tueller}, J., {Mushotzky}, R.~F., {Barthelmy}, S., {et~al.} 2008, ApJ, 681,
  113

\bibitem[{{Ueda} {et~al.}(2003){Ueda}, {Akiyama}, {Ohta}, \&
  {Miyaji}}]{Ueda2003}
{Ueda}, Y., {Akiyama}, M., {Ohta}, K., \& {Miyaji}, T. 2003, \apj, 598, 886

\bibitem[{V\'{e}ron-Cetty \& V\'{e}ron(2003)}]{Veron2003}
V\'{e}ron-Cetty, M.~P. \& V\'{e}ron, P. 2003, A\&A, 412, 399

\bibitem[{{Villata} {et~al.}(2008){Villata}, {Raiteri}, {Larionov},
  {Kurtanidze}, {Nilsson}, {Aller}, {Tornikoski}, {Volvach}, {Aller},
  {Arkharov}, {Bach}, {Beltrame}, {Bhatta}, {Buemi}, {B{\"o}ttcher},
  {Calcidese}, {Carosati}, {Castro-Tirado}, {da Rio}, {di Paola}, {Dolci},
  {Forn{\'e}}, {Frasca}, {Hagen-Thorn}, {Heidt}, {Hiriart}, {Jel{\'\i}nek},
  {Kimeridze}, {Konstantinova}, {Kopatskaya}, {Lanteri}, {Leto}, {Ligustri},
  {Lindfors}, {L{\"a}hteenm{\"a}ki}, {Marilli}, {Nieppola}, {Nikolashvili},
  {Pasanen}, {Ragozzine}, {Ros}, {Sigua}, {Smart}, {Sorcia}, {Takalo},
  {Tavani}, {Trigilio}, {Turchetti}, {Uckert}, {Umana}, {Vercellone}, \&
  {Webb}}]{Villata2008}
{Villata}, M., {Raiteri}, C.~M., {Larionov}, V.~M., {et~al.} 2008, \aap, 481,
  L79

\bibitem[{{Wagner} {et~al.}(1996){Wagner}, {Witzel}, {Heidt}, {Krichbaum},
  {Qian}, {Quirrenbach}, {Wegner}, {Aller}, {Aller}, {Anton}, {Appenzeller},
  {Eckart}, {Kraus}, {Naundorf}, {Kneer}, {Steffen}, \& {Zensus}}]{Wagner1996}
{Wagner}, S.~J., {Witzel}, A., {Heidt}, J., {et~al.} 1996, \aj, 111, 2187

\bibitem[{{Wang} {et~al.}(2016){Wang}, {Liu}, {Qiu}, {Bai}, {Yang}, {Guo}, \&
  {Zhang}}]{Wang2016}
{Wang}, S., {Liu}, J., {Qiu}, Y., {et~al.} 2016, \apjs, 224, 40

\bibitem[{{Wierzcholska} \& {Siejkowski}(2015)}]{Wierzcholska2015}
{Wierzcholska}, A. \& {Siejkowski}, H. 2015, \mnras, 452, L11

\bibitem[{{Wierzcholska} \& {Siejkowski}(2016)}]{Wierzcholska2016}
{Wierzcholska}, A. \& {Siejkowski}, H. 2016, \mnras, 458, 2350

\bibitem[{{Yao} {et~al.}(2015){Yao}, {Yuan}, {Zhou}, {Komossa}, {Zhang},
  {Qiao}, \& {Liu}}]{Yao2015}
{Yao}, S., {Yuan}, W., {Zhou}, H., {et~al.} 2015, \mnras, 454, L16

\end{thebibliography}
}

\small
\newpage
\appendix
\onecolumn
\captionsetup{width=1.0\linewidth}
\renewcommand{\arraystretch}{1.3}

\begin{landscape}
\section{Hard X-ray Data For The MOJAVE-1 Sample}

\begin{longtable}{lllllllllll}
\caption{\label{tab:app} MOJAVE-1 sample with calculated photon indices, fluxes and luminosities in the range of 20\,keV -- 100 keV.}\\
\hline
\hline
Name\tablefootmark{a}	& Common Name	 & RA\tablefootmark{b} & Dec\tablefootmark{b} &  S/N\tablefootmark{c}	& $\Gamma$\tablefootmark{d}	& $\mathrm{F_{20-100\,keV}}$\tablefootmark{e} & $\mathrm{L_{20-100\,keV}}$\tablefootmark{f} & Type\tablefootmark{g}	 & z	& Detections\tablefootmark{h}\\
\hline
\endfirsthead
\caption{continued.}\\
\hline\hline
%%%%%%%%%%%%%%%%%%%%%%%%%%%%%%%%%%%%%%%%%%%%%%%%%%%%%%%%%%%%%%%%%%%%%%
%%                                                                  %%
%%  This is a LaTeX2e table fragment exported from Gnumeric.        %%
%%                                                                  %%
%%%%%%%%%%%%%%%%%%%%%%%%%%%%%%%%%%%%%%%%%%%%%%%%%%%%%%%%%%%%%%%%%%%%%%

Name$^\text{a}$	&Common Name & RA$^\text{b}$& Dec$^\text{b}$	&S/N$^\text{c}$	&$\Gamma^\text{d}$	&F$_\mathrm{{20-100\,keV}}^\text{e}$&L$_\mathrm{{20-100\,keV}}^\text{f}$ &Type$^\text{g}$	&z$^\text{h}$	&Detections$^\text{i}$\\
%& & & & & & & & & & \\
\hline
\endhead
\hline
\endfoot
0003--066 & NRAO 005 &1.558 & -6.393 & 2.34 & $1.04 \pm 1.02$ & $4.49 \pm 1.6$ & $(1.36 \pm 0.53) \cdot 10^{45}$ & B & 0.3467 & $\mathrm{F_{4}}$\\
 0007+106 & III Zw 2 & 2.629 & 10.975 & 13.24 & $1.86 \pm 0.24$ & $16.48 \pm 1.65$ & $(3.24 \pm 0.33) \cdot 10^{44}$ & G & 0.0893 & $\mathrm{B}$ \\
 0016+731 & S5 0016+73 & 4.941 & 73.458 & 3.65 & $1.5 \pm 0.78$ & $4.26 \pm 1.27$ & $(5.53 \pm 1.66) \cdot 10^{46}$ & Q & 1.781 & $\mathrm{F_{4}}$\\
 0048--097 & PKS 0048--09 & 12.672 & -9.485 & 0.99 &   & $< 3.66 ^*$ & $< 5.07 \cdot 10^{45}$ & B & 0.635 &  $\mathrm{F_{3}}$ $\mathrm{F_{4}}$ \\
 0059+581 & TXS 0059+581 & 15.691 & 58.403 & 2.87 & $0.92 \pm 1.14$ & $3.37 \pm 1.29$ & $(3.49 \pm 1.55) \cdot 10^{45}$ & Q & 0.644 &  $\mathrm{F_{3}}$ $\mathrm{F_{4}}$ \\
 0106+013 & 4C +01.02 & 17.162 & 1.583 & 2.17 &   & $< 3.67 ^*$ & $< 6.88 \cdot 10^{46}$ & Q & 2.099 &  $\mathrm{F_{3}}$ $\mathrm{F_{4}}$ \\
 0109+224 & S2 0109+22 & 18.024 & 22.744 & 1.16 &   & $< 3.52 ^*$ & $< 7.04 \cdot 10^{44}$ & B & $0.267442 ^*$ &  $\mathrm{F_{3}}$ $\mathrm{F_{4}}$ \\
 0119+115 & PKS 0119+11 & 20.423 & 11.831 & 0.34 &   & $< 3.73 ^*$ & $< 3.96 \cdot 10^{45}$ & Q & 0.571 &  \\
 0133+476 & DA 55 & 24.244 & 47.858 & 3.0 & $1.87 \pm 0.91$ & $3.9 \pm 1.33$ & $(1.29 \pm 0.45) \cdot 10^{46}$ & Q & 0.859 &  $\mathrm{F_{3}}$ $\mathrm{F_{4}}$ \\
 0202+149 & 4C +15.05 & 31.21 & 15.236 & 0.92 &   & $< 3.74 ^*$ & $< 1.84 \cdot 10^{45}$ & Q & 0.405 &  $\mathrm{F_{3}}$ $\mathrm{F_{4}}$ \\
 0202+319 & B2 0202+31 & 31.271 & 32.208 & --0.6 &   & $< 3.55 ^*$ & $< 3.08 \cdot 10^{46}$ & Q & 1.466 &  $\mathrm{F_{3}}$ $\mathrm{F_{4}}$ \\
 0212+735 & S5 0212+73 & 34.378 & 73.826 & 15.79 & $1.43 \pm 0.17$ & $20.03 \pm 1.4$ & $(4.37 \pm 0.31) \cdot 10^{47}$ & Q & 2.367 & $\mathrm{B}$ $\mathrm{I_{A}}$ $\mathrm{F_{3}}$ $\mathrm{F_{4}}$ \\
 0215+015 & OD 026 & 34.454 & 1.747 & 1.53 &   & $< 2.42$ & $< 2.95 \cdot 10^{46}$ & Q & 1.715 &  $\mathrm{F_{3}}$ $\mathrm{F_{4}}$ \\
 0224+671 & 4C +67.05 & 37.209 & 67.351 & 3.9 & $2.68 \pm 0.9$ & $4.04 \pm 1.26$ & $(5.76 \pm 1.81) \cdot 10^{45}$ & Q & 0.523 & $\mathrm{F_{4}}$\\
 0234+285 & 4C +28.07 & 39.468 & 28.802 & 4.14 &   & $4.53 \pm 1.24$ & $(2.56 \pm 0.7) \cdot 10^{46}$ & Q & 1.206 &  $\mathrm{F_{3}}$ $\mathrm{F_{4}}$ \\
 0235+164 & AO 0235+164 & 39.662 & 16.616 & 2.56 & $1.73 \pm 1.02$ & $4.14 \pm 1.53$ & $(1.56 \pm 0.58) \cdot 10^{46}$ & Q & 0.94 &  $\mathrm{F_{3}}$ $\mathrm{F_{4}}$ \\
 0238--084 & NGC 1052 & 40.27 & -8.256 & 14.36 & $1.86 \pm 0.23$ & $16.23 \pm 1.6$ & $(9.1 \pm 0.9) \cdot 10^{41}$ & G & 0.005037 & $\mathrm{B}$ $\mathrm{I_{A}}$ \\
 0300+470 & 4C +47.08 & 45.897 & 47.271 & 3.57 & $1.53 \pm 1.03$ & $3.77 \pm 1.41$ & \ & B &   &  $\mathrm{F_{3}}$ $\mathrm{F_{4}}$ \\
 0316+413 & 3C 84 & 49.951 & 41.512 & 58.16 & $3.17 \pm 0.12$ & $41.83 \pm 1.53$ & $(2.98 \pm 0.11) \cdot 10^{43}$ & G & 0.0176 & $\mathrm{B}$ $\mathrm{I_{A}}$ $\mathrm{F_{3}}$ $\mathrm{F_{4}}$ \\
 0333+321 & NRAO 140 & 54.125 & 32.308 & 17.4 & $1.71 \pm 0.15$ & $26.35 \pm 1.66$ & $(1.92 \pm 0.12) \cdot 10^{47}$ & Q & 1.259 & $\mathrm{B}$ $\mathrm{I_{A}}$ $\mathrm{F_{3}}$ $\mathrm{F_{4}}$ \\
 0336--019 & CTA 26 & 54.879 & -1.777 & 2.71 & $1.81 \pm 0.95$ & $4.25 \pm 1.49$ & $(1.33 \pm 0.48) \cdot 10^{46}$ & Q & 0.852 &  $\mathrm{F_{3}}$ $\mathrm{F_{4}}$ \\
 0403--132 & PKS 0403-13 & 61.392 & -13.137 & 5.87 & $1.77 \pm 0.6$ & $6.01 \pm 1.52$ & $(7.17 \pm 1.86) \cdot 10^{45}$ & Q & 0.571 & $\mathrm{B}$ $\mathrm{F_{3}}$ $\mathrm{F_{4}}$ \\
 0415+379 & 3C 111 & 64.589 & 38.027 & 52.05 & $2.02 \pm 0.05$ & $73.31 \pm 1.65$ & $(4.18 \pm 0.09) \cdot 10^{44}$ & G & 0.0491 & $\mathrm{B}$ $\mathrm{I_{A}}$ $\mathrm{I_{11}}$ $\mathrm{F_{3}}$ $\mathrm{F_{4}}$ \\
 0420--014 & PKS 0420--01 & 65.816 & -1.343 & 1.33 & $1.53 \pm 1.47$ & $3.09 \pm 1.43$ & $(9.56 \pm 4.62) \cdot 10^{45}$ & Q & 0.9161 &  $\mathrm{F_{3}}$ $\mathrm{F_{4}}$ \\
 0422+004 & PKS 0422+00 & 66.195 & 0.602 & 0.57 & $1.14 \pm 1.6$ & $2.94 \pm 1.48$ & $(5.34 \pm 2.94) \cdot 10^{44}$ & B & 0.268 &  $\mathrm{F_{3}}$ $\mathrm{F_{4}}$ \\
 0430+052 & 3C 120 & 68.296 & 5.354 & 42.95 & $1.99 \pm 0.07$ & $60.34 \pm 1.7$ & $(1.52 \pm 0.04) \cdot 10^{44}$ & G & 0.033 & $\mathrm{B}$ $\mathrm{I_{A}}$ \\
 0446+112 & PKS 0446+11 & 72.282 & 11.358 & 1.5 &   & $< 4.33 ^*$ & $< 8.57 \cdot 10^{46}$ & Q & 2.153 &  $\mathrm{F_{3}}$ $\mathrm{F_{4}}$ \\
 0458--020 & S3 0458--02 & 75.303 & -1.987 & 1.46 &   & $< 3.96 ^*$ & $< 8.88 \cdot 10^{46}$ & Q & 2.286 &  $\mathrm{F_{3}}$ $\mathrm{F_{4}}$ \\
 0528+134 & PKS 0528+134 & 82.735 & 13.532 & 4.88 & $1.06 \pm 0.46$ & $10.18 \pm 1.85$ & $(1.12 \pm 0.21) \cdot 10^{47}$ & Q & 2.07 & $\mathrm{B}$ $\mathrm{I_{A}}$ $\mathrm{F_{3}}$ $\mathrm{F_{4}}$ \\
 0529+075 & OG 050 & 83.162 & 7.545 & 2.38 &   & $< 4.31 ^*$ & $< 2.65 \cdot 10^{46}$ & Q & 1.254 &  $\mathrm{F_{3}}$ $\mathrm{F_{4}}$ \\
 0529+483 & TXS 0529+483 & 83.316 & 48.381 & 1.13 &   & $< 3.97 ^*$ & $< 2.06 \cdot 10^{46}$ & Q & 1.16 &  $\mathrm{F_{3}}$ $\mathrm{F_{4}}$ \\
 0552+398 & DA 193 & 88.878 & 39.814 & 6.03 & $1.78 \pm 0.5$ & $8.27 \pm 1.73$ & $(2.76 \pm 0.58) \cdot 10^{47}$ & Q & 2.363 & $\mathrm{B}$ $\mathrm{F_{4}}$\\
 0605--085 & OC--010 & 91.999 & -8.581 & 2.11 &   & $< 3.94 ^*$ & $< 1.08 \cdot 10^{46}$ & Q & 0.87 &  $\mathrm{F_{3}}$ $\mathrm{F_{4}}$ \\
 0607--157 & PKS 0607--15 & 92.421 & -15.711 & 0.85 &   & $< 3.76 ^*$ & $< 1.12 \cdot 10^{45}$ & Q & 0.3226 &  \\
 0642+449 & OH 471 & 101.633 & 44.855 & 3.3 & $1.57 \pm 0.89$ & $4.58 \pm 1.52$ & $(2.51 \pm 0.84) \cdot 10^{47}$ & Q & 3.396 &  \\
 0648--165 & PKS 0648--16 & 102.602 & -16.628 & 0.68 &   & $< 3.72 ^*$ & \ & U &   &  $\mathrm{F_{3}}$ $\mathrm{F_{4}}$ \\
 0716+714 & TXS 0716+714 & 110.473 & 71.343 & 5.37 & $0.83 \pm 0.46$ & $8.1 \pm 1.43$ & $(2.98 \pm 0.54) \cdot 10^{44}$ & B & $0.127 ^*$ &  $\mathrm{B}$ $\mathrm{F_{3}}$ $\mathrm{F_{4}}$ \\
 0727--115 & PKS 0727--11 & 112.58 & -11.687 & 3.83 & $1.5 \pm 0.64$ & $6.24 \pm 1.66$ & $(6.37 \pm 1.71) \cdot 10^{46}$ & Q & 1.591 &  $\mathrm{F_{3}}$ $\mathrm{F_{4}}$ \\
 0730+504 & TXS 0730+504 & 113.469 & 50.369 & --1.01 &   & $< 3.47 ^*$ & $< 6.19 \cdot 10^{45}$ & Q & 0.72 &  $\mathrm{F_{3}}$ $\mathrm{F_{4}}$ \\
 0735+178 & OI 158 & 114.531 & 17.705 & 0.66 &   & $< 2.11$ & $< 1.34 \cdot 10^{45}$ & B & $0.45 ^*$ &  $\mathrm{F_{3}}$ $\mathrm{F_{4}}$ \\
 0736+017 & OI 061 & 114.825 & 1.618 & 3.68 & $1.52 \pm 0.66$ & $6.35 \pm 1.74$ & $(5.94 \pm 1.67) \cdot 10^{44}$ & Q & 0.1894 &  $\mathrm{F_{3}}$ $\mathrm{F_{4}}$ \\
 0738+313 & OI 363 & 115.295 & 31.2 & 2.0 &   & $< 3.96 ^*$ & $< 5.26 \cdot 10^{45}$ & Q & 0.631 &  \\
 0742+103 & PKS B0742+103 & 116.388 & 10.187 & 3.97 & $2.0 \pm 0.8$ & $5.29 \pm 1.66$ & $(2.96 \pm 0.93) \cdot 10^{47}$ & Q & 2.624 &  \\
 0748+126 & OI 280 & 117.717 & 12.518 & 2.7 & $1.08 \pm 0.97$ & $4.86 \pm 1.65$ & $(1.06 \pm 0.39) \cdot 10^{46}$ & Q & 0.889 &  $\mathrm{F_{3}}$ $\mathrm{F_{4}}$ \\
 0754+100 & PKS 0754+100 & 119.278 & 9.943 & 3.93 & $3.04 \pm 1.56$ & $3.11 \pm 1.47$ & $(8.71 \pm 4.18) \cdot 10^{44}$ & B & 0.266 &  $\mathrm{F_{3}}$ $\mathrm{F_{4}}$ \\
 0804+499 & OJ 508 & 122.165 & 49.843 & 1.2 & $1.04 \pm 1.56$ & $2.65 \pm 1.32$ & $(1.44 \pm 0.79) \cdot 10^{46}$ & Q & 1.436 &  $\mathrm{F_{3}}$ $\mathrm{F_{4}}$ \\
 0805--077 & PKS 0805--07 & 122.065 & -7.853 & 2.76 & $1.2 \pm 1.36$ & $3.5 \pm 1.54$ & $(3.57 \pm 1.63) \cdot 10^{46}$ & Q & 1.837 &  $\mathrm{F_{3}}$ $\mathrm{F_{4}}$ \\
 0808+019 & OJ 014 & 122.861 & 1.781 & --0.6 &   & $< 3.92 ^*$ & $< 2.08 \cdot 10^{46}$ & B & 1.148 &  $\mathrm{F_{3}}$ $\mathrm{F_{4}}$ \\
 0814+425 & OJ 425 & 124.567 & 42.379 & 0.45 &   & $< 3.42 ^*$ & \ & B &   &  $\mathrm{F_{3}}$ $\mathrm{F_{4}}$ \\
 0823+033 & PKS 0823+033 & 126.46 & 3.157 & 2.15 &   & $< 3.93 ^*$ & $< 3.24 \cdot 10^{45}$ & B & 0.505 &  $\mathrm{F_{3}}$ $\mathrm{F_{4}}$ \\
 0827+243 & OJ 248 & 127.717 & 24.183 & 3.38 & $1.07 \pm 0.79$ & $5.2 \pm 1.61$ & $(1.26 \pm 0.42) \cdot 10^{46}$ & Q & 0.94 &  $\mathrm{F_{3}}$ $\mathrm{F_{4}}$ \\
 0829+046 & OJ 049 & 127.954 & 4.494 & 1.72 &   & $< 3.85 ^*$ & $< 3.03 \cdot 10^{44}$ & B & 0.174 &  $\mathrm{F_{3}}$ $\mathrm{F_{4}}$ \\
 0836+710 & 4C +71.07 & 130.352 & 70.895 & 37.52 & $1.65 \pm 0.08$ & $40.73 \pm 1.38$ & $(1.0 \pm 0.03) \cdot 10^{48}$ & Q & 2.218 & $\mathrm{B}$ $\mathrm{I_{A}}$ $\mathrm{I_{11}}$ $\mathrm{F_{3}}$ $\mathrm{F_{4}}$ \\
 0838+133 & 3C 207 & 130.198 & 13.207 & 0.61 &   & $< 3.77 ^*$ & $< 5.92 \cdot 10^{45}$ & Q & 0.68 &  $\mathrm{F_{3}}$ $\mathrm{F_{4}}$ \\
 0851+202 & OJ 287 & 133.704 & 20.109 & 3.82 & $1.31 \pm 0.89$ & $4.68 \pm 1.53$ & $(1.17 \pm 0.41) \cdot 10^{45}$ & B & 0.306 &  $\mathrm{F_{3}}$ $\mathrm{F_{4}}$ \\
 0906+015 & 4C +01.24 & 137.292 & 1.36 & 2.19 &   & $< 3.72 ^*$ & $< 1.46 \cdot 10^{46}$ & Q & 1.024 &  $\mathrm{F_{3}}$ $\mathrm{F_{4}}$ \\
 0917+624 & OK 630 & 140.401 & 62.264 & 2.47 &   & $< 2.94 \dagger $ & $2.48 \cdot 10^{46}$ & Q & 1.447 &  $\mathrm{F_{3}}$ $\mathrm{F_{4}}$ \\
 0923+392 & 4C +39.25 & 141.763 & 39.039 & 3.12 &   & $3.19 \pm 1.05$ & $(5.3 \pm 1.75) \cdot 10^{45}$ & Q & 0.697 &  \\
 0945+408 & 4C +40.24 & 147.231 & 40.662 & 4.59 & $1.87 \pm 0.75$ & $4.49 \pm 1.3$ & $(3.66 \pm 1.06) \cdot 10^{46}$ & Q & 1.25 &  $\mathrm{F_{3}}$ $\mathrm{F_{4}}$ \\
 0955+476 & OK 492 & 149.582 & 47.419 & 1.43 &   & $< 2.98 ^*$ & $< 4.43 \cdot 10^{46}$ & Q & 1.884 &  $\mathrm{F_{3}}$ $\mathrm{F_{4}}$ \\
 1036+054 & PKS 1036+054 & 159.695 & 5.208 & 0.3 &   & $< 3.6 ^*$ & $< 2.51 \cdot 10^{45}$ & Q & 0.473 &  \\
 1038+064 & 4C +06.41 & 160.322 & 6.171 & 3.01 & $2.05 \pm 1.06$ & $3.81 \pm 1.45$ & $(3.71 \pm 1.41) \cdot 10^{46}$ & Q & 1.265 &  \\
 1045--188 & PKS 1045--18 & 162.028 & -19.16 & 0.55 &   & $< 4.02 ^*$ & $< 4.67 \cdot 10^{45}$ & Q & 0.595 &  \\
 1055+018 & 4C +01.28 & 164.623 & 1.566 & 3.69 &   & $4.05 \pm 1.23$ & $(1.17 \pm 0.36) \cdot 10^{46}$ & Q & 0.893 &  $\mathrm{F_{3}}$ $\mathrm{F_{4}}$ \\
 1124--186 & PKS 1124--186 & 171.768 & -18.955 & 0.97 &   & $< 4.06 ^*$ & $< 1.68 \cdot 10^{46}$ & Q & 1.048 &  $\mathrm{F_{3}}$ $\mathrm{F_{4}}$ \\
 1127--145 & PKS 1127--14 & 172.529 & -14.824 & 13.42 & $1.67 \pm 0.24$ & $18.05 \pm 1.81$ & $(1.11 \pm 0.11) \cdot 10^{47}$ & Q & 1.184 & $\mathrm{B}$ $\mathrm{F_{3}}$ $\mathrm{F_{4}}$ \\
 1150+812 & S5 1150+81 & 178.302 & 80.975 & 1.17 &   & $< 2.93 ^*$ & $< 1.79 \cdot 10^{46}$ & Q & 1.25 & $\mathrm{F_{4}}$\\
 1156+295 & 4C +29.45 & 179.883 & 29.246 & 4.4 & $1.55 \pm 0.7$ & $4.65 \pm 1.3$ & $(8.6 \pm 2.49) \cdot 10^{45}$ & Q & 0.725 &  $\mathrm{B}$ $\mathrm{F_{3}}$ $\mathrm{F_{4}}$ \\
 1213--172 & PKS 1213--17 & 183.945 & -17.529 & 2.09 &   & $< 4.09 ^*$ & \ & U &   & $\mathrm{F_{4}}$\\
 1219+044 & 4C +04.42 & 185.594 & 4.221 & 13.09 & $1.34 \pm 0.23$ & $17.63 \pm 1.64$ & $(5.41 \pm 0.52) \cdot 10^{46}$ & Q$\ddagger$ & 0.966 & $\mathrm{B}$ $\mathrm{I_{A}}$ $\mathrm{I_{11}}$ $\mathrm{F_{3}}$ $\mathrm{F_{4}}$ \\
 1222+216 & 4C +21.35 & 186.227 & 21.38 & 11.51 & $1.87 \pm 0.28$ & $12.26 \pm 1.44$ & $(7.98 \pm 0.96) \cdot 10^{45}$ & Q & 0.433 & $\mathrm{B}$ $\mathrm{F_{3}}$ $\mathrm{F_{4}}$ \\
 1226+023 & 3C 273 & 187.278 & 2.052 & 192.37 & $1.75 \pm 0.02$ & $240.34 \pm 1.67$ & $(1.59 \pm 0.01) \cdot 10^{46}$ & Q & 0.1583 & $\mathrm{B}$ $\mathrm{I_{A}}$ $\mathrm{I_{11}}$ $\mathrm{F_{3}}$ $\mathrm{F_{4}}$ \\
 1228+126 & M87 & 187.706 & 12.391 & 2.82 &   & $< 3.17 ^*$ & $< 1.33 \cdot 10^{41}$ & G & 0.00436 &  $\mathrm{F_{3}}$ $\mathrm{F_{4}}$ \\
 1253--055 & 3C 279 & 194.047 & -5.789 & 12.76 & $1.53 \pm 0.21$ & $19.88 \pm 1.76$ & $(1.85 \pm 0.17) \cdot 10^{46}$ & Q & 0.536 & $\mathrm{B}$ $\mathrm{I_{A}}$ $\mathrm{F_{3}}$ $\mathrm{F_{4}}$ \\
 1308+326 & OP 313 & 197.619 & 32.345 & 1.36 &   & $< 2.87 ^*$ & $< 1.06 \cdot 10^{46}$ & Q & 0.997 &  $\mathrm{F_{3}}$ $\mathrm{F_{4}}$ \\
 1324+224 & B2 1324+22 & 201.754 & 22.181 & 0.43 &   & $< 3.11 ^*$ & $< 2.43 \cdot 10^{46}$ & Q & 1.398 &  $\mathrm{F_{3}}$ $\mathrm{F_{4}}$ \\
 1334--127 & PKS 1335--127 & 204.416 & -12.957 & 5.1 & $2.12 \pm 0.68$ & $6.34 \pm 1.72$ & $(7.67 \pm 2.11) \cdot 10^{45}$ & Q & 0.539 & $\mathrm{B}$ $\mathrm{F_{3}}$ $\mathrm{F_{4}}$ \\
 1413+135 & PKS B1413+135 & 213.995 & 13.34 & 1.7 & $1.0 \pm 1.64$ & $2.66 \pm 1.38$ & $(3.94 \pm 2.24) \cdot 10^{44}$ & B & 0.247 &  $\mathrm{F_{3}}$ $\mathrm{F_{4}}$ \\
 1417+385 & B3 1417+385 & 214.944 & 38.363 & 1.37 & $1.76 \pm 1.55$ & $2.22 \pm 1.08$ & $(4.02 \pm 1.95) \cdot 10^{46}$ & Q & 1.831 &  $\mathrm{F_{3}}$ $\mathrm{F_{4}}$ \\
 1458+718 & 3C 309.1 & 224.782 & 71.672 & 3.83 & $2.01 \pm 0.81$ & $3.73 \pm 1.16$ & $(1.53 \pm 0.48) \cdot 10^{46}$ & Q & 0.904 & $\mathrm{B}$ $\mathrm{F_{4}}$\\
 1502+106 & OR 103 & 226.104 & 10.494 & 14.76 &   & $< 17.65 \dagger $ & $< 2.49 \cdot 10^{47}$ & Q & 1.838 &  $\mathrm{F_{3}}$ $\mathrm{F_{4}}$ \\
 1504--166 & PKS 1504--167 & 226.77 & -16.875 & 0.01 &   & $< 4.38 ^*$ & $< 1.21 \cdot 10^{46}$ & Q & 0.876 &  \\
 1510--089 & PKS 1510--08 & 228.211 & -9.1 & 18.74 & $1.24 \pm 0.12$ & $35.88 \pm 1.81$ & $(1.25 \pm 0.07) \cdot 10^{46}$ & Q & 0.36 & $\mathrm{B}$ $\mathrm{F_{3}}$ $\mathrm{F_{4}}$ \\
 1538+149 & 4C +14.60 & 235.206 & 14.796 & 2.34 &   & $< 3.64 ^*$ & $< 4.54 \cdot 10^{45}$ & B & 0.606 &  $\mathrm{F_{3}}$ $\mathrm{F_{4}}$ \\
 1546+027 & PKS 1546+027 & 237.373 & 2.617 & 2.69 &   & $4.29 \pm 1.37$ & $(2.22 \pm 0.71) \cdot 10^{45}$ & Q & 0.414 &  $\mathrm{F_{3}}$ $\mathrm{F_{4}}$ \\
 1548+056 & 4C +05.64 & 237.647 & 5.453 & 1.92 &   & $< 4.03 ^*$ & $< 3.25 \cdot 10^{46}$ & Q & 1.417 &  $\mathrm{F_{3}}$ $\mathrm{F_{4}}$ \\
 1606+106 & 4C +10.45 & 242.193 & 10.485 & 2.33 & $0.81 \pm 0.94$ & $4.69 \pm 1.53$ & $(1.58 \pm 0.61) \cdot 10^{46}$ & Q & 1.232 &  $\mathrm{F_{3}}$ $\mathrm{F_{4}}$ \\
 1611+343 & DA 406 & 243.421 & 34.213 & 1.7 &   & $< 3.18 ^*$ & $< 2.5 \cdot 10^{46}$ & Q & 1.4 &  $\mathrm{F_{3}}$ $\mathrm{F_{4}}$ \\
 1633+382 & 4C +38.41 & 248.815 & 38.135 & 3.92 & $1.59 \pm 0.57$ & $5.76 \pm 1.34$ & $(8.57 \pm 2.0) \cdot 10^{46}$ & Q & 1.814 &  $\mathrm{F_{3}}$ $\mathrm{F_{4}}$ \\
 1637+574 & OS 562 & 249.556 & 57.34 & 3.21 & $1.49 \pm 0.8$ & $4.03 \pm 1.22$ & $(7.81 \pm 2.45) \cdot 10^{45}$ & Q & 0.751 &  $\mathrm{F_{3}}$ $\mathrm{F_{4}}$ \\
 1638+398 & NRAO 512 & 250.123 & 39.779 & 1.15 &   & $< 3.16 ^*$ & $< 3.65 \cdot 10^{46}$ & Q & 1.672 &  $\mathrm{F_{3}}$ $\mathrm{F_{4}}$ \\
 1641+399 & 3C 345 & 250.745 & 39.81 & 7.67 & $1.42 \pm 0.34$ & $9.83 \pm 1.36$ & $(1.09 \pm 0.16) \cdot 10^{46}$ & Q & 0.593 & $\mathrm{B}$ $\mathrm{F_{3}}$ $\mathrm{F_{4}}$ \\
 1655+077 & PKS 1655+077 & 254.538 & 7.691 & 2.65 &   & $< 3.8 ^*$ & $< 4.87 \cdot 10^{45}$ & Q & 0.621 &  \\
 1726+455 & S4 1726+45 & 261.865 & 45.511 & 1.46 & $0.51 \pm 1.08$ & $3.66 \pm 1.3$ & $(3.77 \pm 1.85) \cdot 10^{45}$ & Q & 0.717 &  $\mathrm{F_{3}}$ $\mathrm{F_{4}}$ \\
 1730--130 & NRAO 530 & 263.261 & -13.08 & 3.55 & $1.24 \pm 0.53$ & $6.9 \pm 1.48$ & $(1.72 \pm 0.39) \cdot 10^{46}$ & Q & 0.902 &  $\mathrm{F_{3}}$ $\mathrm{F_{4}}$ \\
 1739+522 & 4C +51.37 & 265.154 & 52.195 & 4.3 & $1.34 \pm 0.61$ & $5.29 \pm 1.31$ & $(3.46 \pm 0.88) \cdot 10^{46}$ & Q & 1.379 &  $\mathrm{B}$ $\mathrm{F_{3}}$ $\mathrm{F_{4}}$ \\
 1741--038 & PKS 1741--03 & 265.995 & -3.835 & 2.92 & $1.15 \pm 0.63$ & $5.79 \pm 1.47$ & $(1.87 \pm 0.5) \cdot 10^{46}$ & Q & 1.054 &  $\mathrm{F_{3}}$ $\mathrm{F_{4}}$ \\
 1749+096 & OT 081 & 267.887 & 9.65 & 3.83 & $1.27 \pm 0.72$ & $4.93 \pm 1.38$ & $(1.36 \pm 0.4) \cdot 10^{45}$ & B & 0.322 &  $\mathrm{F_{3}}$ $\mathrm{F_{4}}$ \\
 1751+288 & B2 1751+28 & 268.427 & 28.801 & 3.76 &   & $< 3.38 ^*$ & $< 1.61 \cdot 10^{46}$ & Q & 1.118 & $\mathrm{F_{4}}$\\
 1758+388 & B3 1758+388B & 270.103 & 38.809 & --0.71 &   & $< 3.24 ^*$ & $< 6.02 \cdot 10^{46}$ & Q & 2.092 &  \\
 1800+440 & S4 1800+44 & 270.385 & 44.073 & 0.49 &   & $< 3.13 ^*$ & $< 4.65 \cdot 10^{45}$ & Q & 0.663 &  $\mathrm{F_{3}}$ $\mathrm{F_{4}}$ \\
 1803+784 & S5 1803+784 & 270.19 & 78.468 & 4.98 & $1.68 \pm 0.6$ & $5.14 \pm 1.29$ & $(8.79 \pm 2.26) \cdot 10^{45}$ & B & 0.6797 &  $\mathrm{B}$ $\mathrm{F_{3}}$ $\mathrm{F_{4}}$ \\
 1807+698 & 3C 371 & 271.711 & 69.824 & 2.72 &   & $3.35 \pm 1.0$ & $(2.02 \pm 0.61) \cdot 10^{43}$ & B & 0.051 &  $\mathrm{F_{3}}$ $\mathrm{F_{4}}$ \\
 1823+568 & 4C +56.27 & 276.029 & 56.85 & 3.31 &   & $< 3.03 ^*$ & $< 4.65 \cdot 10^{45}$ & B & 0.664 &  $\mathrm{F_{3}}$ $\mathrm{F_{4}}$ \\
 1828+487 & 3C 380 & 277.382 & 48.746 & 7.6 & $1.62 \pm 0.34$ & $9.54 \pm 1.35$ & $(1.64 \pm 0.24) \cdot 10^{46}$ & Q & 0.692 & $\mathrm{B}$ $\mathrm{F_{3}}$ $\mathrm{F_{4}}$ \\
 1849+670 & S4 1849+67 & 282.317 & 67.095 & 5.05 & $2.14 \pm 0.85$ & $3.73 \pm 1.19$ & $(7.44 \pm 2.41) \cdot 10^{45}$ & Q & 0.657 &  $\mathrm{B}$ $\mathrm{F_{3}}$ $\mathrm{F_{4}}$ \\
 1928+738 & 4C +73.18 & 291.952 & 73.967 & 7.8 & $2.08 \pm 0.41$ & $7.49 \pm 1.29$ & $(2.24 \pm 0.39) \cdot 10^{45}$ & Q & 0.302 & $\mathrm{B}$ \\
 1936--155 & PKS 1936--15 & 294.861 & -15.429 & --2.86 &   & $< 4.03 ^*$ & $< 4.56 \cdot 10^{46}$ & Q & 1.657 & $\mathrm{F_{4}}$\\
 1957+405 & Cygnus A & 299.868 & 40.734 & 74.71 & $1.92 \pm 0.04$ & $84.77 \pm 1.32$ & $(6.34 \pm 0.1) \cdot 10^{44}$ & G & 0.0561 & $\mathrm{B}$ $\mathrm{I_{A}}$ $\mathrm{I_{11}}$ \\
 1958--179 & PKS 1958--179 & 300.238 & -17.816 & 3.48 &   & $< 4.02 \dagger $ & $< 5.74 \cdot 10^{45}$ & Q & 0.652 &  $\mathrm{F_{3}}$ $\mathrm{F_{4}}$ \\
 2005+403 & TXS 2005+403 & 301.937 & 40.497 & 5.12 & $1.5 \pm 0.44$ & $7.86 \pm 1.41$ & $(9.71 \pm 1.76) \cdot 10^{46}$ & Q & 1.736 &  $\mathrm{B}$ \\
 2008--159 & PKS 2008--159 & 302.815 & -15.778 & 6.7 & $1.96 \pm 0.5$ & $8.19 \pm 1.7$ & $(6.23 \pm 1.3) \cdot 10^{46}$ & Q & 1.18 & $\mathrm{B}$ $\mathrm{F_{4}}$\\
 2021+317 & 4C +31.56 & 305.829 & 31.884 & --0.48 &   & $< 3.51 ^*$ & $< 1.3 \cdot 10^{45}$ & B & 0.356 &  $\mathrm{F_{3}}$ $\mathrm{F_{4}}$ \\
 2021+614 & TXS 2021+614 & 305.528 & 61.616 & 0.97 &   & $< 2.99 ^*$ & $< 4.45 \cdot 10^{44}$ & G & 0.227 &  \\
 2037+511 & 3C 418 & 309.654 & 51.32 & 3.37 & $1.5 \pm 0.93$ & $3.62 \pm 1.25$ & $(4.18 \pm 1.46) \cdot 10^{46}$ & Q & 1.686 &  $\mathrm{F_{3}}$ $\mathrm{F_{4}}$ \\
 2121+053 & OX 036 & 320.935 & 5.589 & 0.56 &   & $< 3.97 ^*$ & $< 6.3 \cdot 10^{46}$ & Q & 1.941 &  $\mathrm{F_{3}}$ $\mathrm{F_{4}}$ \\
 2128--123 & PKS 2128--12 & 322.897 & -12.118 & 2.53 &   & $< 4.3 ^*$ & $< 3.4 \cdot 10^{45}$ & Q & 0.501 &  \\
 2131--021 & 4C --02.81 & 323.543 & -1.888 & 2.15 & $0.72 \pm 1.76$ & $3.06 \pm 1.62$ & $(1.03 \pm 0.71) \cdot 10^{46}$ & Q & 1.284 &  $\mathrm{F_{3}}$ $\mathrm{F_{4}}$ \\
 2134+004 & PKS 2134+004 & 324.161 & 0.698 & 1.04 &   & $< 2.29$ & $< 3.63 \cdot 10^{46}$ & Q & 1.94 & $\mathrm{F_{4}}$\\
 2136+141 & OX 161 & 324.755 & 14.393 & 2.05 &   & $< 3.79 ^*$ & $< 9.63 \cdot 10^{46}$ & Q & 2.427 &  \\
 2145+067 & 4C +06.69 & 327.023 & 6.961 & 9.41 & $1.86 \pm 0.34$ & $12.13 \pm 1.71$ & $(5.73 \pm 0.82) \cdot 10^{46}$ & Q & 0.999 & $\mathrm{B}$ $\mathrm{F_{4}}$\\
 2155--152 & PKS 2155--152 & 329.526 & -15.019 & 1.1 &   & $< 4.22 ^*$ & $< 6.45 \cdot 10^{45}$ & Q & 0.672 &  $\mathrm{F_{3}}$ $\mathrm{F_{4}}$ \\
 2200+420 & BL Lac & 330.68 & 42.278 & 16.79 & $1.69 \pm 0.16$ & $20.87 \pm 1.41$ & $(2.34 \pm 0.16) \cdot 10^{44}$ & B & 0.0686 & $\mathrm{B}$ $\mathrm{I_{A}}$ $\mathrm{F_{3}}$ $\mathrm{F_{4}}$ \\
 2201+171 & PKS 2201+171 & 330.862 & 17.43 & 1.59 &   & $< 3.71 ^*$ & $< 1.63 \cdot 10^{46}$ & Q & 1.076 &  $\mathrm{F_{3}}$ $\mathrm{F_{4}}$ \\
 2201+315 & 4C +31.63 & 330.812 & 31.761 & 5.89 & $1.94 \pm 0.54$ & $6.33 \pm 1.44$ & $(1.73 \pm 0.4) \cdot 10^{45}$ & Q & 0.2947 & $\mathrm{B}$ $\mathrm{F_{3}}$\\
 2209+236 & PKS 2209+236 & 333.025 & 23.928 & 2.5 & $1.91 \pm 1.71$ & $2.43 \pm 1.27$ & $(1.58 \pm 0.84) \cdot 10^{46}$ & Q & 1.125 &  $\mathrm{F_{3}}$ $\mathrm{F_{4}}$ \\
 2216--038 & PKS 2216--03 & 334.717 & -3.594 & 2.77 &   & $< 3.98 ^*$ & $< 1.17 \cdot 10^{46}$ & Q & 0.901 & $\mathrm{F_{4}}$\\
 2223--052 & 3C 446 & 336.447 & -4.95 & 1.86 & $1.11 \pm 1.59$ & $3.14 \pm 1.59$ & $(1.74 \pm 0.95) \cdot 10^{46}$ & Q & 1.404 &  $\mathrm{F_{3}}$ $\mathrm{F_{4}}$ \\
 2227--088 & PHL 5225 & 337.417 & -8.548 & 6.06 & $1.13 \pm 0.47$ & $9.73 \pm 1.81$ & $(6.75 \pm 1.3) \cdot 10^{46}$ & Q & 1.56 & $\mathrm{B}$ $\mathrm{F_{3}}$ $\mathrm{F_{4}}$ \\
 2230+114 & CTA 102 & 338.152 & 11.731 & 10.11 & $1.42 \pm 0.26$ & $15.89 \pm 1.67$ & $(6.01 \pm 0.65) \cdot 10^{46}$ & Q & 1.037 & $\mathrm{B}$ $\mathrm{F_{3}}$ $\mathrm{F_{4}}$ \\
 2243--123 & PKS 2243--123 & 341.576 & -12.114 & 0.75 &   & $< 3.96 ^*$ & $< 5.28 \cdot 10^{45}$ & Q & 0.632 &  \\
 2251+158 & 3C 454.3 & 343.491 & 16.148 & 55.16 & $1.52 \pm 0.05$ & $74.34 \pm 1.56$ & $(1.98 \pm 0.04) \cdot 10^{47}$ & Q & 0.859 & $\mathrm{B}$ $\mathrm{I_{A}}$ $\mathrm{I_{11}}$ $\mathrm{F_{3}}$ $\mathrm{F_{4}}$ \\
 2331+073 & TXS 2331+073 & 353.553 & 7.608 & 1.76 &   & $< 3.75 ^*$ & $< 1.81 \cdot 10^{45}$ & Q & 0.401 &  $\mathrm{F_{3}}$ $\mathrm{F_{4}}$ \\
 2345--167 & PKS 2345--16 & 357.011 & -16.52 & --0.14 &   & $< 3.65 ^*$ & $< 3.94 \cdot 10^{45}$ & Q & 0.576 &  $\mathrm{F_{3}}$ $\mathrm{F_{4}}$ \\
 2351+456 & 4C +45.51 & 358.59 & 45.885 & 0.81 &   & $< 3.25 ^*$ & $< 5.41 \cdot 10^{46}$ & Q & 1.986 &  $\mathrm{F_{3}}$ $\mathrm{F_{4}}$ \\

\end{longtable}

\tablefoot{
\tablefoottext{a}{Name in IAU B1950 format}
\tablefoottext{b}{Equatorial coordinates (J2000) in degrees.}
\tablefoottext{c}{\textit{Swift}/BAT S/N (105 month survey data)}
\tablefoottext{d}{Photon Index for 20\,keV -- 100\,keV}
\tablefoottext{e}{X-ray flux in $10^{-12}\mathrm{ergs\,s^{-1}\,cm^{-2}}$ (* indicates fluxes calculated by assuming a photon index based on sub-sample of bright sources; $\dagger$ for upper limits caused by contamination of a nearby source)}
\tablefoottext{f}{X-ray luminosity in $\mathrm{erg\,s^{-1}}$}
\tablefoottext{g}{Optical classification \citep{Veron2003}: Q: Flat Spectrum Radio Quasar, B: BL Lac, G: Radio Galaxy, U: Unidentified}
\tablefoottext{h}{Redshift, taken from the MOJAVE webpage (http://www.physics.purdue.edu/astro/MOJAVE/), * indicates the redshift value from \citep{Lister2015}}
\tablefoottext{i}{Published hard X-ray / gamma-ray detection in: $\mathrm{B}$: \textit{Swift}/BAT 105-month survey catalog \citep{Oh2018}, $\mathrm{I_{A}}$: \textit{INTEGRAL}/IBIS AGN catalog \citep{Malizia2012}, $\mathrm{I_{11}}$: \textit{INTEGRAL} 11-year Hard X-ray Survey \citep{Krivonos2015}, $\mathrm{F_{3}}$: \textit{Fermi}/LAT 3LAC catalog \citep{Ackermann2015}, and $\mathrm{F_{4}}$: 4FGL catalog \citep{4FGL2019}. $\ddagger$: the source has been re-classified as a NLSY1 galaxy by \citet{Yao2015}. We keep the original classification of Q (see Sect.~\ref{sec:sam})}
}
\end{landscape}

\twocolumn
\end{document}